\documentclass[review]{elsarticle}

\makeatletter
\def\ps@pprintTitle{%
 \let\@oddhead\@empty
 \let\@evenhead\@empty
 \def\@oddfoot{\centerline{\thepage}}%
 \let\@evenfoot\@oddfoot}
\makeatother

\AtBeginDocument{%
  \def\thefnote{\myfnsymbol{fnote}}}
\makeatletter
\def\myfnsymbol#1{\expandafter\@myfnsymbol\csname c@#1\endcsname}
\def\@myfnsymbol#1{\ifcase #1\or $\dagger$\or $\#\#$\else \@ctrerr\fi}
\def\fntext[#1]#2{\g@addto@macro\@fnotes{%
   \refstepcounter{fnote}\elsLabel{#1}%
   \def\thefootnote{\thefnote}
   \global\setcounter{footnote}{\c@fnote}%
   \footnotetext{#2}}}
\makeatother

\pdfoutput=1
\usepackage{graphicx}
\usepackage{amsfonts, amsmath, amssymb}
\usepackage{mathabx}
\usepackage{booktabs,siunitx}
\usepackage[svgnames,table]{xcolor}
\usepackage[tableposition=above]{caption}
\usepackage{pifont}
\usepackage{bm}
\usepackage{cancel}
\usepackage{caption}
\usepackage{color}
\usepackage{enumerate}
\usepackage{float}
\usepackage{hyperref}
\usepackage[english]{babel}
\usepackage[margin=1in]{geometry}
\usepackage{mathtools}
\usepackage{multirow}
\usepackage{setspace}
\usepackage{subfigure}
\usepackage{lineno}
\usepackage[ruled,linesnumbered]{algorithm2e}
\RestyleAlgo{ruled}
\SetKwComment{Comment}{/* }{ */}

\SetCommentSty{mycommfont}

\renewcommand \d [2]{\frac{{\rm d} #1}{{\rm d} #2}}
\renewcommand \d [1]{\text{d} #1}
\newcommand \D [2]{\frac{\partial #1}{\partial #2}}

\renewcommand{\vec}[1]{\bm{\mathrm{#1}}}
\newcommand{\V}[1]{\bm{\mathrm{#1}}}

\def \div{\nabla \cdot \mbox{}}
\def \grad{\nabla}

\def \x{\vec{x}}

\def \n{\vec{n}}

\def \u{\vec{u}}

\def \I{\vec{I}}

\def \U{\vec{U}}
\def \L{\vec{L}}

\def \cM{\V{\mathcal{M}}}

 \def \defn{{\overset{\triangle}{=}}}

\def \A{\vec{A}}
\def \b{\vec{b}}

\def \C{\vec{C}}
\def \ddt{\frac{\text{d}}{\text{d}t}}
\def \dOmega{\text{dV}}
\def \dS{\text{dS}}
\def \vD{\vec{D}}
\def \depsdt{\frac{\text{d}\varepsilon}{\text{d}t}}
\def \cF{\vec{\mathcal{F}}}
\def \vecPsi{{\bf{\Psi}}}

\def \g{\vec{g}}
\def \G{\vec{G}}

\def \I{\V{I}}
\def \Ib{\I_{\text{b}}}

\def \Lmu{\vec{L_{\mu}}}
\def \vrho{\vec{\rho}}

\def \M{\vec{M}}
\def \Mb{\text{M}_{\text{b}}}

\def \Nx{N_x}
\def \Ny{N_y}
\def \Nz{N_z}

\def \OmegaFt{\Omega^\text{F}(t)}

\def \OmegaFt{\Omega^\text{F}(t)}
\def \OmegaF{\Omega^\text{F}}
\def \dOmegaFt{\partial \Omega^\text{F}(t)}
\def \dOmegaF{\partial \Omega^\text{F}}
\def \dOmegaSt{\partial \Omega^\text{S}(t)}
\def \dOmegaS{\partial \Omega^\text{S}}

\def \U{\vec{U}}

\def \Ur{\U_{\text{r}}}
\def \W{\vec{W}}

\def \Wr{\W_{\text{r}}}

\def \X{\vec{X}}
\def \Xcom{\X_{\text{COM}}}

\def \f{\vec{f}}

\def \fc{\f_{\text{c}}}

\def \fst{\f_{\text{st}}}

\def \half{\frac{1}{2}}
\def \3half{\frac{3}{2}}
\def \5half{\frac{5}{2}}

\def \n{\vec{n}}

\def \ns{\n_{\text{s}}}

\def \ncells{n_{\text{cells}}}
\def \ncycles{n_{\text{cycles}}}

\def \dphidt{\frac{\partial \phi}{\partial t}}
\def \mV{\mathcal{V}}
\def \mA{\mathcal{A}}

\def \rhol{\rho_{\text{l}}}

\def \rhog{\rho_{\text{g}}}

\def \u{\vec{u}}
\def \ub{\u_{\text{b}}}

\def \us{\vec{u}_{\text{s}}}

\def \varn{\hat{\delta}}

\def \bu{\breve{\u}}

\def \vs{\vec{v}_{\text{s}}}

\def \x{\vec{x}}
\def \xu{\vec{x_u}}
\def \xp{\vec{x_p}}

\def \dzetadt{\frac{\partial \zeta}{\partial t}}

\def \div{\nabla \cdot \mbox{}}
\def \grad{\nabla}

\def \dt{\Delta t}
\def \dx{\Delta x}
\def \dy{\Delta y}
\def \dz{\Delta z}

\newcommand{\upperRomannumeral}[1]{\uppercase\expandafter{\romannumeral#1}}

\newcommand{\KK}[1]{{ #1}}
\newcommand{\ADDITION}[1]{{ #1}}

\begin{document}
\let\today\relax

\begin{frontmatter}
	
\title{Preventing mass loss in the standard level set method: New insights from variational analyses}
\author[SDSU]{Kaustubh Khedkar\fnref{eq_contrib}}
\author[SDSU]{Amirreza Charchi Mamaghani}
\author[LBNL]{Pieter Ghysels}
\author[Northwestern]{Neelesh A. Patankar\corref{mycorrespondingauthor}}
\ead{n-patankar@northwestern.edu}
\author[SDSU]{Amneet Pal Singh Bhalla\corref{mycorrespondingauthor}\fnref{eq_contrib}}
\ead{asbhalla@sdsu.edu}

\address[SDSU]{Department of Mechanical Engineering, San Diego State University, San Diego, CA}
\address[Northwestern]{Department of Mechanical Engineering, Northwestern University, Evanston, IL}
\address[LBNL]{Scalable Solvers Group, Lawrence Berkeley National Laboratory, Berkeley, CA}
\cortext[mycorrespondingauthor]{Corresponding author}
\fntext[eq_contrib]{These authors contributed equally}

\begin{abstract}
For decades, the computational multiphase flow community has grappled with mass loss in the level set method. Numerous solutions have been proposed, from fixing the reinitialization step to combining the level set method with other conservative schemes. However, our work reveals a more fundamental culprit: the smooth Heaviside and delta functions inherent to the standard formulation. Even if reinitialization is done exactly, i.e., the zero contour interface remains stationary, the use of smooth functions lead to violation of mass conservation. 
We propose a novel approach using variational analysis to incorporate a mass conservation constraint.  This introduces a Lagrange multiplier that enforces overall mass balance. Notably, as the delta function sharpens, i.e., approaches the Dirac delta limit, the Lagrange multiplier approaches zero. However, the exact Lagrange multiplier method disrupts the signed distance  property of the level set function. This motivates us to develop an approximate version of the Lagrange multiplier that preserves both overall mass and signed distance property of the level set function. Our framework even recovers existing mass-conserving level set methods, revealing some inconsistencies in prior analyses. We extend this approach to three-phase flows for fluid-structure interaction (FSI) simulations. We present variational equations in both immersed and non-immersed forms, demonstrating the convergence of the former formulation to the latter when the body delta function sharpens. Rigorous test problems confirm that the FSI dynamics produced by our simple, easy-to-implement immersed formulation with the approximate Lagrange multiplier method are accurate and match state-of-the-art solvers. 
\end{abstract}


\end{frontmatter}

\section{Introduction}
Following Sussman and colleagues' seminal work~\cite{Sussman1994} in the early 1990s, the level set method for modeling multiphase flows gained popularity. As an alternative to the volume of fluid (VOF) method that was prevalent at the time, it offered many advantages. It provides a continuous representation of the interface, accurate computation of geometric quantities such as the curvature and normal of the interface, as well as ease of implementation, especially in three dimensions.  The level set method lends itself naturally to finite element~\cite{Kees2011, Hysing2012, Nagrath2005} and higher order discretization schemes~\cite{Nourgaliev2007} because it is based on partial differential equations. VOF, on the other hand, is mostly limited to codes using second order finite difference/volume schemes. Level set method produces spurious mass gain/loss of phases over time, which is its main disadvantage.   

There has been extensive research on the mass loss issue associated with the level set method. The main culprit has been identified as the level set reinitialization equation, which restores the signed distance function (SDF) property. The SDF property is disrupted by the advection of the level set function. Many fixes have been proposed in the literature to limit the motion of the interface (represented as the zero contour of the level set) during reinitialization. Solomenko et al.~\cite{Solomenko2017} compare many of these fixes proposed in the literature for the reinitialization equation on benchmarking problems. While some fixes perform better than others, the level set method still loses a substantial amount of mass over time, particularly as complex motions occur at the interface. 

Another approach involves hybridizing the level set method with mass-conserving schemes like VOF~\cite{Sussman2000} (or its advanced variant, the moment of fluid (MOF) method~\cite{Dyadechko2008, Ahn2009, Taek2009}) or Lagrangian particles~\cite{Enright2002}. These methods track the interface using both the level set function and a conservative method, which corrects the former for mass loss. However, such hybrid methods increase complexity, computational cost, and lose the original simplicity of the level set method.

We revisit the root cause of mass loss in the level set method. While reinitialization can contribute, our findings show that the primary culprit lies in the use of smooth Heaviside and delta functions within the standard formulation. Even if reinitialization is done exactly, i.e., the zero contour interface remains stationary, the standard level set method violates the mass conservation principle. 
Based on this insight, we propose a novel approach – incorporating a mass conservation constraint into the level set equation using variational analysis. This approach introduces a Lagrange multiplier to enforce mass/volume conservation within the two-phase level set framework. As the smooth delta function becomes sharper, the Lagrange multiplier approaches zero. However, the exact Lagrange multiplier method disrupts the signed distance property to conserve mass. This motivates us to develop an approximate Lagrange multiplier method that preserves both properties. Although they are derived differently, exact and approximate Lagrange multipliers are related. Furthermore, our framework can be used to derive existing mass-conserving two-phase level set methods~\cite{Wen2023,Kees2011}, revealing some inconsistencies in the previous analyses~\cite{Wen2023}.

We extend our variational analysis to three-phase flows, enabling fluid-structure interaction (FSI) simulations in two-fluid systems using the fictitious domain Brinkman penalization (FD/BP) technique. We present variational equations for three-phase flows in both non-immersed and immersed forms. As the diffuse body delta function in the immersed formulation sharpens, it converges to the non-immersed form. In this case, both formulations lead to identical exact and approximate Lagrange multipliers. To assess the immersed formulation's practical performance compared to the non-immersed one, we design rigorous test problems for three-phase flows. The immersed formulation is tested using the FD/BP method, while the non-immersed approach utilizes inherently mass-conserving techniques such as the geometric VOF technique with moving unstructured grids and cut-cell methods, and particle-based hydrodynamics methods. We demonstrate that our (simple to implement) immersed formulation in conjunction with the (approximate) Lagrange multiplier method produces FSI dynamics that match very well with the other state-of-the-art solvers.

\section{Mathematical framework}
\label{sec_cont_eqs}


 We begin by stating the continuous equations of motion for the multiphase flow system. This includes a continuous description of the level-set interface tracking method and its reasons for mass loss in the context of two phase flows. Next, a new variational analysis is presented, which introduces a Lagrange multiplier to impose mass/volume conservation constraints with the level set method. Additionally, we extend the two phase variational analysis to three phase flows, which allows us to simulate fluid structure interaction (FSI) in the presence of two fluids using fictitious domain Brinkman penalization (FD/BP) technique. Variational equations for three-phase flows are presented in non-immersed and immersed forms.

\subsection{Continuous equations of motion}

We use the single fluid formulation~\cite{tryggvason2011direct} for multiphase flows, which considers a single viscous incompressible fluid with spatially and temporally varying density $\rho(\x,t) $ and viscosity $\mu(\x,t)$ in a fixed region of space $\Omega \subset \mathbb{R}^d$, where $d = 3$ represents a three-dimensional region in space. The equations of motion for an incompressible fluid are given by the Navier-Stokes equations,  which in conservative form read as

\begin{align}
	 \D{\rho \u(\x,t)}{t} + \div (\rho \, \u(\x,t) \otimes \u(\x,t)) &= -\grad p(\x,t) + \div \left[\mu \left(\grad \u(\x,t) + \grad \u(\x,t)^\intercal\right) \right]+ \fst + \fc + \rho\g, \label{eqn_momentum}\\
	 \div \u(\x,t) &= 0. 
 \label{eqn_continuity} 
\end{align}

Eq.~\eqref{eqn_momentum} describes the momentum of the system and Eq.~\eqref{eqn_continuity} expresses the incompressibility of various phases present in the system. In the above equations, $\u(\x,t)$ and $p(\x,t)$ denote the Eulerian velocity and pressure fields, respectively,  $\x = (x,y,z) \in \mathbb{R}^3$, $\fst$ denotes the continuous surface tension force along the gas-liquid interface, and $\fc$ is the Brinkman penalty term that imposes a rigid body velocity in the solid domain. For two phase problems the 
$\fc$ term is absent from Eq.~\eqref{eqn_momentum}. The specific forms of $\fst$ and $\fc$ will be provided later in Sec.~\ref{sec_temp_disc}. The acceleration due to gravity is directed towards the negative z-direction, $\g = (0, 0, -g)$.

\subsection{Interface tracking}
\label{subsec_interface_track}

\begin{figure}[]
   \centering
   	\includegraphics[scale= 0.4]{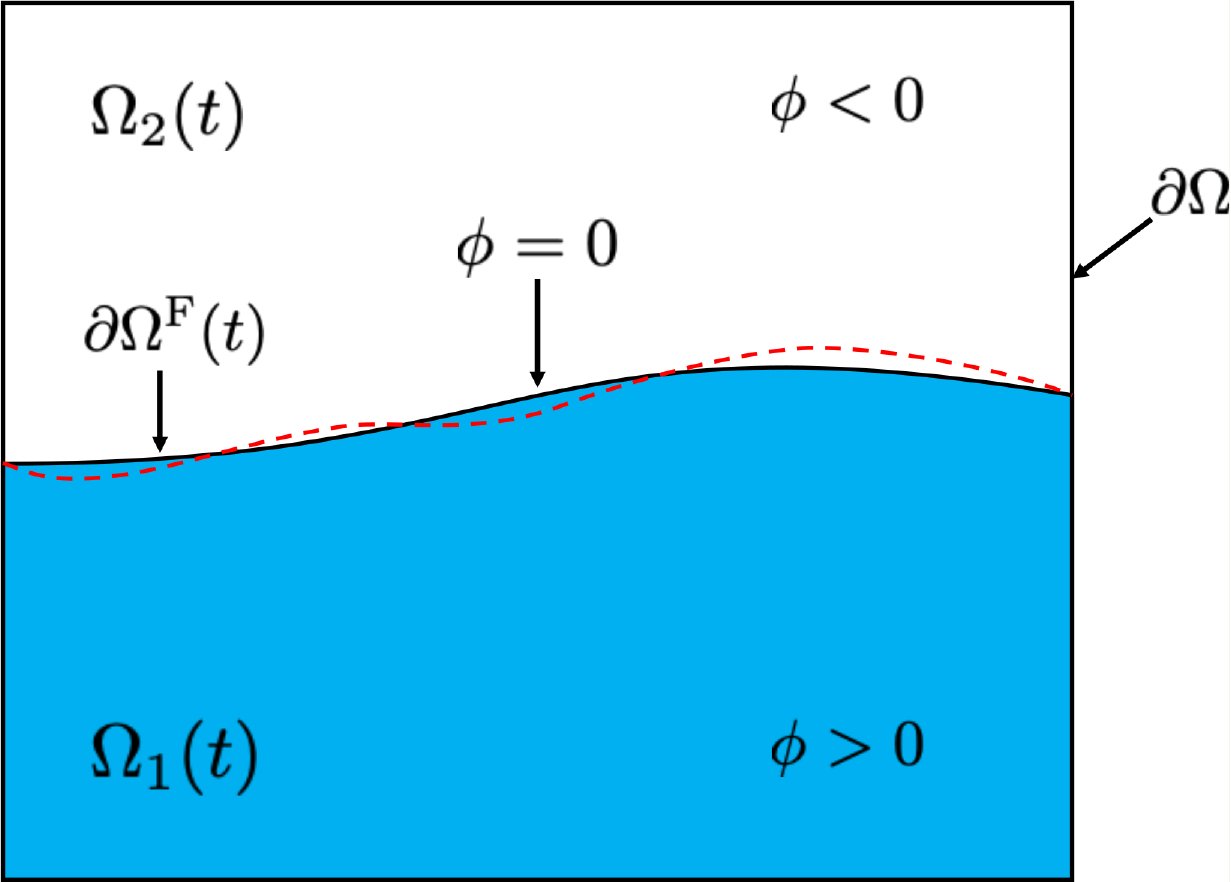}
    \caption{Schematic of the computational domain $\Omega$ for two phase flows depicting the time evolving interface $\dOmegaF(t)$. The variation of the two-fluid interface $\dOmegaF$ is illustrated by the red dashed line. For two phase flows $\Omega_1(t) \cup \Omega_2(t) = \Omega$.}
   \label{fig_schematic_2_phase_flows}
\end{figure}

The interface $\dOmegaFt$ between two fluids is captured implicitly by the zero-contour of the level set function $\phi(\x,t)$, which denotes the distance of a fixed location $\x \in \Omega$ from the time-evolving interface with a sign. At time $t$, fluid-1 and fluid-2 occupy non-overlapping regions $\Omega_1(t)$ and $\Omega_2(t)$, respectively, such that $\Omega_1(t) \cup \Omega_2(t) = \Omega$; see Fig.~\ref{fig_schematic_2_phase_flows}. In our sign convection, $\phi > 0$ in $\Omega_1(t)$ and $\phi<0$ in $\Omega_2(t)$. In the absence of mass transfer across the interface, the interface moves with the local fluid velocity $\u(\x,t)$, which can be described by an advection equation of the form

\begin{equation}
	\dphidt + \u \cdot \grad \phi = 0.
	\label{eq_phi_advection}
\end{equation}

The density and viscosity in the computational domain is set using the signed distance function $\phi(\x,t)$ 

\begin{align}
	\rho(\phi) &= \rho_1 H(\phi) + \rho_2\left(1 - H(\phi) \right), \label{eq_den_mixture_model} \\
	\mu(\phi) &= \mu_1 H(\phi) + \mu_2\left(1 - H(\phi) \right), \label{eq_vis_mixture_model}
\end{align}
in which, $\rho_1$ and $\rho_2$, and $\mu_1$ and $\mu_2$, are the density and viscosity of fluid-1 and fluid-2, respectively.  $H(\phi)$ is a smoothed Heaviside function, which allows material properties to vary smoothly over $n_\text{cells}$ grid cells for numerical stability, see Eqs.~\eqref{eqn_Hflow} and~\eqref{eqn_Hbody}. The derivative of the smoothed Heaviside function produces a smoothed delta function $\delta(\phi)$. 




It is well known that the level set field does not maintain the signed distance property after advection~\cite{Sussman1994, Sussman1998}. \ADDITION{Although this fact is widely known, its proof is not readily available in the literature (to our knowledge). The following derivation shows that $\phi$'s signed distance property is lost at the continuous level under advection. This is not purely a numerical artifact. 

By taking gradient of Eq.~\eqref{eq_phi_advection} we get

\begin{align}   
    &\frac{\partial \grad \phi}{\partial t} + \grad \left(\u \cdot \grad \phi \right) = 0,  \nonumber \\
    \hookrightarrow &\frac{\partial \grad \phi}{\partial t} + \grad \u \cdot \grad \phi + \u \cdot \grad \left(\grad \phi \right) = 0.
    \label{eq_grad_phi_advection}
\end{align}
Defining $\vecPsi \equiv \grad \phi$ in Eq.~\eqref{eq_grad_phi_advection}, and taking a dot product with $\vecPsi$ on both sides, we obtain
\begin{align}   
    &\vecPsi \cdot \left[\frac{\partial \vecPsi}{\partial t} + \grad \u \cdot \vecPsi + \u \cdot \grad \vecPsi \right] = 0, \nonumber \\
    \hookrightarrow & \vecPsi \cdot \frac{\partial \vecPsi}{\partial t} + \vecPsi \cdot \grad \vecPsi \cdot \u = -\vecPsi \cdot \grad \u \cdot \vecPsi.
\end{align}
Using the relations $\vecPsi \cdot \partial \vecPsi / \partial t = \frac{1}{2}  \frac{\partial (\vecPsi \cdot \vecPsi)} {\partial t}   = |\grad \phi| \frac{\partial |\grad \phi|} {\partial t} $ and $\vecPsi \cdot \grad \vecPsi \cdot \u = \frac{\u \cdot \grad(\vecPsi \cdot \vecPsi)}{2}  = |\grad \phi| \u \cdot \grad(|\grad \phi|) $, and simplifying further yields a transport equation for $|\grad \phi|$
\begin{equation}   
    \frac{\partial |\grad \phi|}{\partial t} + \u \cdot \grad \left(|\grad \phi| \right) =  \frac{-\grad \phi \cdot \grad \u \cdot \grad \phi}{|\grad \phi|}. \label{eq_sdf_change} 
\end{equation}
The right hand side of Eq.~\eqref{eq_sdf_change} has a non-zero value if the symmetric part of the velocity gradient tensor, $\V{S} = \frac{1}{2}(\grad \u + \grad \u^\intercal)$, is not zero. When the velocity field is a rigid body motion (translation and rotation), the right hand side of Eq.~\eqref{eq_sdf_change} will be zero, and $\phi$ will preserve its signed distance property under advection. However, in general, an advective transport of $\phi$ disrupts its signed distance property. 
}

Retaining the signed distance property is numerically advantageous as the advected $\phi$ is used to prescribe various material properties for the one-fluid model and to calculate geometric quantities like the interface curvature and surface normals. To regain the signed distance property of $\phi$, a reinitialization step is typically performed after the advection step. The reinitialization step involves time-advancing the Hamilton-Jacobi equation to steady state
\begin{equation}
	\frac{\partial \phi}{\partial \tau} + \text{sgn}(\tilde{\phi})\left( |\grad \phi| - 1 \right) = 0.
	\label{eq_HJ}
\end{equation}
Here, $\tilde{\phi}$ denotes the level set field before reinitialization and $\tau$ is the pseudo time used for time-marching Eq.~\eqref{eq_HJ}. It is widely believed that the reinitialization step leads to mass loss because it artificially shifts the zero-contour or the two phase interface $\dOmegaF$.  In the next section, we show that is not entirely correct: the level set method would still lead to mass loss even if the interface remains static during the reinitialization step. 

\subsection{Precise reason for mass loss with the standard level set method}
\label{subsec_reasons_for_mass_loss}

It is helpful to note down the Leibniz integral rule and Reynolds transport theorem (RTT) for performing differentiation under the integral sign for a general integrand $f$ before working out the details of mass loss. We use both these rules several times in this and other sections. The Leibniz integral rule reads as
\begin{subequations} \label{eq_Leibniz}
\begin{alignat}{2}
&\text{Leibniz rule for a moving domain:} \qquad && 	\ddt \int_{\mV(t)} f~\text{d} \mV = \int_{\mV(t)} \frac{\partial f}{\partial t}~\d \mV +  \int_{\mA (t) } \mathit{f}~(\us \cdot \n)~\text{d} \mA,  \label{eq_Leibniz_1} \\
&\text{Leibniz rule for a static domain:} \qquad && \ddt \int_{\mV} f~\text{d} \mV = \int_{\mV} \frac{\partial f}{\partial t}~\d \mV, \label{eq_Leibniz_2}
\end{alignat}
\end{subequations} 
in which, $\n$ and $\us$ are the outward unit normal vector and velocity of the surface $\mA(t)$ enclosing the time-varying domain $\mV(t)$, respectively. When the domain is stationary $\mV(t) \equiv \mV$, Eq.~\eqref{eq_Leibniz_1} becomes Eq.~\eqref{eq_Leibniz_2}. Note that the region $\mV(t)$ is a geometric region that moves with a kinematic velocity $\us$, which in general is different from the fluid/material velocity $\u$. The Leibniz integral rule should not be confused with the Reynolds transport theorem. The two are related, but different because the latter uses a material control volume instead of a geometric one. A material control volume is defined as the one whose surface moves with the fluid/material velocity $\u$. Thus, a material control volume is a special case of a geometric volume. It is more restrictive because the amount of matter contained inside it cannot change over time as long as the fluid at every point within the volume moves with material velocity $\u$ governed by the conservation laws (mass and momentum). In general, a geometric control volume does not conserve the amount of matter within itself. The two theorems/rules are related to each other as
\begin{align}
&\text{RTT for a material volume:} && \ddt \int_{\mV(t)} f~\text{d} \mV = \int_{\mV(t)} \frac{\partial f}{\partial t}~\d \mV +  \int_{\mA(t)} \mathit{f}~(\u \cdot \n)~\d \mA. \label{eq_RTT}
\end{align} 
Comparing Eq.~\eqref{eq_RTT} to~\eqref{eq_Leibniz_1}  it can be seen that the RTT is a Leibniz integral rule applied to a material control volume (whose surface moves with the material velocity) to express the rate of change of a conserved quantity. In this work, we leverage both the RTT and the Leibniz integral rule to express the rate of change of conserved and non-conserved quantities, respectively.  This distinction clarifies the context and avoids potential confusion, as the terms RTT and Leibniz integral rule are sometimes used interchangeably in the literature.       

To see why the standard level set method leads to mass loss, consider the rate of change of mass (which is a conserved quantity) contained within a closed domain $\Omega$ 

\begin{subequations} 
\begin{alignat}{2}
 \ddt \int_{\Omega} \rho~\dOmega  &= \ddt \int_{\Omega} \bigl\{\rho_1 H(\phi) + \rho_2 (1 - H(\phi)) \bigr\}~\dOmega,  \\
                                                        & = (\rho_1 - \rho_2)  \int_{\Omega} \delta(\phi) \dphidt~\dOmega,    \\
                                                        & = (\rho_1 - \rho_2)  \int_{\Omega} \delta(\phi) (-\bu \cdot \grad \phi)~\dOmega,  \label{eqn_ml_proof}
\end{alignat}
\end{subequations}  
 in which we used the mixture model of density (Eq.~\eqref{eq_den_mixture_model}),  RTT (Eq.~\eqref{eq_RTT}) with zero material velocity at the domain boundary, and the level set advection equation (Eq.~\eqref{eq_phi_advection}) in arriving at Eq.~\eqref{eqn_ml_proof}. During the reinitialization process, the contours of the level set function surrounding the interface $\dOmegaF$ move with velocity $\u^{\rm reinit}  = \text{sgn}(\tilde{\phi}) \frac{\grad \phi}{|\grad \phi|}$ and adjust themselves to satisfy the Eikonal property $|\grad \phi| = 1$. Overall $\phi$ gets advected with a combination of $\u$ and $\u^{\rm reinit}$, which is denoted by $\bu$ in Eq.~\eqref{eqn_ml_proof}. 
 
 
 To ensure mass conservation, the term on the right hand side (RHS) of Eq.~\eqref{eqn_ml_proof} must equal zero. This is possible if and only if  (i) $\delta(\phi)$ is the Dirac/sharp delta function and (ii) $\u^{\rm reinit} \equiv 0$ for $\dOmegaF$. In this case $\int_{\Omega} \delta(\phi)(\bu \cdot \grad \phi)~\dOmega   = \int_{\dOmegaF} \u \cdot |\grad \phi | \n~\dS =    \int_{\dOmegaF} \u \cdot \n~\dS$ represents the net normal (advective) velocity of the interface. For a closed domain, this surface integral is zero. We prove this identity at the end of this section. However, if  $\delta$ is smooth, which is almost always the case in numerical simulations, the RHS of Eq.~\eqref{eqn_ml_proof} is non-zero, even when $\dOmegaF$ stays stationary during the reinitialization process. Thus, the smooth Heaviside and delta functions  are the ``main culprits" that lead to spurious mass loss/gain in the level set method from continuous equations point of view. This also explains why previous methods that have solely aimed to ``fix" the reinitialization equation have not yielded satisfactory results for curbing the mass loss with the level set method.   
 
 \begin{figure}[]
  \centering
  \subfigure[A closed interface]{
  	\includegraphics[scale= 0.41]{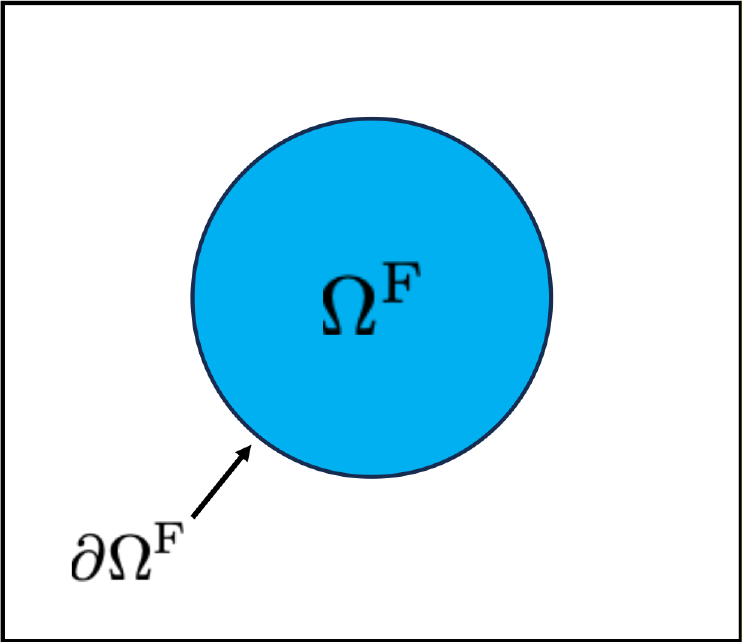}
	\label{fig_closed_interface}
  }
   \subfigure[An open interface]{
  	\includegraphics[scale= 0.3]{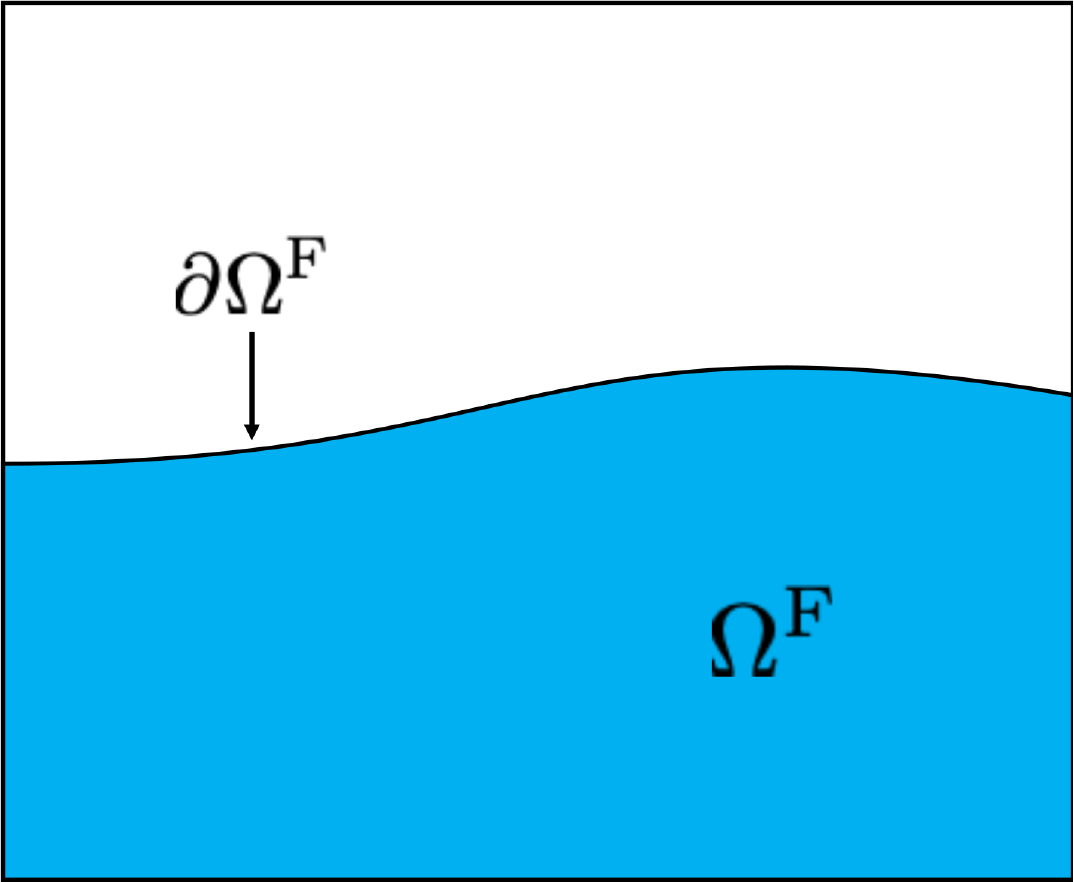}
	\label{fig_open_interface}
  }  
  \caption{Two possible configurations of a two-phase interface $\dOmegaF$ in a closed domain $\Omega$.}
  \label{fig_closed_open_interface}
\end{figure}
 
Eq.~\eqref{eqn_ml_proof} helps explain why level set method continues to lose/gain mass over time. To see this integrate Eq.~\eqref{eqn_ml_proof} over a time period $\Delta t = t_{\rm f} - t_{\rm i}$. A straightforward integration shows that the spurious mass $\Delta M$ gained/lost over a period $\Delta t$ is proportional to the interfacial region displacement:
\begin{align}
\Delta M & = (\rho_1 - \rho_2) \int_{\Delta t} \int_{\Omega} \delta(\phi) (-\bu \cdot \grad \phi)~\dOmega~\d t, \nonumber \\
         & = (\rho_1 - \rho_2) \int_{\Delta t} \int_{\Omega} \delta(\phi) (-\bu  \cdot |\grad \phi| \n)~\dOmega~\d t,\nonumber \\
         & = -(\rho_1 - \rho_2) \int_{\Delta t} \int_{\Omega} \delta(\phi)\,  \breve{u}_\text{n}~\dOmega~\d t. \label{eq_deltaM_slsm}
\end{align}
In numerical experiments, it is commonly observed that problems with large interface motion perform poorly in terms of mass conservation. The culprit is the smeared delta function that picks up the ``residual" displacement around $\dOmegaF$. In Sec.~\ref{sec_2phase_cases} we devise a special test problem that verifies Eq.~\eqref{eq_deltaM_slsm}.    
 
To prove that $\int_{\dOmegaF} \u \cdot \n~\dS = 0$ holds for a two-phase interface, consider Fig.~\ref{fig_closed_open_interface}, which depicts two possible configurations of the interface $\dOmegaF$ in a closed domain $\Omega$. An application of the Gauss-divergence theorem to a closed interface (shown in Fig.~\ref{fig_closed_interface}) results in $\int_{\dOmegaF} \u \cdot \n~\dS = \int_{\OmegaF}\div \u~\dOmega$. The identity is proved by the fact that velocity $\u$ is divergence-free. An open interface is only possible if its two ends touch the domain boundary; see Fig.~\ref{fig_open_interface}. Applying the Gauss-divergence theorem to the integrand $\div \u$ defined over the domain and interface bounded region $\OmegaF$ proves the identity.  

\subsection{Conserving mass for two phase flows with the standard level set method: a variational/constraint formulation}
\label{subsec_mass_conservation_LSM}


To conserve mass with the standard level set method we need to explicitly add in a constraint to the level set advection equation. A variational approach is followed that introduces a Lagrange multiplier that enforces mass conservation. The mass of fluid-1 and its rate of change can be computed from $\phi$ as 
\begin{subequations} 
\begin{alignat}{2}
 & M_1(t) &&= \int_{\Omega} \rho_1 H(\phi)~\dOmega, \label{eq_M1} \\
& \frac{\text{d}M_1}{\text{d}t} &&= \ddt \int_{\Omega} \rho_1 H(\phi)~\dOmega = \int_{\Omega} \rho_1 \delta(\phi) \dphidt~\dOmega.  \label{eq_dM1dt}
\end{alignat}
\end{subequations} 
Here, we made use of the RTT in deriving Eq.~\eqref{eq_dM1dt}.  If the mass of fluid-1 is conserved then $M_1(t) \equiv M_1^0 \; \forall \; t$, in which $M_1^0$ is the initial mass of fluid-1.  For simplicity we have assumed that the domain is closed and there is no inflow and outflow of the phases. Thus, conserving mass of fluid-1 is equivalent to imposing a constraint on $\phi$ of the form
\begin{equation}
\mathcal{C}(\phi)~~\defn~~\int_{\Omega} \left( \rho_1 H(\phi) -  \frac{M_1^0}{V} \right)~\dOmega = 0, 
	\label{eq_M1_mass_loss_fix}
\end{equation}
 in which, $V =  \int_{\Omega}  \dOmega$ is the volume of the computational domain.

The dynamical equation of the level set function $\phi$ in strong and weak form reads as
 \begin{subequations}  \label {eq_phi_sf_wf}
\begin{alignat}{2}
&\text{Strong form:}  \qquad && \mathcal{S}(\phi)~~\defn~~\frac{\text{D}\phi}{\text{D}t} = \dphidt + \bu \cdot \grad \phi = 0,~~\text{at}~\forall~\x \in \Omega,  \label {eq_phi_sf}\\
&\text{Weak form:} \qquad && \mathcal{W}(\phi, \varn \phi)~~\defn~~\int_{\Omega} \left[ \dphidt + \bu \cdot \grad \phi \right] \varn \phi~\dOmega = 0,~~\phi \in H^1(\Omega)^3,~\forall~\varn \phi \in H^1(\Omega)^3. \label {eq_phi_wf}
\end{alignat}
\end{subequations}
The strong form of the level set Eq.~\eqref{eq_phi_sf} considers the motion of $\phi$  due to both advection and reinitialization velocity fields (\emph{cf.} Sec.~\ref{subsec_reasons_for_mass_loss}). For the purposes of this derivation, it is convenient to consider that $\bu$ is known \textit{a priori} using which $\phi$ can be computed (Eq.~\eqref{eq_phi_sf}). This assumption is consistent with how incompressible multiphase flow simulations work in practice---level set (or for that matter VOF) advection and reinitialization are decoupled from momentum and continuity equations through operator splitting. The weak form of the equation is obtained by multiplying the strong form with variation\footnote{We use the symbol $\varn$ to denote variation of a quantity and $\delta$ to denote the Dirac/smooth delta function in this work.} in $\phi$ (also referred to as test function) and integrating it over the domain. Neither the strong nor the weak form of the equation guarantees mass conservation. In other words, $\bu$ is such that mass conservation is not guaranteed by the resulting $\phi$ field, which in turn is used to determine the density field. However, overall mass conservation constraint can be included through the use of a Lagrange multiplier $\lambda_1$. Adding this constraint does not change $\bu$ but it should modify the equation for $\phi$. To derive the modified equation for $\phi$, we define the Lagrangian $\mathcal{L}$ of the constraint $\mathcal{C}(\phi)$ and take its variation denoted by $\varn \mathcal{L}$
 \begin{subequations} 
\begin{alignat}{2}
\mathcal{L}(\Lambda_1, \phi)~~&\defn~~\lambda_1 \int_{\Omega} \left( \rho_1 H(\phi) - \frac{M_1^0}{V} \right)~\dOmega,~~\lambda_1 \in L^2(\Omega)^3 \\
\varn \mathcal{L}(\Lambda_1, \phi)~~&\defn~~\varn \lambda_1 \int_{\Omega} \left( \rho_1 H(\phi) - \frac{M_1^0}{V} \right)~\dOmega +   \lambda_1 \int_{\Omega}  \rho_1 \delta(\phi) \varn \phi~\dOmega,~~\varn \lambda_1 \in L^2(\Omega)^3.
\end{alignat}
\end{subequations}
Adding the variation of the Lagrangian of the constraint to the weak form of the level set Eq.~\eqref{eq_phi_wf} yields the overall weak form of the constrained level set function $\phi$
\begin{align}
 \mathcal{T}(\varn \phi, \varn \lambda_1)~~&\defn~~\int_{\Omega} \left[ \dphidt + \bu \cdot \grad \phi \right] \varn \phi~\dOmega + \lambda_1 \int_{\Omega}  \rho_1 \delta(\phi) \varn \phi~\dOmega + \varn \lambda_1 \int_{\Omega}  \left( \rho_1 H(\phi) - \frac{M_1^0}{V} \right)~\dOmega=0, \nonumber\\
 ~&\forall~\varn \phi \in H^1(\Omega)^3,~\varn \lambda_1 \in L^2(\Omega)^3. \label{eq_comb_weak}
\end{align}  
Collecting terms in $\varn \lambda_1$ and $\varn \phi$, and equating them to zero separately,  yields the original constraint (Eq.~\eqref{eq_M1_mass_loss_fix}) and a new constrained dynamical equation for $\phi$ that aims to conserve the mass of fluid-1, respectively. The latter equation reads as
\begin{equation}
	\dphidt + \bu \cdot \grad \phi = -\lambda_1 \rho_1 \delta(\phi).
	\label{eq_modified_LS_eqn1}
\end{equation}     
The value of the Lagrange multiplier $\lambda_1$ is obtained by substituting $\dphidt = -  \bu \cdot \grad \phi  -\lambda_1 \rho_1 \delta(\phi)$ into Eq.~\eqref{eq_dM1dt} and setting $ \frac{\text{d}M_1}{\text{d}t} = 0$. This yields
\begin{align}
      & \frac{\text{d} M_1}{\text{d} t} = \int_{\Omega} \rho_1 \delta(\phi) \left( -\bu \cdot \grad \phi - \lambda_1 \rho_1 \delta(\phi) \right)~\dOmega = 0, \nonumber \\
      \hookrightarrow & \lambda_1 = \frac{-\int_{\Omega} \delta(\phi)(\bu \cdot \grad \phi) \dOmega}{\int_{\Omega} \rho_1 \delta^2(\phi) \dOmega}.  \label{eq_lambda1}
\end{align}
Note that Eq.~\eqref{eq_lambda1} is not an explicitly solvable equation for $\lambda_1$. This is because the $\phi$ field used on the right hand side of the equation depends on the value of $\lambda_1$ itself as seen in Eq.~\eqref{eq_modified_LS_eqn1}. Thus, a fully implicit numerical implementation would require iterations. Due to the decoupling of level set advection from the incompressible Navier-Stokes equation, $\bu$ should remain fixed while iterating for $\lambda_1$. This means momentum, advection, and reinitialization equations are not solved during iterations.

Proceeding analogously, one can conserve the mass of fluid-2 $M_2 = \int_{\Omega} \rho_2 (1 - H(\phi))~\dOmega$ by imposing a constraint of the form $\mathcal{C}(\phi) \overset{\triangle}{=} \int_{\Omega} \rho_2(1-H(\phi)) - M^0_2/V~\dOmega = 0$ with the help of Lagrange multiplier $\lambda_2 \in L^2(\Omega)^3$. In this case the dynamical equation for the level set reads as
\begin{equation}
		\dphidt + \bu \cdot \grad \phi = \lambda_2 \rho_2 \delta(\phi). \label{eq_modified_LS_eq2} 
\end{equation}
The Lagrange multiplier $\lambda_2$ is obtained analogously as
\begin{equation}
		\lambda_2 = \frac{\int_{\Omega} \delta(\phi)(\bu \cdot \grad \phi)~\dOmega}{\int_{\Omega} \rho_2 \delta^2(\phi)~\dOmega}.  \label{eq_lambda2}
\end{equation}

From Eqs.~\eqref{eq_modified_LS_eqn1}-\eqref{eq_lambda2}, it is easy to verify that $-\lambda_1 \rho_1 = \lambda_2 \rho_2$. In light of this, we can see that Eqs.~\eqref{eq_modified_LS_eqn1} and~\eqref{eq_modified_LS_eq2} are essentially the same, which in expanded form reads as 
\begin{align}
		\dphidt + \bu \cdot \grad \phi = \beta \delta(\phi) =  \frac{\int_{\Omega} \delta(\phi)(\bu \cdot \grad \phi)~\dOmega}{\int_{\Omega} \delta^2(\phi)~\dOmega} \delta(\phi).
\label{eq_exact_Lagrange_multiplier_method}
\end{align}
Physically speaking, for two phase flows, conserving the mass of phase-1 leads to mass conservation for the other phase automatically. Thus, imposing a single constraint suffices. 

It is instructive to analyze Eq.~\eqref{eq_exact_Lagrange_multiplier_method} in, both, the continuous differential equation form (with a sharp delta) and the discrete form. In the RHS of Eq.~\eqref{eq_exact_Lagrange_multiplier_method} $\beta$ is a constant. When $\delta$ is the Dirac/sharp delta function then $\beta \rightarrow 0$. This is because $\beta$'s numerator tends to zero, while its denominator $\int_\Omega \delta^2~\dOmega \rightarrow \infty$.  As analyzed in Sec.~\ref{subsec_reasons_for_mass_loss}, the integral term in the numerator of $\beta$,  $\int_{\Omega} \delta(\phi)(\bu \cdot \grad \phi)~\dOmega   = \int_{\dOmegaF} \u \cdot |\grad \phi | \n~\dS  =    \int_{\dOmegaF} \u \cdot \n~\dS$ represents the net normal (advective) velocity of the interface. For a closed domain, this is zero. Thus, from a continuous point of view (when $\delta$ is sharp)  $\beta$, $\lambda_1$, and $\lambda_2$ are zero. However, the RHS of Eq.~\eqref{eq_exact_Lagrange_multiplier_method} is $\beta \delta(\phi)$, which has a zero times infinity form. It is seen from Eq.~\eqref{eq_exact_Lagrange_multiplier_method} that $\beta \delta(\phi) \sim \frac{0}{\delta(0)} \delta(0)$, implying that  $\beta \delta(\phi)=0$. At the discrete level $\beta$ is non-zero because neither its denominator nor its numerator evaluate to zero due to the finite width of the smooth delta function. 

It is also informative to see how the non-linear Eq.~\eqref{eq_exact_Lagrange_multiplier_method} would be implemented in practice. The most natural way to implement a constrained equation is to employ some sort of operator-splitting technique. For example, in the first step, the level set field is advected. In the second step $\phi$ is reinitialized to restore its signed distance property, and in the final third step, $\phi$ is corrected (for mass loss errors) by evaluating the RHS of Eq.~\eqref{eq_exact_Lagrange_multiplier_method}. It is easy to see that the third step essentially disrupts the signed distance property of $\phi$ near the interface because the corrective term has a non-constant gradient magnitude, i.e., $|\grad (\beta \delta(\phi))|$ varies spatially. The exact Lagrange multiplier destroys the signed distance property of $\phi$ at the cost of conserving mass. This motivates the development of an approximate Lagrange multiplier that leads to both mass-conservation and signed distance retention for $\phi$. We discuss the approximate Lagrange multiplier technique in Sec.~\ref{subsec_approx_lm}.

\subsection{Overall versus pointwise mass conservation}

The Lagrange multiplier approach of Eq.~\eqref{eq_exact_Lagrange_multiplier_method} conserves mass in the domain discretely (i.e., when delta function is smooth and $\u^{\rm reinit} \ne 0$) in an integral sense. This can be proved as follows
\begin{align}
 \ddt \int_{\Omega} \rho~\dOmega  &= \ddt \int_{\Omega} \bigl\{\rho_1 H(\phi) + \rho_2 (1 - H(\phi)) \bigr\}~\dOmega  \nonumber  \\
                                                        & = (\rho_1 - \rho_2)  \int_{\Omega} \delta(\phi) \dphidt~\dOmega \nonumber  \\
                         & =  (\rho_1 - \rho_2)  \int_{\Omega} \delta(\phi) \Bigl\{  -  \bu \cdot \grad \phi +   \delta(\phi) \frac{\int_{\Omega} \delta(\phi)(\bu \cdot \grad \phi)~\dOmega}{\int_{\Omega} \delta^2(\phi)~\dOmega} \Bigr\}~\dOmega  \nonumber   \\ 
                         & =  (\rho_1 - \rho_2)  \Bigl( \int_{\Omega} -\delta(\phi) \bu \cdot \grad \phi~\dOmega   +  \int_{\Omega} \delta(\phi) \bu \cdot \grad \phi~\dOmega   \Bigr) \nonumber \\
                         & = 0.  \label{eq_2phase_integral_consv_proof}
\end{align}     
In proving Eq.~\eqref{eq_2phase_integral_consv_proof} mass of both phases is used. This is different from the derivation of $\lambda_1$ or $\lambda_2$ in the previous section which involved only one phase. Although Eq.~\eqref{eq_exact_Lagrange_multiplier_method} conserves mass in an integral sense, it does not guarantee pointwise mass conservation. To demonstrate this consider the pointwise conservation of mass equation
\begin{align}
\frac{\partial \rho}{\partial t} + \div (\u \rho) &= \frac{\partial \rho}{\partial t} + \u \cdot \grad \rho \nonumber \\
                &= (\rho_1 - \rho_2) \delta(\phi) \left(\dphidt + \u \cdot \grad \phi  \right) \nonumber\\
                &= (\rho_1 - \rho_2) \delta(\phi) \left( -\bu \cdot \grad \phi +\beta \delta(\phi) + \u \cdot \grad \phi  \right) \nonumber \\
        &= (\rho_1 - \rho_2) \delta(\phi) \left( [\u -\bu] \cdot \grad \phi +\beta \delta(\phi) \right). \label{eq_2phase_pw_nconsv_proof}
\end{align}
Eq.~\eqref{eq_2phase_pw_nconsv_proof} can be analyzed for two cases -- the continuous form (with sharp $\delta$ and an exact reinitialization step for $\phi$) and the discrete form (with a smooth delta). In either case the RHS of Eq.~\eqref{eq_2phase_pw_nconsv_proof} is zero wherever $\delta$ is zero. Thus, only at the interfacial region (sharp or smooth) needs to be checked. 

If the reinitialization step is exact then the $\phi = 0$ interface will not move in this step, implying $\u = \bu$ on the interface. Additionally, if $\delta$ is sharp then $\beta \delta(\phi) = 0$, as discussed above. Furthermore to understand pointwise mass conservation at the interface in this case, integrate Eq.~\eqref{eq_2phase_pw_nconsv_proof} over a pillbox at point on the $\phi = 0$ contour. The integral of the RHS of Eq.~\eqref{eq_2phase_pw_nconsv_proof} over the pillbox will be zero because both $\u - \bu$ and $\beta \delta(\phi)$ are zero in this case. This implies that the jump of $\rho \u \cdot \n$ across the interface is zero which confirms pointwise mass conservation on the interface (note that integral of the time derivative term is zero since the volume of the pillbox is zero).  

In the smeared/smoothed interfacial region case Eq.~\eqref{eq_2phase_pw_nconsv_proof} suggests that mass is generated/lost in the interfacial region due to: (i) the motion of level set contours at a velocity different than the material velocity, i.e., $\bu \ne \u$; and (ii) the Lagrange multiplier. The latter term counteracts the effects of reinitialization and the smooth delta function to conserve mass in an average/integral sense, as shown in Eq.~\eqref{eq_2phase_integral_consv_proof}. Thus, Eq.~\eqref{eq_2phase_pw_nconsv_proof}  highlights the limitation of the level set method to achieve pointwise or local mass conservation with a smooth $\delta$ function. This is because there will always be errors associated with $\bu \ne \u$ in the smeared interface. In other words, a signed distance function based on level set reinitialization is fundamentally incompatible with pointwise mass conservation in the discrete case. However, for many practical applications, including those considered in this work, integral/global mass conservation is sufficient. For problems requiring local mass conservation, a different interface tracking method (e.g., geometric VOF) would be necessary. Another implication of the lack of pointwise mass conservation in the level set method is that a fully coupled (fully implicit) system that arises from the discretization of incompressible Navier-Stokes and level set equations will require future investigation to check for solvability and consistency. The operator-splitting approach is more ``forgiving."

\subsection{Conserving mass is the same as conserving volume for incompressible flows}
\label{subsec_conserve_mass_and_volume}

Though it appears obvious that conserving mass and volume are equivalent for an incompressible fluid, a recent paper by Wen et al.~\cite{Wen2023} on a mass-preserving level set method has claimed otherwise. Later in Sec.~\ref{subsec_comparison_LSM} we show where specifically the authors went wrong.  We demonstrate in this section that we will obtain the same dynamical equation for the level set method as we did in the previous section if we impose constraints on conserving the volume of the two phases instead of mass. 

Specially, consider $V_1$ as the volume of fluid-1 ($\phi > 0$ region) in the domain $\Omega$ that needs to be  conserved. $V_1$ and its rate of change can be obtained from $\phi$ as
\begin{subequations} 
\begin{alignat}{2}
 & V_1(t) &&= \int_{\Omega}  H(\phi)~\dOmega, \\
& \frac{\d V_1}{\d t} &&= \ddt \int_{\Omega}  H(\phi)~\dOmega = \int_{\Omega}  \delta(\phi) \dphidt~\dOmega.  \label{eq_dV1dt}
\end{alignat}
\end{subequations} 

We follow the same procedure outlined in Sec.~\ref{subsec_mass_conservation_LSM} to derive the equations for conserving the volume of fluid-1 by imposing a constraint of the form $\mathcal{C}(\phi) \defn \int_{\Omega} H(\phi)~\dOmega - V^0_1 = 0$. Here, $V^0_1$ is the initial volume of fluid-1 in the domain. The dynamical equation for $\phi$ in this case reads as
\begin{subequations}
\begin{alignat}{2}
		\dphidt &+ \bu \cdot \grad \phi = - \lambda_3 \delta(\phi), \label{eq_modified_LS_eqn3} \\
		\lambda_3 &= \frac{-\int_{\Omega} \delta(\phi) (\bu \cdot \grad \phi)~\dOmega}{\int_{\Omega} \delta^2(\phi)~\dOmega}. \label{eq_lambda3}
\end{alignat}
\end{subequations}

We could also impose a constraint of the form $\mathcal{C}(\phi) \overset{\triangle}{=} \int_{\Omega} (1-H(\phi))~\dOmega - V^0_2 = 0$ to conserve the volume of fluid-2 ($\phi < 0$ region). In this case the dynamical equation for $\phi$ reads as
\begin{subequations}
\begin{alignat}{2}
		\dphidt &+ \bu \cdot \grad \phi = \lambda_4 \delta(\phi), \label{eq_modified_LS_eqn4} \\
		\lambda_4 &= \frac{\int_{\Omega} \delta(\phi) (\bu \cdot \grad \phi)~\dOmega}{\int_{\Omega} \delta^2(\phi)~\dOmega}. \label{eq_lambda4}
	\end{alignat}
\end{subequations}

Comparing Eqs.~\eqref{eq_modified_LS_eqn3}-\eqref{eq_lambda4}, we can observe the relation $-\lambda_3 = \lambda_4$. This implies that Eqs.~\eqref{eq_modified_LS_eqn3} and \eqref{eq_modified_LS_eqn4} are the same. For two-phase flows, if we satisfy volume conservation for one fluid, the other fluid's volume is automatically conserved. Moreover, substituting  $\lambda_3$ in Eq.~\eqref{eq_modified_LS_eqn3} or $\lambda_4$ in Eq.~\eqref{eq_modified_LS_eqn4}, we obtain the same dynamical equation for $\phi$ as written in Eq.~\eqref{eq_exact_Lagrange_multiplier_method}.



\subsection{Towards an approximate Lagrange multiplier method to prevent mass loss with the standard level set method}
\label{subsec_approx_lm}

With the exact Lagrange multipliers $\lambda_1$-$\lambda_4$, mass can be conserved discretely, but there are two ``issues":
\begin{enumerate}
\item Near the interface, the level set function does not remain a signed distance function. As discussed near the end of Sec.~\ref{subsec_mass_conservation_LSM}, this is due to the RHS of Eq.~\eqref{eq_exact_Lagrange_multiplier_method}, which is of the form $\beta \delta(\phi)$. As a corrective term, this disrupts the signed distance property of $\phi$ derived from the reinitialization equation. As a result, the Lagrange multiplier ``undoes" the reinitialization equation. 

 
 \item Computing $\beta$, specifically its numerator $\int_{\Omega} -\delta(\phi) \bu \cdot \grad \phi~\dOmega$ is not straightforward or convenient.
\end{enumerate}
Despite these challenges, implementing the exact Lagrange multiplier approach is still feasible. With respect to the first issue, although the ``quality" of the mixture model given by Eqs.~\eqref{eq_den_mixture_model} and~\eqref{eq_vis_mixture_model} gets deteriorated, it is unlikely that the effect will be significant because the Lagrange multiplier magnitude will not be very large (mass loss per time step will be small especially for resolved simulations). In other words, $\phi$ is unlikely to deviate too much from a signed distance function. Nonetheless, when computing geometric quantities such as the normal to the interface, care must be exercised: $\grad \phi$ must be explicitly normalized by its magnitude to obtain the unit normal $\n$ to the interface. The second issue can be addressed by an indirect estimation of $\beta$. This approach is suggested in Wen et al.~\cite{Wen2023}, where a similar integral term also appears. The dimension of this integral is rate of volume change---in Wen et al. the dimension is rate of mass change. The authors in~\cite{Wen2023} approximated the integral as $(M_1(t) - M_1^0)/\Delta t$, in which $M_1(t)$ represents the mass of fluid-1, which is estimated from the reinitialized level set function at time $t$ and $\Delta t$ is the current time step size. In Sec.~\ref{subsec_comparison_LSM} we critically analyze Wen et al's mass-conserving technique using our variational framework. Next, we present an alternative method that circumvents both issues while conserving mass with the standard level set method. 

Consider the reinitialized level set function $\hat{\phi}$ that is obtained by solving the reinitialization Eq.~\eqref{eq_HJ} at time $t$. At this stage $\hat{\phi}$ is a signed distance function that does not satisfy the mass constraint $\mathcal{C}(\hat{\phi})$ of Eq.~\eqref{eq_M1_mass_loss_fix}.  We seek a spatially uniform correction $\varepsilon$ for the reinitialized level set $\hat{\phi}$ such that
\begin{equation}
	\mathcal{C}(\hat{\phi} + \varepsilon) = f(\varepsilon) = \int_{\Omega} \rho_1 H(\hat{\phi} + \varepsilon)~\dOmega - M^0_1 = 0.
	\label{eq_f_epsilon}
\end{equation}
The spatially uniform correction $\varepsilon$ guarantees that the corrected level set $\phi = \hat{\phi} + \varepsilon$ is a mass-conserving signed distance function. It will be seen below that the correction $\varepsilon$ is proportional to an approximate Lagrange multiplier that enforces mass conservation with the level set method through a predictor-corrector type of a scheme
\begin{subequations} \label{eq_approx_lm}
\begin{alignat}{2} 
\frac{\hat{\phi} - \phi^n}{\Delta t} + \bu \cdot \grad \hat{\phi} &=  0, \qquad [\text{predictor step}]\\
\frac{\phi - \hat{\phi}}{\Delta t} = \frac{\varepsilon - 0}{\Delta t}&=\frac{\Delta \varepsilon}{\Delta t}. \qquad [\text{corrector step}]   
\end{alignat}
\end{subequations}
The equation set~\eqref{eq_approx_lm} can be thought of as a representation of a ``continuous" equation of the form
\begin{equation}
 \frac{\partial \phi}{\partial t} + \bu \cdot \grad \phi =  \frac{\d \varepsilon}{ \d{t}}. \label{eq_LS_RHS_constant} 
 \end{equation}
To relate the informal  Eq.~\eqref{eq_LS_RHS_constant} and the formal  Eq.~\eqref{eq_exact_Lagrange_multiplier_method}, we first consider an intuitive argument and then present a mathematical derivation. Note that the ``forcing" that induces a shift in the contour levels of $\phi$ is the RHS of Eq.~\eqref{eq_LS_RHS_constant}. This term is taken to be uniform in the entire domain leading to the same correction to every contour level. The corresponding RHS of the formal Eq.~\eqref{eq_exact_Lagrange_multiplier_method} is $\beta\delta(\phi)$, where $\beta$ is uniform everywhere. However, the $\delta(\phi)$ term makes the ``forcing" in the formal equation non-uniform in the domain. If one picks a contour level, say $\phi =0$, then the ``forcing" does have the same value on that entire contour and this will lead to the same correction to $\phi$ on that contour. Similarly, each contour level has its own constant value of correction to $\phi$. This correction is maximum on $\phi = 0$ contour and decreases away from it until it is zero outside the smeared interfacial region where $\delta(\phi) = 0$. Since, in the formal method of Eq.~\eqref{eq_exact_Lagrange_multiplier_method} each contour level is corrected by a different amount, the distance function property is lost. The only way to preserve the distance function property of $\phi$ is to correct all contour levels by the same amount. This is the assumption made in the informal approach of Eq.~\eqref{eq_LS_RHS_constant}. Thus, Eq.~\eqref{eq_LS_RHS_constant} would arise from Eq.~\eqref{eq_exact_Lagrange_multiplier_method} under the assumption that $\frac{\d \varepsilon}{\d t} = \beta \delta(\phi_m)$, which means that the correction of an appropriately chosen $\phi_m$ contour in the formal approach is uniformly applied in the entire domain to obtain the informal approach.

We now proceed with the mathematical derivation by obtaining an expression for the rate of change of correction $\frac{\d \varepsilon}{\d t}$. This can be obtained by differentiating Eq.~\eqref{eq_f_epsilon} with respect to time
\begin{subequations} 
\begin{alignat}{2}
			  & \frac{\text{d} f(\varepsilon)}{\text{d} t} = \int_{\Omega} \rho_1 \delta(\hat{\phi} + \varepsilon) \left( \frac{\partial \hat{\phi}}{\partial t} + \depsdt \right)~\dOmega = 0, \label{eq_dfdt_a} \\
\hookrightarrow   & \int_{\Omega} \rho_1 \delta(\hat{\phi} + \varepsilon) \frac{\partial \hat{\phi}}{\partial t}~\dOmega = - \int_{\Omega} \rho_1 \delta(\hat{\phi} + \varepsilon) \depsdt~\dOmega, \label{eq_dfdt_b} \\
	\hookrightarrow & \depsdt = \frac{\int_{\Omega} \delta(\hat{\phi} + \varepsilon) (\bu \cdot \grad \hat{\phi})~\dOmega}{\int_{\Omega} \delta(\hat{\phi} + \varepsilon)~\dOmega}  = \frac{\int_{\Omega} \delta(\hat{\phi} + \varepsilon) (\bu \cdot \grad (\hat{\phi} + \varepsilon))~\dOmega}{\int_{\Omega} \delta(\hat{\phi} + \varepsilon)~\dOmega} = \frac{\int_{\Omega} \delta(\phi) (\bu \cdot \grad \phi )~\dOmega}{\int_{\Omega} \delta(\phi)~\dOmega}. \label{eq_dfdt_c}  
\end{alignat} 
\end{subequations} 
In Eq.~\eqref{eq_dfdt_a} we used the Leibniz rule to carry out the differentiation of the non-conserved quantity $f(\varepsilon)$ defined over the static region $\Omega$.  Eq.~\eqref{eq_dfdt_c} is arrived at by using the relations  $\grad \hat{\phi} = \grad (\hat{\phi} + \varepsilon)$ and $\phi = \hat{\phi} + \varepsilon$. Comparing the right hand sides of Eqs.~\eqref{eq_exact_Lagrange_multiplier_method} and~\eqref{eq_LS_RHS_constant}, we see that 
\begin{align}
\depsdt &= \frac{\int_{\Omega} \delta(\phi) (\bu \cdot \grad \phi )~\dOmega}{\int_{\Omega} \delta(\phi)~\dOmega} = \delta(\phi_m) \; \frac{  \int_{\Omega} \delta(\phi) (\bu \cdot \grad \phi )~\dOmega}{\delta(\phi_m)  \int_{\Omega} \delta(\phi)~\dOmega}  =  \delta (\phi_m) \widetilde{\beta}. 
\end{align}
Eqs.~\eqref{eq_exact_Lagrange_multiplier_method} and~\eqref{eq_LS_RHS_constant} are therefore similar with a slight difference in $\beta$ and $\widetilde{\beta}$: the latter is an approximation to $\beta$ in which one of the smooth delta function terms (in the denominator) is evaluated at the $m^\text{th}$ contour. A uniform value is used with the approximate method to correct all $\phi$ contours. This correction corresponds to the shift/correction in the $m^\text{th}$ contour of $\phi$ in the exact case. We demonstrate this by taking a commonly used smooth delta function in the level set literature
\begin{equation}
\delta(\phi) =  \frac{1}{2\Delta} + \frac{1}{2\Delta} \cos\left(\frac{\pi \phi}{\Delta} \right)  \quad \hookrightarrow \int_{-\Delta}^{\Delta} \delta(x) \, \d x = 1~~\text{and}~~\int_{-\Delta}^{\Delta} \delta^2(x) \, \d x = \frac{3}{4\Delta},  
\end{equation}
in which $\Delta$ denotes the half-width of the interfacial region. The $m^\text{th}$ contour of $\phi$ can be identified as follows:
\begin{subequations}
\begin{alignat}{2}
& \beta \delta(\phi_m) = \depsdt = \widetilde{\beta} \delta(\phi_m) \implies \beta = \widetilde{\beta} \\
\hookrightarrow  & \frac{  \int_{\Omega} \delta(\phi) (\bu \cdot \grad \phi )~\dOmega}{\int_{\Omega} \delta^2(\phi)~\dOmega}  =  \frac{  \int_{\Omega} \delta(\phi) (\bu \cdot \grad \phi )~\dOmega}{\delta(\phi_m)  \int_{\Omega} \delta(\phi)~\dOmega}    \\
\hookrightarrow  & \delta(\phi_m) = \frac{\int_{\Omega} \delta^2(\phi)~\dOmega}{\int_{\Omega} \delta(\phi)~\dOmega}  \\
\hookrightarrow  & \frac{1}{2\Delta} + \frac{1}{2\Delta} \cos\left(\frac{\pi \phi_m}{\Delta} \right) =  \frac{3}{4\Delta} \\
\hookrightarrow  & \phi_m = \pm \frac{\Delta}{3}.
\end{alignat}
\end{subequations}
The uniform correction applied by the approximate Lagrange multiplier corresponds to the exact correction at the contour $\Delta/3$ distance away from the interface.

Computing $\varepsilon$ that preserves mass for the level set method is a simple root-finding problem, $f(\varepsilon) = 0$, for which we can use Newton's method:
\begin{equation}
	f(\varepsilon_{k+1}) = f(\varepsilon_k) + \frac{\text{d}f}{\text{d}\varepsilon} {\Bigg|}_{\varepsilon_k}(\varepsilon_{k+1}-\varepsilon_k). \label{eq_Newton_method}
\end{equation}
Setting  $f(\varepsilon_{k+1}) = 0$ in the equation above yields
\begin{subequations}
	\begin{alignat}{2}
		&\varepsilon_{k+1} && = \varepsilon_k - \frac{f(\varepsilon_k)}{{\frac{\text{d}f}{\text{d}\varepsilon}}\Big|_{\varepsilon_k}}, \\
		\hookrightarrow &\Delta \varepsilon_{k+1} && =  -\frac{ \int_{\Omega} \rho_1 H(\hat{\phi} + \varepsilon_k)~\dOmega - M^0_1}{ \int_{\Omega} \rho_1 \delta(\hat{\phi} + \varepsilon_k)~\dOmega} \nonumber \\
		& && = -\frac{ \int_{\Omega} H(\hat{\phi} + \varepsilon_k)~\dOmega - M^0_1/\rho_1}{ \int_{\Omega} \delta(\hat{\phi} + \varepsilon_k)~\dOmega}  \nonumber \\
		& && = -\frac{ \int_{\Omega} H(\hat{\phi} + \varepsilon_k)~\dOmega - V^0_1}{ \int_{\Omega} \delta(\hat{\phi} + \varepsilon_k)~\dOmega}. \label{eq_delta_eps}
	\end{alignat} 
\end{subequations}
Here, $k$ represents the (Newton) iteration counter. With $k = 0$, we start with a zero correction, i.e., $\varepsilon_0 = 0$ and iterate until mass is conserved to machine accuracy. The correction value $\varepsilon$ will remain the same even when we impose a volume conservation constraint. This can be seen in the steps leading up to Eq.~\eqref{eq_delta_eps}. 

\vspace{1em}
\noindent \textbf{\underline{Summary of exact vs. approximate Lagrange multiplier approaches:}} The inexact approach overcomes the two problems of the exact Lagrange multiplier equation, but its main drawback is that it is post-hoc in nature; Eq.~\eqref{eq_LS_RHS_constant} is formally not a continuous equation. The inexact technique, however, closely mimics the exact Lagrange multiplier approach (implemented via operator splitting). Another key difference between exact and approximate methods lies in how they adjust the level set contours to achieve mass balance. The exact approach prioritizes the zero-contour (interface), moving it the most. This movement gradually diminishes for contours farther away from the interface, with those outside the interfacial region remaining entirely static. In contrast, the approximate Lagrange multiplier approach applies a uniform adjustment across all contours throughout the domain. This acts as a long-range correction mechanism.          

\subsection{Comparison with other mass-preserving level set methods}
\label{subsec_comparison_LSM}

Although there are many techniques available in the literature for conserving mass using the level set method, we highlight two studies that have applied continuous formulations in place of pure numerical ones (i.e., combining level set with volume of fluid/moment of fluid methods, particle level set methods, etc.) to address mass loss problems. The conservative level set (CLS) method~\cite{olsson2005conservative} is a continuous formulation, but we exclude it from this discussion since, despite its name, it also leads to mass loss---for example, see the dam break problem simulated in Parameswaran and  Mandal~\cite{parameswaran2023stable} with CLS wherein mass losses up to 50\% are reported (refer Fig. 22 of their work).  

\subsubsection{Mass preserving formulation of Wen et al.} \label{sec_wen}

A recent work by Wen et al.~\cite{Wen2023} describes a method of conserving mass with the standard level set method by including an additional source term in the equation. The source term form is selected ad-hoc with a free parameter $\eta$, whose value is determined through fluid-1's mass balance. In addition, the authors used a non-standard definition of mass in their derivation, which results in several inconsistencies. In this section, we present Wen et al.'s mass-preserving level set method based on the variational framework of Sec.~\ref{subsec_mass_conservation_LSM}. In our derivation, we will continue to use their mass definition to obtain a similar level set equation as in~\cite{Wen2023}. Their equation reads as  
\begin{subequations} \label{eqn_Wen_et_al}
\begin{alignat}{2}
	\dphidt + \bu \cdot \grad \phi = \eta \delta(\phi) |\grad \phi|, \label{eq_Wen_et_al_LSeqn} \\
	\eta = \frac{\int_{\Omega} L(\phi) \mu_\phi~\dOmega}{\int_{\Omega} L(\phi) \delta(\phi) |\grad \phi|~\dOmega} \;. \label{eq_Wen_et_al_eta}
	\end{alignat}
\end{subequations}
Here, $L(\phi) = \delta(\phi) \left[ 2(\rho_1 - \rho_2)H(\phi) + \rho_2 \right]$ (see Eq. (26) in~\cite{Wen2023}) and $\mu_\phi$ represents the ``discretization error" in approximating $-\bu \cdot \grad \phi$. Though the authors did not justify the form of the RHS of  Eq.~\eqref{eq_Wen_et_al_LSeqn}, it resembles the RHS of our Eq.~\eqref{eq_exact_Lagrange_multiplier_method}. Note that for level set methods, $|\grad \phi| = 1$ thanks to the reinitialization process. Therefore, the use of $|\grad \phi|$ in Eqs.~\eqref{eqn_Wen_et_al} is redundant.

In~\cite{Wen2023}, $\eta$ is arrived at by defining mass of fluid-1 as   
 \begin{equation}
	M_1(\phi) = \int_{\Omega} \rho(\phi) H(\phi)~\dOmega,
	\label{eq_Wen_et_al_M1}
\end{equation}
which is different from the standard definition of mass used in Eq.~\eqref{eq_M1_mass_loss_fix}. The implications of defining $M_1$ through Eq.~\eqref{eq_Wen_et_al_M1}  will be discussed later. For now we will continue using Eq.~\eqref{eq_Wen_et_al_M1} to derive Wen et al.'s Eq.~\eqref{eq_Wen_et_al_LSeqn}  utilizing our variational framework. The rate of change of mass for fluid-1 can be computed using RTT as
\begin{align}
	\frac{\text{d}M_1}{\text{d}t} &= \ddt \int_{\Omega} \rho(\phi) H(\phi)~\dOmega, \nonumber \\
	&=\int_{\Omega} \left[ H(\phi)\frac{\partial \rho(\phi)}{\partial t} + \rho(\phi) \frac{\partial H(\phi)}{\partial t} \right]~\dOmega,  \nonumber  \\
	&=  \int_{\Omega} \left[ 2(\rho_1 - \rho_2)H(\phi) + \rho_2 \right] \delta(\phi) \dphidt~\dOmega, \nonumber \\
	& = \int_{\Omega} L(\phi) \dphidt~\dOmega. \label{eq_dM1_dt_Wen}
\end{align}
The strong and weak form of the level set equation remains the same as the equation set~\eqref{eq_phi_sf_wf}. Based on the definition of mass of fluid-1, a constraint of the form  
\begin{equation}
	\mathcal{C}(\phi) = \int_{\Omega} \left( \rho(\phi) H(\phi) - \frac{M_1^0}{V} \right)~\dOmega = 0, \label{eq_Wen_et_al_constraint}
\end{equation}
is imposed with the help of the Lagrange multiplier $\Gamma \in L^2(\Omega)^3$. Following the derivation steps of Sec.~\ref{subsec_mass_conservation_LSM} we obtain an equation for the mass-conserving level set field
\begin{subequations}
\begin{alignat}{2}
	&\dphidt + \bu \cdot \grad \phi = -\Gamma L(\phi),  \label{eq_Wen_LS_eq1} \\
	& \Gamma = \frac{-\int_{\Omega}L(\phi) (\bu \cdot \grad \phi)~\dOmega}{\int_{\Omega}L^2(\phi)~\dOmega}.
\end{alignat}	
\end{subequations}
The Lagrange multiplier $\Gamma$ is obtained by substituting $\partial \phi/\partial t = -\Gamma L(\phi) -\bu \cdot \grad \phi$ from Eq.~\eqref{eq_Wen_LS_eq1} into Eq.~\eqref{eq_dM1_dt_Wen} and setting $\frac{\text{d}M_1}{\text{d}t} = 0$. Overall, Eq.~\eqref{eq_Wen_LS_eq1} reads as
\begin{equation}
\dphidt + \bu \cdot \grad \phi  = 	L(\phi) \frac{\int_{\Omega}L(\phi) (\bu \cdot \grad \phi)~\dOmega}{\int_{\Omega}L^2(\phi)~\dOmega}. \label{eq_Wen_LS_eq1_full}
\end{equation}
Aside from the (redundant) $|\grad \phi|$ terms, Eq.~\eqref{eq_Wen_LS_eq1_full} is similar, but not the same as Eq.~\eqref{eq_Wen_et_al_LSeqn}. Accordingly, Wen et al.'s Eq.~\eqref{eq_Wen_et_al_LSeqn} does not follow the constraint formulation. Additionally, the discretization error $\mu_\phi$ has not been quantified in~\cite{Wen2023}. If $\mu_\phi = 0$ for some advective scheme, then $\eta = 0$. In this case Eq.~\eqref{eq_Wen_et_al_LSeqn} reverts back to the standard level set equation, which does not conserve mass. The interpretation of $\mu_\phi$  as a discretization error is therefore not correct. Furthermore, the use of a non-standard definition of mass also leads to several inconsistencies. The inconsistencies can best be described by determining the constrained level equation that aims to conserve the mass of fluid-2 $M_2(\phi) = \int_{\Omega} \rho(\phi) [1-H(\phi)]~\dOmega$. Following the same procedure as above, we can derive fluid-2's mass-conserving equation, which reads as  
\begin{equation}
		\dphidt + \bu \cdot \grad \phi  = L_{*}(\phi) \frac{\int_{\Omega} L_{*}(\phi) (\bu \cdot \grad \phi)~\dOmega}{\int_{\Omega} L_{*}^2(\phi)~\dOmega}. \label{eq_Wen_LS_eq2} \\
\end{equation}
Here, $L_{*}(\phi) = [(\rho_1-2\rho_2) + 2(\rho_2-\rho_1)H(\phi)]\delta(\phi)$. Note that Eq.~\eqref{eq_Wen_LS_eq2} differs from Eq.~\eqref{eq_Wen_LS_eq1_full}. Each fluid has its own governing equation. As part of their numerical experiments, Wen et al. (arbitrarily) chose the level set equation to preserve fluid-1's mass. The only way to conserve both fluids' masses simultaneously with this approach is to have $L(\phi) = L_{*}(\phi)$. This condition simplifies to yield $4(\rho_1 - \rho_2)H(\phi) = \rho_1 - 3\rho_2$. Considering that the RHS is constant, and the LHS varies spatially, this is a contradiction. There is also an unphysical condition $\rho_1 = -\rho_2$ at the interface where $\phi = 0$ and $H(\phi=0) = 1/2$.

While Wen et al's method is inconsistent, their numerical results for two-phase flows match well with existing literature. We attribute this fortunate match to the fact that for well-resolved simulations, $\eta$ (or the Lagrange multipliers) do not cause a substantial change in $\phi$.

\subsubsection{Mass preserving formulation of Kees et al.} \label{sec_kees}

Next, we describe the mass-conserving level set approach of Kees et al.~\cite{Kees2011} that can be considered an extension of the approximate Lagrange multiplier approach of Sec.~\ref{subsec_approx_lm}. Their approach uses an advected volume fraction field $\widehat{H}$ as a target field to modify the reinitialized level set function $\widehat{\phi}$ directly to correct for spurious mass loss/gain. Their mass-preserving level set approach can be described by the following equations:
\begin{subequations}
\begin{alignat}{2}
        &\text{volume fraction advection:} \qquad &&  \frac{\partial \widehat{H}}{\partial t} + \div (\u \widehat{H}) = 0, \label{eq_hatH_adv}\\
        &\text{level set advection:} \qquad && \frac{\partial \tilde{\phi}}{\partial t} + \u \cdot \grad \tilde{\phi} = 0, \\
        &\text{level set reinitialization:} \qquad &&  \frac{\partial \widehat{\phi}}{\partial \tau} + \text{sgn}(\tilde{\phi})\left( |\grad \widehat{\phi}| - 1 \right) = 0, \\
	&\text{level set mass correction:} \qquad && \gamma \nabla^2 \varepsilon(\x) = H(\widehat{\phi} + \varepsilon(\x)) - \widehat{H}~~\text{with}~\grad \varepsilon(\x) \cdot {\n} = 0~\text{on}~\partial \Omega. \label{eq_reaction_diffusion} 
\end{alignat}	
\end{subequations}
Assuming no numerical diffusion errors, advecting fluid-1's volume fraction $\widehat{H}$ in a closed domain yields $\int_{\Omega} \widehat{H}(t)~\dOmega = V_1^0$. The main idea behind the mass conserving approach of Kees et al.~\cite{Kees2011} is to solve (the nonlinear reaction-diffusion) Eq.~\eqref{eq_reaction_diffusion} with homogenous Neumann boundary conditions for a spatially varying level set correction field $\varepsilon(\x)$. Integrating Eq.~\eqref{eq_reaction_diffusion} over the computational domain $\Omega$ reveals the constraint that Eq.~\eqref{eq_reaction_diffusion} imposes
\begin{align}
& \int_{\Omega}H(\hat{\phi} + \varepsilon(\x))~\dOmega -  \int_{\Omega} \widehat{H}~\dOmega  = \gamma  \int_{\Omega} \nabla^2 \varepsilon(\x)~\dOmega, \nonumber \\
\hookrightarrow  & \int_{\Omega}H(\hat{\phi} + \varepsilon(\x))~\dOmega -  \int_{\Omega} \widehat{H}~\dOmega =  \gamma  \int_{\partial \Omega} \grad \varepsilon(\x) \cdot {\n}~\dS  = 0, \nonumber \\
\hookrightarrow  & \int_{\Omega}H(\hat{\phi} + \varepsilon(\x))~\dOmega =  \int_{\Omega} \widehat{H}~\dOmega = V_1^0. \label{eq_vol_const_kees}
\end{align}	
 Eq.~\eqref{eq_vol_const_kees} defines essentially the same constraint as Eq.~\eqref{eq_f_epsilon} with the difference that the corrective field $\varepsilon$ is allowed to vary spatially. In their work, Kees et al. mention ``\emph{$\gamma$ is a parameter that penalizes the deviation of $\varepsilon(\x)$ from a global constant.} " In their numerical experiments, the authors take a large value of the penalty parameter $\gamma$. According to their results (see Figs. 5 and 6, and Table 1), $\varepsilon$ remains essentially constant throughout the domain. In practice, Kees et al.'s approach is the same as the approximate Lagrange multiplier approach introduced in Sec.~\ref{subsec_approx_lm}. The approximate Lagrange multiplier method is computationally more efficient than Kees et al.'s approach because it does not require maintaining an additional advective field $\widehat{H}$. It also avoids inverting a large system of equations. The need for the latter arises due to the presence of a Laplacian operator in Eq.~\eqref{eq_reaction_diffusion}.   

\subsection{Extension to three phase flows}
\label{subsec_3phase_flows}

\begin{figure}[]
  \centering
  \subfigure[Non-immersed formulation]{
  	\includegraphics[scale= 0.37]{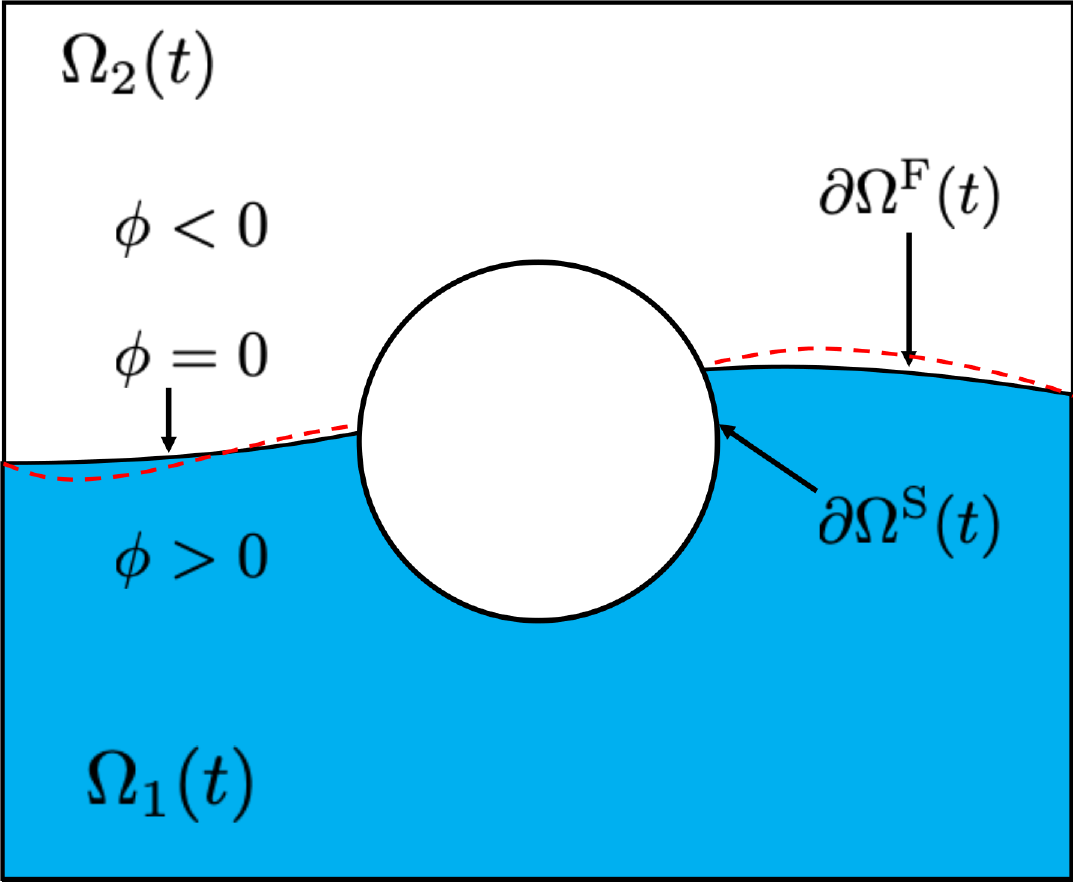}
	\label{fig_schematic_non_immersed_formulation}
  }
   \subfigure[Immersed formulation]{
  	\includegraphics[scale= 0.37]{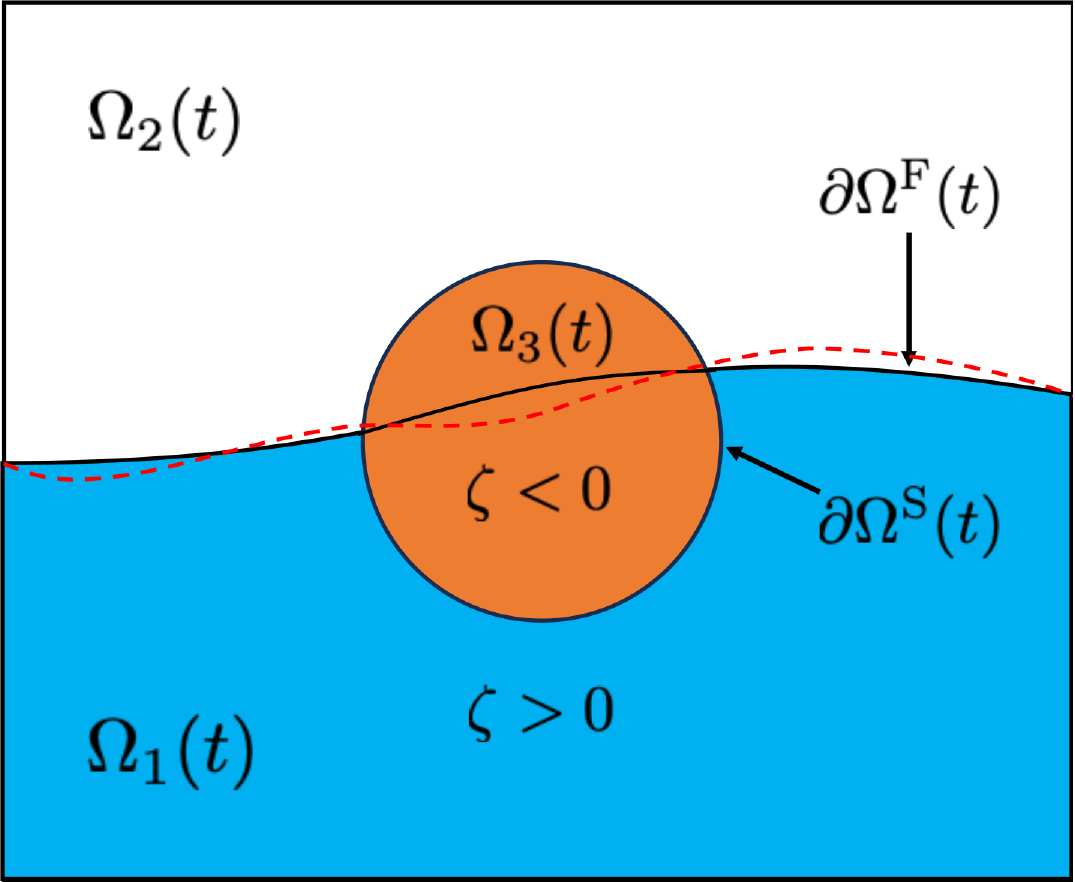}
	\label{fig_schematic_immersed_formulation}
  }  
  \caption{Schematic of the computational domain $\Omega$ for three phase flows considered in  \subref{fig_schematic_non_immersed_formulation} non-immersed and \subref{fig_schematic_immersed_formulation} immersed formulations. The variation of the two-fluid interface $\dOmegaF$ is illustrated by the red dashed line. }
  \label{fig_schematic_non_immersed_immersed_formulation}
\end{figure}

In this section, we extend mass conserving level set techniques for two-phase flows, i.e., exact and approximate Lagrange multiplier methods, to three-phase flows, by including a (moving) solid phase in the domain; see Fig.~\ref{fig_schematic_non_immersed_formulation}. Fluid-structure interactions are modeled using the fictitious domain Brinkman Penalization (FD/BP) method, which is an immersed boundary method. The FD/BP method solves a single momentum equation in the entire computational domain $\Omega$, including the immersed solid region $\Omega_3(t) \subset \Omega$. The momentum of the solid body is accounted for by the penalty term $\fc(\x, t)$ in the momentum Eq.~\eqref{eqn_momentum}, whose form reads as
\begin{align}  
	\fc(\x,t) = \frac{\chi(\x,t)}{\kappa}\left(\ub(\x,t) - \u(\x,t)\right).
\label{eqn_brinkman_force}
\end{align}
The penalty term $\fc$ ensures that the velocity inside the structure region $\Omega_3(t)$ is a rigid body velocity $\ub(\x,t)$. $\ub$ is determined by hydrodynamic and gravity forces acting on the body~\cite{Bhalla2020}. An indicator function $\chi(\x,t)$ tracks the location of a solid body within $\Omega$. $\chi$ is non-zero only within $\Omega_3(t)$. The Brinkman penalization method treats the solid body as a porous region with vanishing permeability $\kappa \ll 1$. At the fluid-solid interface, the penalty force $\fc$ can also be treated differently in the normal and tangential directions. This possibility is explored in Sec.~\ref{sec_penalty_split}.  


The interface $\dOmegaSt$ between fluid and solid domains is tracked by a level set function $\zeta(\x, t)$: $\zeta > 0$ in $ \OmegaFt$, $\zeta < 0$ in $\Omega_3(t)$, and $\zeta = 0$ on $\dOmegaSt$. The solid level set is advected using 
\begin{equation}
	\frac{\partial \zeta}{\partial t} + \u \cdot \grad \zeta = 0.
	\label{eq_zeta_advection}
\end{equation}

In what follows next, we assume that the solid domain does not loose mass/volume due to the motion/advection of $\zeta(\x,t)$ within $\Omega$, and all mass/volume issues stem from the level set function $\phi(\x,t)$ which defines the interface $\dOmegaFt$ between the two fluid phases. 
\subsubsection{Conserving mass/volume for three phase flows - Non-immersed formulation}
\label{subsec_3phase_non_immersed_formulation}

\vspace{1em}
\noindent \textbf{\underline{Exact Lagrange multiplier approach:}} We first present the non-immersed formulation, in which the level set function $\phi(\x,t)$ does not exist within the structure region. $\phi(\x,t)$ is assumed to satisfy appropriate boundary conditions on $\partial \Omega$ and $\dOmegaSt$. The non-immersed scenario is akin to a deforming/moving mesh that has a hole (to represents the body) inside it. We aim to conserve volume $V_1$ of fluid-1, which can be expressed in terms of $\phi$ as
\begin{equation}
	V_1 = \int_{\OmegaFt} H(\phi)~\dOmega.
\end{equation}
The rate change of volume $V_1$ (a conserved quantity) can be expressed using the RTT Eq.~\eqref{eq_RTT} as
\begin{equation}
	\frac{\text{d}V_1}{\text{d} t} = \int_{\OmegaFt} \delta(\phi) \dphidt~\dOmega + \int_{\dOmegaSt} H(\phi) (\us \cdot \n)~\dS, \label{eq_dv1dt_nib}
\end{equation}
In Eq.~\eqref{eq_dv1dt_nib} $\n$ represents a unit normal vector to the fluid-solid interface that points away from the fluid and into the solid. The no-slip condition on fluid-structure interface implies $\u = \us = \ub(\dOmegaSt,t)$. 

The strong and weak form of the dynamical equation for $\phi(\x,t)$, $\mathcal{S}(\phi)$ and $\mathcal{W}(\phi)$, respectively,  the volume constraint $\mathcal{C}(\phi)$, the Lagrangian of the constraint  and its variation, $\mathcal{L}(\Lambda_1, \phi)$ and $\varn \mathcal{L}(\Lambda_1, \phi)$. respectively, in the time-varying fluid domain $\OmegaFt$ read as
\begin{subequations}
	\begin{alignat}{2}
		 \mathcal{S}(\phi)~~&\defn~~\dphidt + \bu \cdot \grad \phi = 0,~~\text{at}~\forall~\x \in \OmegaFt, \label{eq_sf_nib} \\
		\mathcal{W}(\phi, \varn \phi)~~&\defn~~\int_{\OmegaFt} \left[ \dphidt + \bu \cdot \grad \phi \right] \varn \phi~\dOmega,~~\phi \in  H^1(\OmegaFt)^3,~\forall~\varn \phi \in  H^1(\OmegaFt)^3, \label{eq_wf_nib} \\ 
		\mathcal{C}(\phi)~~&\defn~~\int_{\OmegaFt}  \left(H(\phi) - \frac{V^0_1}{V} \right)~\dOmega = 0, \label{eq_const_nib} \\
		\mathcal{L}(\Lambda_1, \phi)~~&\defn~~\Lambda_1 \int_{\OmegaFt}  \left(H(\phi) - \frac{V^0_1}{V} \right)~\dOmega,~~\Lambda_1 \in  L^2(\OmegaFt)^3,  \label{eq_lag_nib} \\
		 \varn \mathcal{L}(\Lambda_1, \phi)~~&\defn~~\varn  \Lambda_1 \int_{\OmegaFt}  \left(H(\phi) - \frac{V^0_1}{V} \right)~\dOmega +   \Lambda_1 \int_{\OmegaFt}  \delta(\phi) \varn \phi~\dOmega,~~\varn \Lambda_1 \in  L^2(\OmegaFt)^3. \label{eq_var_lag_nib}
	\end{alignat}
\end{subequations}
Note that in Eq.~\eqref{eq_var_lag_nib} the variation in the movement of the fluid domain $\OmegaFt$ is not included. This is because the solid interface moves with the material velocity field. As noted earlier we are assuming that the velocity field is known (operator splitting with an explicit level set approach). The effect of domain variation would have to be probed if the velocity field was also an unknown. In that case, probing the effect of variation in the velocity field would cause variation in the movement of the fluid domain which in turn would lead to an extra term in the variation of the Lagrangian of the volume conservation constraint. The total variation, including the effect of variation in the domain movement on the Lagrangian of the constraint, would be: 
\begin{align*}
\varn^\text{total} \mathcal{L}(\Lambda_1, \phi) 
= \varn \mathcal{L}(\Lambda_1, \phi) + \Lambda_1 \int_{\dOmegaSt} \left(H(\phi) - \frac{V^0_1}{V} \right) (\vs \d t  \cdot \n)~\dS, 
\end{align*}
in which $\vs$ is the variation in velocity. The last term in the equation above arises due to variation in velocity (and the corresponding variation in domain movement) and is similar to that in the Leibniz theorem. If it is assumed that the velocity field is known \textit{a priori} when solving for $\phi$, then the last term is dropped because $\vs$ is zero.

Adding variation of the Lagrangian $\varn \mathcal{L}(\Lambda_1, \phi)$ to the weak form $\mathcal{W}(\phi, \varn \phi)$, collecting terms in $\varn  \Lambda_1$ and $\varn \phi$, and equating them to zero separately, yields the original constraint (Eq.~\eqref{eq_const_nib}) and a new dynamical equation for $\phi$ that reads as
\begin{subequations} \label{eq_LS_non_immersed_set1}
	\begin{alignat}{2} 
		\dphidt &+ \bu \cdot \grad \phi = - \Lambda_1 \delta(\phi), \label{eq_LS_non_immersed_1} \\
		\Lambda_1 &= \frac{-\int_{\OmegaFt} \delta(\phi) (\bu \cdot \grad \phi)~\dOmega + \int_{\dOmegaSt} H(\phi) (\us \cdot \n)~\dS}{\int_{\OmegaFt} \delta^2(\phi)~\dOmega}.
	\end{alignat}
\end{subequations}
The value of the Lagrange multiplier $\Lambda_1$ is obtained by substituting $\dphidt = -  \bu \cdot \grad \phi  -\Lambda_1 \delta(\phi)$ into Eq.~\eqref{eq_dv1dt_nib} and setting $ \frac{\text{d}V_1}{\text{d}t} = 0$.

Proceeding analogously, we can conserve the volume of fluid-2 $V_2 = \int_{\OmegaFt} (1-H(\phi))~\dOmega$ by imposing a constraint of the form $\mathcal{C}(\phi) \defn \int_{\OmegaFt} (1-H(\phi)) - V^0_2/V~\dOmega = 0$ with the help of Lagrange multiplier $\Lambda_2 \in L^2(\Omega)^3$. In this case the dynamical equation for the level set reads as
\begin{subequations} \label{eq_LS_non_immersed_set2}
	\begin{align}
		&\dphidt + \bu \cdot \grad \phi = \Lambda_2 \delta(\phi), \label{eq_LS_non_immersed_2} \\
		& \Lambda_2 = \frac{\int_{\OmegaFt} \delta(\phi) (\bu \cdot \grad \phi)~\dOmega - \int_{\dOmegaSt} H(\phi) (\us \cdot \n)~\dS}{\int_{\OmegaFt} \delta^2(\phi)~\dOmega}. \label{eq_nib_Lambda2}
	\end{align}
\end{subequations}
In arriving at the RHS of Eq.~\eqref{eq_nib_Lambda2}, we used $\frac{\text{d}V_2}{\text{d}t} = 0 $ and the relation $\int_{\dOmegaSt} (\us \cdot \n)~\dS  =  \int_{\Omega_3} \div \ub~\dOmega =  0$. The latter holds because $\ub$ is a volume-preserving rigid body velocity field.

Comparing the two level set equation sets~\eqref{eq_LS_non_immersed_set1} and~\eqref{eq_LS_non_immersed_set2}, we observe the relation $-\Lambda_1 = \Lambda_2$. Therefore,  Eqs.~\eqref{eq_LS_non_immersed_1} and \eqref{eq_LS_non_immersed_2} are the same, given by
\begin{equation}
		\dphidt + \bu \cdot \grad \phi = \alpha \delta(\phi)  =  \frac{ \int_{\OmegaFt} \delta(\phi) (\bu \cdot \grad \phi)~\dOmega - \int_{\dOmegaSt} H(\phi) (\us \cdot \n)~\dS}{\int_{\OmegaFt} \delta^2(\phi)~\dOmega} \,  \delta(\phi).		\label{eq_LS_non_immersed_3}
\end{equation}
Thus, conserving volume of fluid-1 automatically conserves the volume of fluid-2, even in the moving domain $\OmegaFt$. Furthermore, it is straightforward to show that the equation for $\phi$ remains the same (as  Eq.~\eqref{eq_LS_non_immersed_3}), if we impose mass conservation constraints instead of volume ones. We omit the derivation steps for brevity.

\vspace{1em}
\noindent \textbf{\underline{Approximate Lagrange multiplier approach:}}  Here we follow the derivation procedure of Sec.~\ref{subsec_approx_lm} to derive an approximate Lagrange multiplier to conserve mass/volume of the two fluid phases in the moving domain $\dOmegaFt$. Consider the reinitialized level set function $\widehat{\phi}$ at time $t$ that is a signed distance function, but it does not satisfy the constraint Eq.~\eqref{eq_const_nib} yet. We seek a spatially uniform correction $\varepsilon$ that corrects $\widehat{\phi}$ to $\phi = \widehat{\phi} + \varepsilon$ while maintaining its signed distance property $|\grad \phi | = |\grad(\widehat{\phi} + \epsilon)| = 1$. This is achieved by finding the root of the nonlinear equation
\begin{equation}
	\mathcal{C}(\widehat{\phi} + \varepsilon) = f(\varepsilon) = \int_{\OmegaFt} H(\widehat{\phi} + \varepsilon)~\dOmega - V^0_1= 0,
	\label{eq_f_epsilon_nonimmersed}
\end{equation}
using Newton's method. The correction at the $k^{\rm th}$ Newton iteration becomes  
\begin{subequations}
	\begin{align}
		& \varepsilon_{k+1} = \varepsilon_k - \frac{f(\varepsilon_k)}{{\frac{\text{d}f}{\text{d}\varepsilon}}\Big|_{\varepsilon_k}}, \\
 \hookrightarrow & \Delta \varepsilon_{k+1} = -\frac{ \int_{\OmegaFt} H(\widehat{\phi} + \varepsilon_k)~\dOmega - V^0_1}{ \int_{\OmegaFt} \delta(\widehat{\phi} + \varepsilon_k)~\dOmega},
	\end{align} 
\end{subequations}
which is iterated till the relative error ($\Delta = f(\varepsilon_k)/V^0_1$) drops to machine accuracy. To show the similarity between the exact and approximate Lagrange multiplier approaches for three phase flows, we need an expression for level set correction per unit time $\frac{\d \varepsilon}{\d t}$ instead of the total one: $\varepsilon$. This can obtained by differentiating $f(\varepsilon)$, which is a non conserved quantity defined over a moving domain, with respect to time using the Leibniz integral rule   
\begin{subequations}
	\begin{alignat}{2}
		& \frac{\text{d}f(\varepsilon)}{\text{d}t} = \int_{\OmegaFt} \delta(\widehat{\phi} + \varepsilon) \left( \frac{\partial \widehat{\phi}}{\partial t} + \depsdt \right)~\dOmega + \int_{\dOmegaSt} H(\widehat{\phi} + \varepsilon) (\us \cdot \n)~\dS = 0, \\
		\hookrightarrow & \depsdt = \frac{\int_{\OmegaFt} \delta(\widehat{\phi} + \varepsilon) (\bu \cdot \grad(\widehat{\phi} + \varepsilon))~\dOmega - \int_{\dOmegaSt} H(\widehat{\phi} + \varepsilon) (\us \cdot \n)~\dS}{\int_{\OmegaFt} \delta(\widehat{\phi} + \varepsilon)~\dOmega}, \label{eq_non_immersed_dedt1} \\
		\hookrightarrow & \depsdt =  \frac{ \int_{\OmegaFt} \delta(\phi) (\bu \cdot \grad\phi)~\dOmega - \int_{\dOmegaSt} H(\phi) (\us \cdot \n) \dS}{\delta(\phi_m)\int_{\OmegaFt} \delta(\phi)~\dOmega}\, \delta(\phi_m)  =   \widetilde{\alpha} \, \delta(\phi_m).\label{eq_non_immersed_dedt2}
	\end{alignat}
\end{subequations}
Here, again it can be seen that the RHS of Eq.~\eqref{eq_LS_non_immersed_3} is similar to $\depsdt$ with a slight difference in $\alpha$ and $\widetilde{\alpha}$:  the latter is an approximation to $\alpha$ in which one of the smooth delta function terms (in the denominator) is evaluated at the $m^\text{th}$ contour. Overall, there is no spatial variation in the correction and $\phi$ retains its signed distance property. Moreover, the root finding technique avoids calculating the numerator of $\alpha$ or $\widetilde{\alpha}$.

\subsubsection{Conserving mass/volume for three phase flows - Immersed formulation}
\label{subsec_3phase_Immersed_formulation}

For non-immersed formulations with Cartesian grids, significant bookkeeping and stencil modifications are required to avoid solving the level set equation inside the (moving) structure domain $\Omega_3$. With unstructured grids, when a solid body moves, the mesh can deform significantly, causing several problems. An alternate approach is to use an immersed formulation and allow the two-fluid interface to pass through the solid body as shown in Fig~\ref{fig_schematic_immersed_formulation}. The implementation of the level set method is greatly simplified as a result. Because the two phases existing inside the solid body are fictitious, care must be taken when imposing constraints. To calculate the actual volume (or mass) of the three phases, we introduce a second level set function $\zeta$ (outlined in Sec.~\ref{subsec_3phase_flows}), which can be expressed as follows:     
\begin{subequations} \label{eq_V1V2V3}
	\begin{align}
		V_1 &= \int_{\Omega} H(\phi) H(\zeta)~\dOmega, \label{eq_M1_V1_3phase} \\
		V_2 &= \int_{\Omega} (1-H(\phi)) H(\zeta)~\dOmega, \label{eq_M2_V2_3phase} \\
		V_3 &= \int_{\Omega} (1-H(\zeta))~\dOmega. 	\label{eq_M3_V3_3phase}
	\end{align}
\end{subequations}
In Eqs.~\eqref{eq_M1_V1_3phase} and~\eqref{eq_M2_V2_3phase} the inclusion of $H(\zeta)$ term ensures that the fluid volume is considered only outside the body. We will assume that the solid volume $V_3$ remains conserved, and the issue of mass/volume loss pertains to only fluids 1 and 2.  

Using the RTT (Eq.~\eqref{eq_RTT}), the time derivatives of $V_1$ and $V_2$ (conserved quantities) in the closed and stationary computational domain $\Omega$ are obtained as:
\begin{subequations}
\begin{align}
	\frac{\text{d}V_1}{\text{d}t} &= \int_{\Omega} \left[ \delta(\phi) H(\zeta) \dphidt + H(\phi) \delta(\zeta) \dzetadt \right]~\dOmega,  \label{eq_dV1dt_3phase} \\
	 \frac{\text{d}V_2}{\text{d}t} &= \int_{\Omega} \left[ -\delta(\phi) H(\zeta) \dphidt + (1-H(\phi)) \delta(\zeta) \dzetadt \right]~\dOmega. \label{eq_dV2dt_3phase} 
\end{align}
\end{subequations}
The strong and weak form of the level set equation remains the same as Eq.~\eqref{eq_sf_nib} and Eq.~\eqref{eq_wf_nib}, respectively, with the difference that they are now defined over the entire (static) domain $\Omega$. That is $\phi \in H^1(\Omega)^3$ and $\varn \phi \in H^1(\Omega)^3$. Working with the constraint of conserving fluid-1 volume, the Lagrangian of the constraint and its variation reads as
\begin{subequations}
	\begin{alignat}{2}
		\mathcal{L}(\Lambda_3, \phi)~~&\defn~~\Lambda_3 \int_{\Omega}  \left(H(\phi)H(\zeta) - \frac{V^0_1}{V} \right)~\dOmega,~~\Lambda_3 \in  L^2(\Omega)^3,  \label{eq_lag_ib} \\
		 \varn \mathcal{L}(\Lambda_3, \phi)~~&\defn~~\varn  \Lambda_3 \int_{\Omega}  \left(H(\phi)H(\zeta) - \frac{V^0_1}{V} \right)~\dOmega +   \Lambda_3 \int_{\Omega}  \delta(\phi)H(\zeta) \varn \phi~\dOmega,~~\varn \Lambda_3 \in  L^2(\Omega)^3. \label{eq_var_lag_ib}
	\end{alignat}
\end{subequations}
Adding variation of the Lagrangian to the weak form, collecting terms in $\varn \Lambda_3$  and $\varn \phi$, and equating them to zero separately, yield the original constraint and a new dynamical equation for $\phi$ that reads as
\begin{subequations} \label{eq_LS_immersed_set1}
	\begin{align}
		&\dphidt + \bu \cdot \grad \phi = -\Lambda_3 \delta(\phi) H(\zeta), \label{eq_LS1_immersed_eq1} \\
		&\Lambda_3 = \frac{-\int_{\Omega} \delta(\phi) H(\zeta) (\bu \cdot \grad \phi)~\dOmega - \int_{\Omega} H(\phi) \delta(\zeta) (\u \cdot \grad \zeta)~\dOmega}{\int_{\Omega} \delta^2(\phi) H^2(\zeta)~\dOmega} \label{eq_Lambda_3}, \\
		\hookrightarrow & \Lambda_3  = \frac{-\int_{\Omega} \delta(\phi) H(\zeta) (\bu \cdot \grad \phi)~\dOmega +  \int_{\dOmegaSt} H(\phi) (\us \cdot \n)~\dS}{\int_{\Omega} \delta^2(\phi) H^2(\zeta)~\dOmega}. \label{eq_Lambda_3_final}
	\end{align}
\end{subequations}	
The value of the Lagrange multiplier $\Lambda_3$ in Eq.~\eqref{eq_Lambda_3} is obtained by substituting $\dphidt = -\bu \cdot \grad \phi  -\Lambda_3 \delta(\phi) H(\zeta)$ from Eq.~\eqref{eq_LS1_immersed_eq1} and $\partial \zeta/\partial t = -\u \cdot \grad \zeta$ from Eq.~\eqref{eq_zeta_advection} into Eq.~\eqref{eq_dV1dt_3phase} and setting $\frac{\d V_1}{\d t} = 0$. If the body delta function $\delta(\zeta)$ is sharp, then the expression for $\Lambda_3$ in Eq.~\eqref{eq_Lambda_3} can be further simplified to Eq.~\eqref{eq_Lambda_3_final} by using the relation $\int_{\Omega} H(\phi) \delta(\zeta) (\u \cdot \grad \zeta) =     \int_{\dOmegaSt} H(\phi) (\us \cdot \grad \zeta)~\dS =  \int_{\dOmegaSt} H(\phi) (\us \cdot \n_{\rm s})~\dS  = - \int_{\dOmegaSt} H(\phi) (\us \cdot \n)~\dS$. Here, $\n_{\rm s} = -\n$ is the outward unit normal to the solid surface, which can be obtained from the signed distance function $\zeta$ as $\n_{\rm s} =  \grad \zeta / |\grad \zeta| = \grad \zeta$.

Proceeding analogously, we can conserve the volume of fluid-2 with the help of the Lagrange multiplier $\Lambda_4 \in L^2(\Omega)^3$. In this case the level set equation and the value of the Lagrange multiplier reads as 
\begin{subequations} \label{eq_LS_immersed_set2}
	\begin{align}
		& \dphidt + \bu \cdot \grad \phi = \Lambda_4  \delta(\phi) H(\zeta), \label{eq_LS2_immersed_eq1} \\
		& \Lambda_4 = \frac{\int_{\Omega} \delta(\phi) H(\zeta) (\bu \cdot \grad \phi)~\dOmega - \int_{\Omega} (1-H(\phi)) \delta(\zeta) (\u \cdot \grad \zeta)~\dOmega}{\int_{\Omega} \delta^2(\phi) H^2(\zeta)~\dOmega}, \label{eq_Lambda_4} \\
		\hookrightarrow & \Lambda_4 = \frac{\int_{\Omega} \delta(\phi) H(\zeta) (\bu \cdot \grad \phi)~\dOmega - \int_{\dOmegaSt} H(\phi) (\us \cdot \n)~\dS}{\int_{\Omega} \delta^2(\phi) H^2(\zeta)~\dOmega}. \label{eq_Lambda_4_final}
	\end{align}
\end{subequations}
Comparing equations sets~\eqref{eq_LS_immersed_set1} and \eqref{eq_LS_immersed_set2}, we observe the relation $-\Lambda_4 = \Lambda_3$. Therefore,  the level set equations~\eqref{eq_LS1_immersed_eq1} and \eqref{eq_LS2_immersed_eq1} are the same.  More importantly, the immersed formulation of the mass/volume conserving level set equation converges to the non-immersed one as the smeared body delta function $\delta(\zeta)$ becomes sharp. This can be observed by comparing equation sets~\eqref{eq_LS_non_immersed_set1} and~\eqref{eq_LS_non_immersed_set2}, and equation sets~\eqref{eq_LS_immersed_set1} and~\eqref{eq_LS_immersed_set2}. Outside the solid region where $H(\zeta) = 1$, these equations are exactly the same.  The equations remain the same if we impose mass conservation constraints instead of volume ones. The derivation steps are omitted for brevity.    

\vspace{1em}
\noindent \textbf{\underline{Approximate Lagrange multiplier approach:}} Here we derive an approximate Lagrange multiplier to conserve mass/volume of the two fluid phases in an immersed sense. The nonlinear equation to solve for the spatially uniform correction $\varepsilon$ in this case is 
\begin{equation}
	\mathcal{C}(\widehat{\phi} + \varepsilon) = f(\varepsilon) = \int_{\Omega} H(\widehat{\phi} + \varepsilon) H(\zeta)~\dOmega - V_1^0 = 0
	\label{eq_f_epsilon_immersed}
\end{equation}
Eq.~\eqref{eq_f_epsilon_immersed} can be solved to machine accuracy using Newton's method. As done before, the connection between the approximate and exact Lagrange multipliers emerges through $\depsdt$. Differentiating the non-conserved quantity $f(\varepsilon)$ defined over a static domain $\Omega$ with respect to time (using the Leibniz integral rule) yields  
\begin{subequations}
	\begin{align}
		& \depsdt = \frac{\int_{\Omega} \delta(\widehat{\phi} + \varepsilon) H(\zeta) (\bu \cdot \grad (\widehat{\phi} + \varepsilon))~\dOmega + \int_{\Omega} H(\widehat{\phi} + \varepsilon) \delta(\zeta) (\us \cdot \grad \zeta) \dOmega}{\int_{\Omega} \delta(\widehat{\phi} + \varepsilon) H(\zeta)~\dOmega}, \label{eq_immersed_dedt1} \\
		\hookrightarrow & \depsdt = \widetilde{\Lambda} \delta(\phi_m)H(\zeta_n) =  \frac{\left[ \int_{\Omega} \delta(\phi) H(\zeta) (\bu \cdot \grad \phi)~\dOmega - \int_{\dOmegaSt} H(\phi) (\us \cdot  \n)~\dS \right]}{\delta(\phi_m) H(\zeta_n) \int_{\Omega} \delta(\phi) H(\zeta)~\dOmega} \; \delta(\phi_m)H(\zeta_n). \label{eq_immersed_dedt2}
	\end{align}
\end{subequations}
Comparing Eqs.~\eqref{eq_Lambda_4_final} and~\eqref{eq_immersed_dedt2}, it can be seen that $\widetilde{\Lambda}$ is approximately equal to the exact Lagrange multiplier $\Lambda_4$: the former is obtained by evaluating one of the smooth delta function terms and one of the smooth Heaviside function terms (in the denominator) at the $m^\text{th}$ and $n^\text{th}$ contour, respectively. Overall, there is no spatial variation in the correction and $\phi$ retains its signed distance property.

Enforcing the constraint defined by Eq.~\eqref{eq_f_epsilon_immersed} using Newton's method ensures that the volume of phase-1 (defined by Eq.~\eqref{eq_M1_V1_3phase}) is preserved to machine accuracy. Since the sum of the volumes for all three phases (Eqs.~\eqref{eq_M1_V1_3phase} - \eqref{eq_M3_V3_3phase}) equals the total volume of the domain, this constraint guarantees a key property for closed domains: if the volume of phase-1 is conserved to machine precision, the combined volumes of phase-2 and phase-3 will also be conserved to machine precision irrespective of the smeared body Heaviside function. The numerical simulations of Sec.~\ref{sec_results_and_discussion} confirm this property. 

\subsubsection{Contact angle conditions}
\label{sec_triple_pts}
Material triple points are the points (or lines in 3D) where the two-fluid interface $\dOmegaF$ intersects the solid surface $\dOmegaS$. Under equilibrium/static conditions, $\dOmegaF$ pins at $\dOmegaS$ at an angle $\theta_\text{s}$ according to the Young-Laplace equation. A dynamic contact angle $\theta_\text{d}$ condition is better suited for transient conditions. For problems at the capillary length scale, the contact angle condition is relevant. The capillary length scale\footnote{The capillary length scale can be estimated as $l_\text{c} \, \propto \, \sqrt{\frac{\sigma}{(\rhol-\rhog)g}}$.} in an air-water system is about $2.7$ mm. The contact angle condition is not necessary for many fluid-structure problems in ocean engineering, since the relevant length scales are much larger than capillary ones. 

If a specific contact angle $\theta$ needs to be imposed at the triple points, an equation similar to the reinitialization equation Eq.~\eqref{eq_HJ} can be used. This idea is proposed by Jettestuen et al.~\cite{jettestuen2013level} who suggest using an equation of the form 
\begin{equation}
\frac{\partial \phi}{\partial \tau} + \text{sgn}(\zeta)\left(\grad \zeta \cdot \grad \phi - \cos(\theta) |\grad \phi| |\grad \zeta| \right) = 0. \label{eq_contact_angle}
\end{equation}
As with the reinitialization equation, the contact angle Eq.~\eqref{eq_contact_angle} is also solved till steady state to obtain the desired geometric relation between the fluid and solid level sets at the triple points
\begin{equation}
\frac{\grad \zeta}{|\grad \zeta|} \cdot \frac{\grad \phi}{|\grad \phi|} = \cos(\theta).  
\end{equation}

Eq.~\eqref{eq_contact_angle} implies that the level set function $\phi$ moves with a velocity $\u^\text{cont} = \text{sgn}(\zeta) \left( \grad \zeta - \cos(\theta) \frac{\grad \phi}{|\grad \phi|}\right)$\footnote{This additional motion also relaxes the no-slip condition on the fluid-solid interface.}. Overall $\phi$ moves with a combination of fluid velocity $\u$, reinitialization velocity $\u^\text{reinit}$ and contact angle imposing velocity $\u^\text{cont}$. If we denote the overall velocity by $\breve{\bu}$, then the analysis presented in Sec.~\ref{subsec_3phase_non_immersed_formulation} still applies. The terms involving $\bu$ are replaced by $\breve{\bu}$. Additionally, it can be seen that $\Lambda_3$ and $\Lambda_4$ will disrupt the signed distance property of $\phi$, as well as change the contact angle at the triple points. In contrast, the approximate Lagrange multiplier approach will respect the contact angle condition since the shifted contours of $\phi$ retain their original geometric shape. 

Due to the length scales of FSI problems consider in this work, we do not consider the contact angle condition. This will be examined with appropriate problems in a future study. 
\subsubsection{Differential treatment of the Brinkman penalty force}
\label{sec_penalty_split}

In the Brinkman penalty method the permeability of the body needs to be low $\kappa \ll 1$ to represent a non-porous body. In practice, while $\kappa$ value is small, it does not approach machine zero. This implies that the no-slip condition on the fluid-solid interface is only partially enforced. The approach of $\kappa$ to machine zero poses two issues:
\begin{enumerate}
\item the system of equations becomes stiff to solve; and 
\item a no-slip condition in the tangential direction implies stress singularity at the material triple points.
\end{enumerate}
 
Our recent paper~\cite{Rama2023} describes a preconditioning strategy that overcomes the numerical stiffness issue of the Brinkman penalization method: the fluid solver converges rapidly regardless of $\kappa$. Therefore, issue \#1 is not a concern with our implementation. The second issue is unavoidable, and one must allow for tangential slip at the triple points. This is necessary for the solid to move across/pierce the two fluid interfaces. The classical no-slip condition at the material triple point breaks down as discussed in Huh and Scriven~\cite{huh1971hydrodynamic}. A question naturally arises in light of issue \#2 for the Brinkman penalization method: is it possible to impose different amounts of slip in the normal and tangential directions? For example, no slip in the normal direction and some slip in the tangential direction. Using numerical experiments, this section explores this possibility. 

\begin{figure}[]
 \centering
 \subfigure[t = 0 s]{
 	\includegraphics[scale= 0.335]{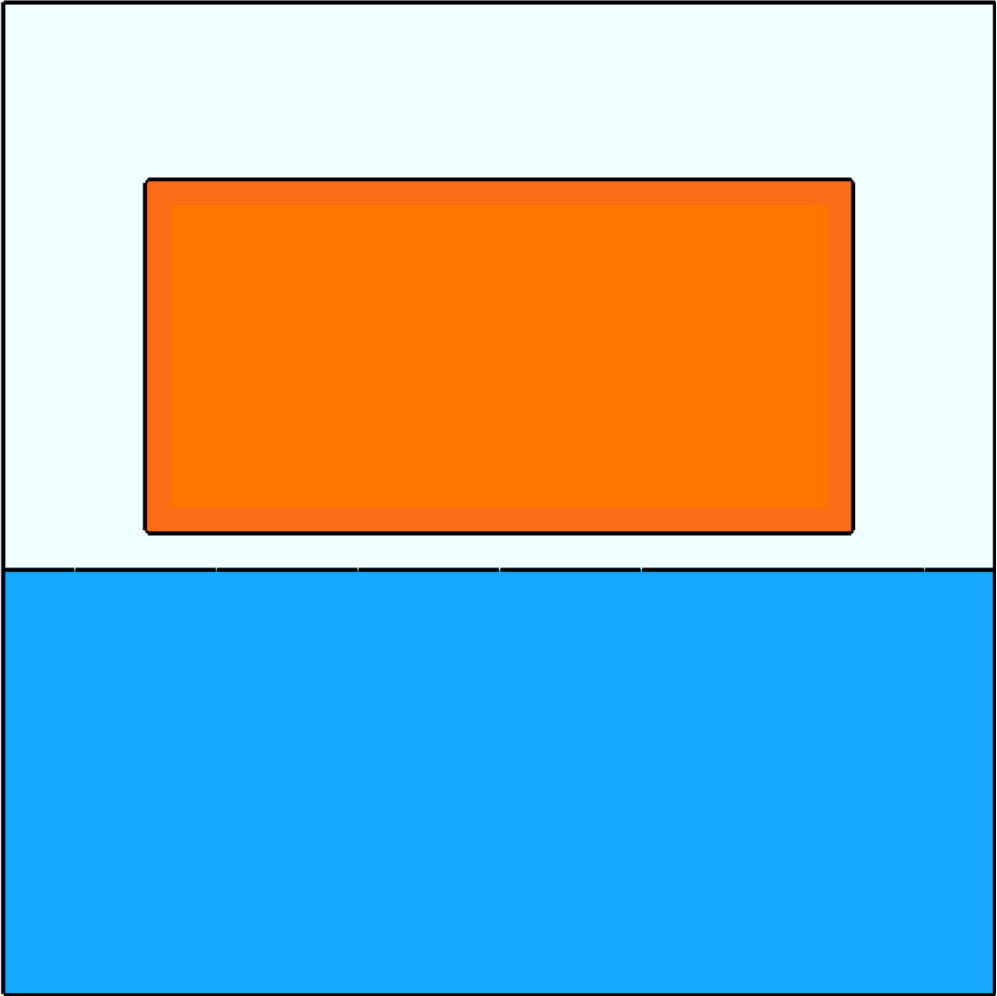}
	\label{fig_rect_kn50_T0}
 }
  \subfigure[t = 0.13679 s]{
 	\includegraphics[scale= 0.335]{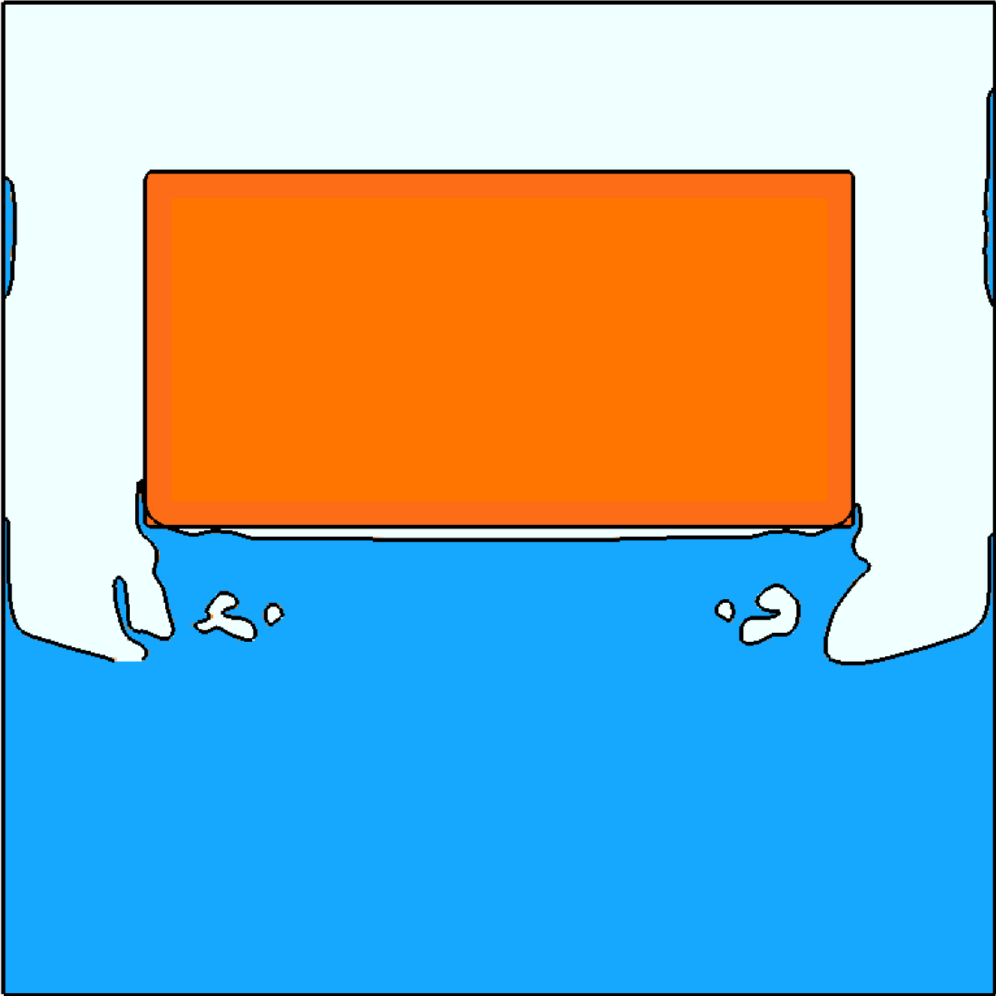}
	\label{fig_rect_kn50_T1}
 }    
 \subfigure[t = 0.136794 s]{
 	\includegraphics[scale= 0.27]{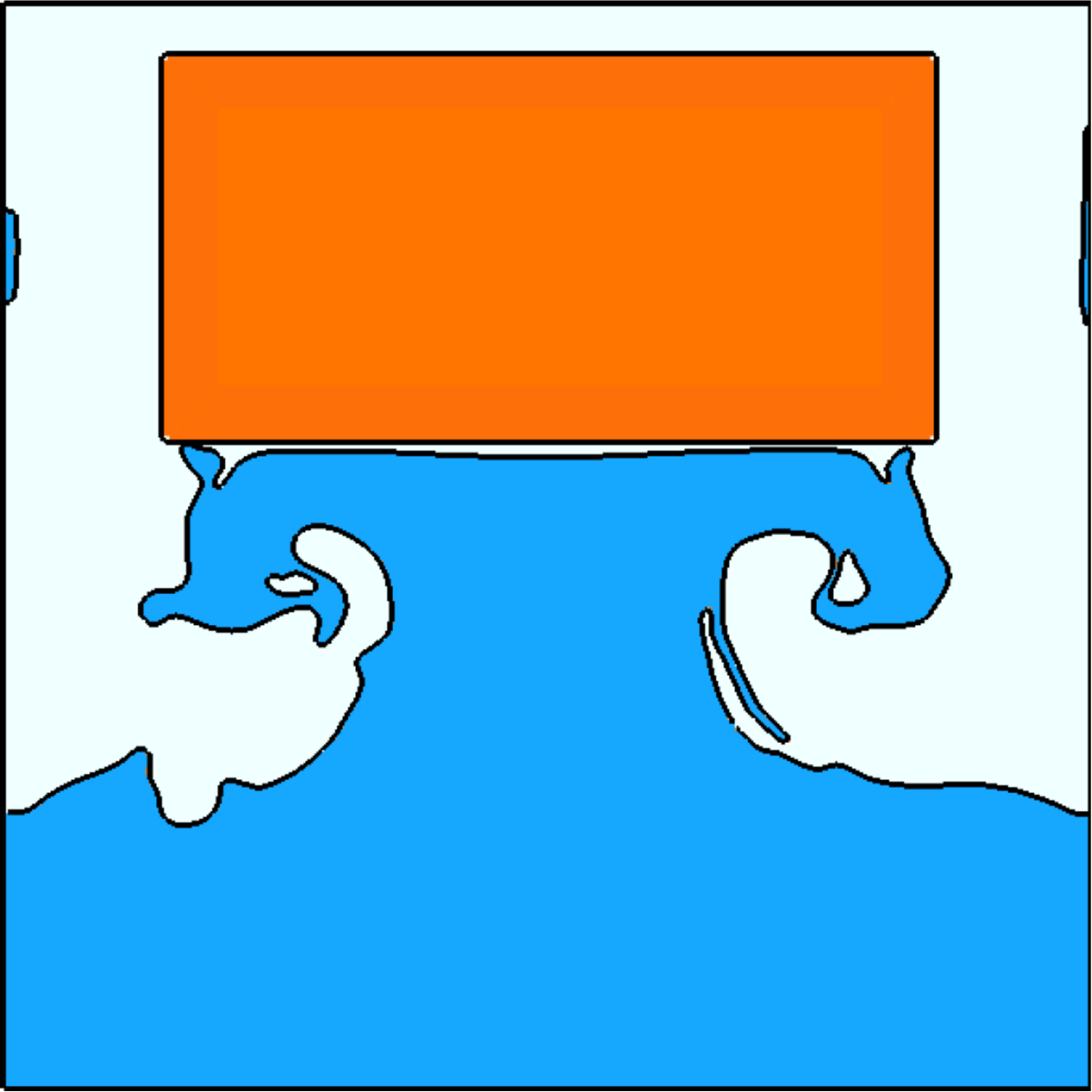}
	\label{fig_rect_kn50_T2}
 }    
 \subfigure[t = 0.13717 s]{
 	\includegraphics[scale= 0.335]{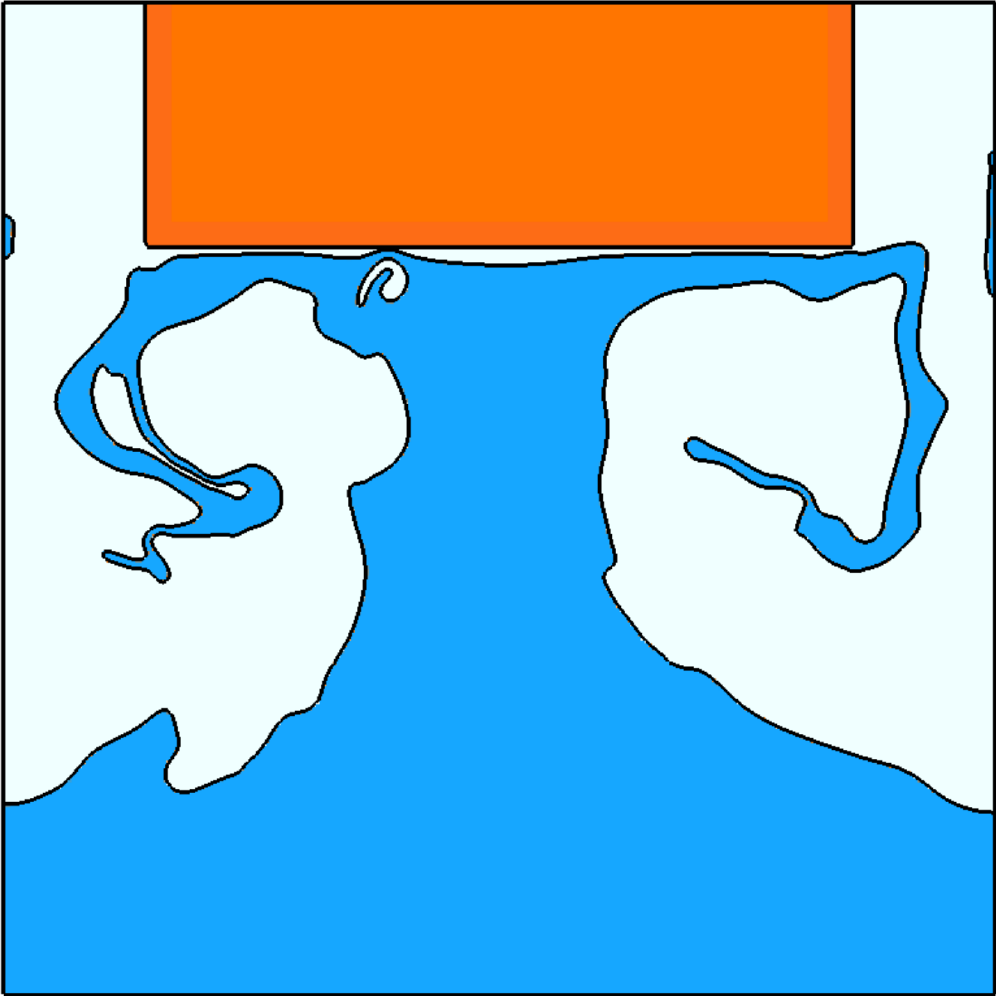}
	\label{fig_rect_kn50_T3}
 }  
 \caption{Three phase fluid-structure interaction of a rectangular block when the air-water interface is not allowed to enter the solid body ($\mathcal{K}_{\rm n} = 50, \mathcal{K}_{\rm t} = 1$).}
 \label{fig_rect_kn50}
\end{figure}

A three phase FSI problem is simulated in which a rectangular block is released from a small height above the air-water interface. Detailed information on the problem setup is provided in Sec.~\ref{subsubsec_floating_rect}, where we focus on the mass loss issues with the level set method. We focus here on the dynamics of the block and the air-water interface as a function of differential slip. This case is studied by splitting the Brinkman penalty force, described in Eq.~\eqref{eqn_brinkman_force} into normal and tangential components as
\begin{equation}
	\f_c(\x, t) = \frac{\chi(\x,t)}{\kappa}\left[ \mathcal{K}_{\rm n} (\u_b(\x,t) - \u(\x,t)) \cdot (\ns \otimes \ns) + \mathcal{K}_{\rm t} (\u_b(\x,t) - \u(\x,t)) \cdot (\I - (\ns \otimes \ns)) \right],
	\label{eq_Brinkman_split}
\end{equation}
in which, $\mathcal{K}_{\rm n}$ and $\mathcal{K}_{\rm t}$ are the normal and tangential penalty factors, respectively, and $\ns$ is the unit normal to the solid surface. The splitting of the Brinkman force into normal and tangential components is done in the fluid-solid interfacial zone. Inside the body, the penalty force reverts to its original form. When $\mathcal{K}_{\rm n} = 1$ and $\mathcal{K}_{\rm t} = 1$, Eq.~\eqref{eq_Brinkman_split} reverts back to Eq.~\eqref{eqn_brinkman_force}. We impose a no-normal penetration condition by increasing the normal penalty factor to $\mathcal{K}_{\rm n} = 50$. The tangential penalty factor is kept at $\mathcal{K}_{\rm t} = 1$. The large value of the normal penalty factor restricts the air-water interface within the solid body. Fig.~\ref{fig_rect_kn50} displays very unphysical FSI dynamics in this case, where the block appears to ``ride the waves". After being released, the rectangular block bounces to a higher elevation and continues to rise. 

Fig.~\ref{fig_rect_FSI} of Sec.~\ref{subsubsec_floating_rect} shows physically correct dynamics when both $\mathcal{K}_{\rm n}$ and $\mathcal{K}_{\rm t}$ are set to 1. The air-water interface is thus allowed to enter the rectangle block. As long as the mass/volume conservation constraint (of the actual fluid) is maintained, it is not a problem for the fluid to penetrate into the solid region. It corresponds to the immersed formulation of the mass-preserving level set method.

\section{Discretization}
\label{sec_discretization}
\subsection{Spatial discretization} 
\label{sec_spatial_discretization}

\begin{figure}
  \centering
  \subfigure[2D staggered Cartesian grid]{
    \includegraphics[scale = 0.45]{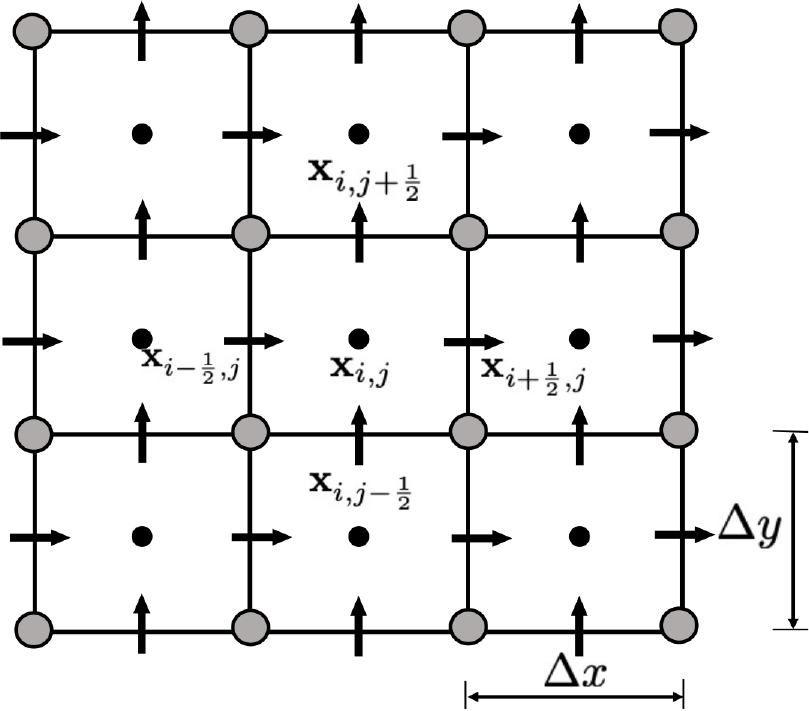} 
    \label{fig_discretized_staggered_grid}
  }
   \subfigure[A single grid cell]{
    \includegraphics[scale = 0.35]{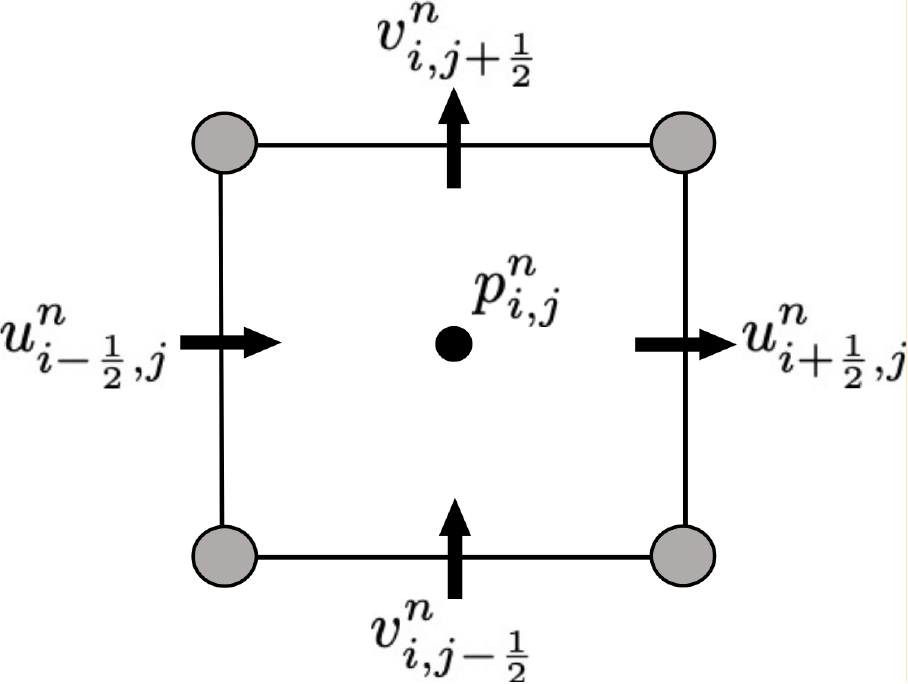}
    \label{fig_single_cell}
  }
  \caption{Schematic representation of a 2D staggered Cartesian grid. \subref{fig_discretized_staggered_grid} A staggered grid coordinate system, and \subref{fig_single_cell} a single grid cell with velocity components $u$ and $v$ approximated at the cell faces (${\bf{\rightarrow}}$) and scalars pressure $p$ and level set function $\phi$ approximated at the cell center ($\bullet$) at $n^\text{th}$ time step.}
\label{fig_cfd_domains}
\end{figure}

The continuous equations of motion for the incompressible multiphase flow system written in Eqs.~\eqref{eqn_momentum} and \eqref{eqn_continuity} are discretized on a locally refined staggered Cartesian grid. The coarsest grid level covers the entire domain $\Omega$ with $\Nx \times \Ny \times \Nz$ rectangular cells. The cell size in $x$, $y$, and $z$ directions is  $\dx$, $\dy$, and $\dz$, respectively. Unless stated otherwise, a uniform grid spacing  $\Delta x = \Delta y = \Delta z = h$ is used for all simulations in this work. Without any loss of generality, the lower left corner of the domain is taken as the origin $(0, 0, 0)$. The center of the cell has a position $\x_{i,j,k} = \left((i + \half)\dx,(j + \half)\dy,(k + \half)\dz\right)$ for $i = 0, \ldots, \Nx - 1$, $j = 0, \ldots, \Ny - 1$, and $k = 0, \ldots, \Nz - 1$. The cell face location that is half a grid cell away from $\x_{i,j,k}$ in the $x$-direction is at $\x_{i-\half,j,k} = \left(i\dx,(j + \half)\dy,(k + \half)\dz\right)$. Similarly, for the location of a cell face that is half a grid cell away from $\x_{i,j,k}$ in the $y$-directions is $\x_{i,j-\half,k} =\left((i + \half)\dx,j\dy,(k + \half)\dz\right)$ and in the $z$-direction it is $\x_{i,j,k-\half} =\left((i + \half)\dx,(j + \half)\dy,k\dz\right)$. Fig.~\ref{fig_discretized_staggered_grid} illustrates the staggered arrangement of the variables on a 2D grid. The simulation time at time step $n$ is denoted by $t^n$. The scalar quantities: level set functions, pressure, and the material properties (density and viscosity) are approximated at cell centers and are denoted $\phi_{i,j,k}^{n} \approx \phi \left(\x_{i,j,k}, t^n\right)$, $\zeta_{i,j,k}^{n} \approx \zeta\left(\x_{i,j,k}, t^n\right)$, $p_{i,j,k}^{n} \approx p\left(\x_{i,j,k},t^{n}\right)$, $\rho_{i,j,k}^{n} \approx \rho\left(\x_{i,j,k},t^{n}\right)$ and  $\mu_{i,j,k}^{n} \approx \mu\left(\x_{i,j,k},t^{n}\right)$, respectively. See also Fig.~\ref{fig_single_cell}. Scalar quantities are interpolated onto the required degrees of freedom when required, see Nangia et al.~\cite{Nangia2019MF} for further details. The vector velocity is approximated on the cell face as $u_{i-\half,j,k}^{n} \approx u\left(\x_{i-\half,j,k}, t^{n}\right)$, $v_{i,j-\half,k}^{n} \approx v\left(\x_{i,j-\half,k}, t^{n}\right)$,  $w_{i,j,k-\half}^{n} \approx w\left(\x_{i,j,k-\half}, t^{n}\right)$. The body force terms in the momentum equation are also approximated at the cell faces. Second-order finite differences are used for all spatial derivatives.  For the ease of readability, the discretized version of the differential operators are denoted with a $h$ subscript, e.g., $\grad \approx \grad_h$. For further details on the spatial discretization and boundary conditions on adaptively refined meshes, refer our prior works~\cite{Nangia2019MF, Nangia2019, Bhalla13}.

\subsection{Density and viscosity specification}

Smooth Heaviside functions are used to vary material properties at the fluid-fluid and fluid-solid interfaces $\dOmegaFt$ and $\dOmegaSt$, respectively. A given material property $\Im$, say density or viscosity, is set in the computational domain by first calculating the \emph{flowing} phase property as
\begin{equation}
	\Im^{\text{flow}}_{i,j,k} = \Im_l + (\Im_g - \Im_l) \widetilde{H}^{\text{flow}}_{i,j,k},
\label{eqn_ls_flow}
\end{equation}
and later correcting $\Im^{\text{flow}}$ to account for the solid body by 
\begin{equation}
	\Im_{i,j,k}^{\text{full}} = \Im_s + (\Im^{\text{flow}}_{i,j,k} - \Im_s) \widetilde{H}^{\text{body}}_{i,j,k}.
\label{eqn_ls_solid}
\end{equation}
Here, $\Im^{\text{full}}$ is the final scalar material property (density or viscosity) field throughout $\Omega$. For the transition specified by Eqs.~\eqref{eqn_ls_flow} and~\eqref{eqn_ls_solid}, the usual numerical Heaviside functions are used: 
\begin{align}
\widetilde{H}^{\text{flow}}_{i,j,k} &= 
\begin{cases} 
       0,  & \phi_{i,j,k} < -\ncells \, h,\\
        \frac{1}{2}\left(1 + \frac{1}{\ncells \, h} \phi_{i,j,k} + \frac{1}{\pi} \sin\left(\frac{\pi}{ \ncells \, h} \phi_{i,j,k}\right)\right) ,  & |\phi_{i,j,k}| \le \ncells \, h,\\
        1,  & \textrm{otherwise}.
\end{cases}       \label{eqn_Hflow} \\
\widetilde{H}^{\text{body}}_{i,j,k} &= 
\begin{cases} 
       0,  & \zeta_{i,j,k} < -\ncells \, h,\\
        \frac{1}{2}\left(1 + \frac{1}{\ncells \, h} \zeta_{i,j,k} + \frac{1}{\pi} \sin\left(\frac{\pi}{ \ncells \, h} \zeta_{i,j,k}\right)\right) ,  & |\zeta_{i,j,k}| \le \ncells \, h,\\
        1,  & \textrm{otherwise}.  \label{eqn_Hbody}
\end{cases}
\end{align}

All simulations performed in this study use $\ncells = 1$ for two-phase flows and $\ncells = 2$ (for fluid-solid and fluid-fluid interfaces) for three-phase flows. 

\subsection{Temporal discretization} \label{sec_temp_disc}

We employ a fixed-point iteration time stepping scheme with $\ncycles$ cycles per time step to evolve quantities from time level $t^n$ to time level $t^{n+1} = t^n + \Delta t$. Superscript $k$ denotes the cycle number for the fixed-point iteration. At the beginning of each time step, the solutions from the previous time step are used to initialize cycle $k = 0$: $\u^{n+1,0} = \u^{n}$, $p^{n+\half,0} = p^{n-\half}$, $\phi^{n+1,0} = \phi^{n}$ and $\zeta^{n+1,0} = \zeta^{n}$. Initial conditions are prescribed for physical quantities at the initial time $n = 0$. A larger number of cycles in the simulation allows a larger, more stable time step size. In this work, we limit $\ncycles = 2$ for 2D cases and $\ncycles = 1$ for 3D cases to reduce linear solver computational costs.

\subsubsection{Level set advection}

We evolve the two level set/signed distance functions $\phi$ and $\zeta$ using a standard explicit advection scheme as follows
\begin{align}
\frac{\phi^{n+1,k+1} - \phi^{n}}{\dt} + Q\left(\u^{n+\half,k}, \phi^{n+\half,k}\right) &= 0, \label{eq_linear_advection_phi} \\
\frac{\zeta^{n+1,k+1} - \zeta^{n}}{\dt} + Q\left(\u^{n+\half,k}, \zeta^{n+\half,k}\right) &= 0 , \label{eq_linear_advection_zeta}
\end{align}
in which $Q(\cdot,\cdot)$ denotes the cubic upwind interpolation (CUI) approximation to the linear advection terms on cell centers. Let $\widetilde{\phi}^{n+1}$ denote the level set function following an advection procedure after time stepping through the interval
  $[t^n, t^{n+1}]$. We aim to reinitialize it to obtain a signed distance function. As proposed by Sussman et al.~\cite{Sussman1994}, this can be achieved by computing a steady-state solution to the Hamilton-Jacobi Eq.~\eqref{eq_HJ}. The spatial gradients involved in the Hamilton-Jacobi equation are typically discretized using high-order essentially non-oscillatory (ENO) or weighted ENO (WENO) schemes. 

With the solid body geometries considered in this study, we are able to reset the solid level set function $\zeta$ analytically. The analytical resonstruction preserves the mass/volume of the body while not distorting $\zeta$'s signed distance property following the linear advection Eq.~\eqref{eq_linear_advection_zeta}.

\subsubsection{Multiphase incompressible Navier-Stokes solution}

The multiphase incompressible Navier-Stokes Eqs.~\eqref{eqn_momentum} and \eqref{eqn_continuity} is discretized in conservative form as
\begin{align}
	\frac{\breve{\V \rho}^{n+1,k+1} \u^{n+1,k+1} - { \V \rho}^{n} \u^n}{\dt} + \C^{n+1,k} &= -\grad_h \, p^{n+\half, k+1}
	+ \left(\L_{\mu} \u\right)^{n+\half, k+1}
	+  \V \wp^{n+1,k+1}\g +  \f_c^{n+1,k+1} + \fst^{n+\half,k+1}, \label{eqn_c_discrete_momentum}\\
	 \grad \cdot \u^{n+1,k+1} &= \V{0}, \label{eqn_c_discrete_continuity}
\end{align}
in which, $\C^{n+1,k}$ is the discrete version of the convective term $\div (\V{m_\rho} \otimes \u)$. A consistent mass/momentum transport scheme is used to compute the density $\breve{\V \rho}^{n+1,k+1}$ and mass flux $\V{m_\rho} \equiv \rho \u$ approximations, ensuring numerical stability even under high density contrasts. Our prior works~\cite{Nangia2019MF,Bhalla2020} provide a detailed description of the consistent mass/momentum transport scheme. The viscous strain rate in Eq.~\eqref{eqn_c_discrete_momentum} is calculated using the Crank-Nicolson approximation: $\left(\L_{\mu} \u\right)^{n+\half, k+1} =  \half\left[\left(\L_{\mu} \u\right)^{n+1,k+1} + \left(\L_{\mu} \u\right)^n\right]$, in which $\left(\L_{\mu}\right)^{n+1} = \grad_h \cdot \left[\mu^{n+1} \left(\grad \u + \grad \u^T\right)^{n+1}\right]$. The newest approximation to viscosity $\mu^{n+1,k+1}$ is obtained using the two-stage process described by Eqs.~\eqref{eqn_ls_flow} and~\eqref{eqn_ls_solid}. 
\KK{To avoid generating spurious currents due to large density variations near the fluid-solid interface~\cite{Nangia2019}, the gravitational body force term $\V \wp \g = \V{\rho}^{\text{flow}} \g$ is calculated using the flow density field; see also Eq.~\eqref{eqn_ls_flow}.}

\subsubsection{Fluid-structure coupling}
\label{sec_fsi_coupling}

The Brinkman penalization term $\f_c$ written in Eq.~\eqref{eqn_brinkman_force} is treated implicitly in the momentum equation
\begin{align}
	\f_c^{n+1,k+1} = \frac{\widetilde{\chi}}{\kappa}\left(\u_b^{n+1,k+1} - \u^{n+1,k+1}\right).  \label{eqn_bp_discrete}
\end{align}
The indicator function $\widetilde{\chi}$ is defined using the body Heaviside function (Eq.~\eqref{eqn_Hbody}) as $\widetilde{\chi} = 1 - \widetilde{H}^{\text{body}}$. $\widetilde{\chi} = 1$ inside the solid region and zero outside. 
The rigid body velocity $\u_b$ is expressed as the sum of translational $\U_r$ and rotational $\W_r$ velocities:
\begin{equation}
	\u_b = \U_r + \W_r \times \left(\x-\Xcom\right),
\label{eq_ub_velocity}
\end{equation}
in which $\Xcom$ is the center of mass position of the body. The rigid body velocity is obtained by integrating Newton’s second law of motion
\begin{align}
	\Mb \frac{\Ur^{n+1,k+1} - \Ur^n}{\dt} &=  \cF^{n+1,k} + \Mb \g,  \label{eq_newton_u} \\
	 \frac{\Ib^{n+1,k+1} \Wr^{n+1,k+1} - \Ib^{n}\Wr^n}{\dt} &=  \cM^{n+1,k}  \label{eq_newton_w}
\end{align}
in which $\Mb$ is the mass of the body, $\Ib$ is its moment of inertia tensor, and $\cF$ is the net hydrodynamic force, $\cM$ is the net hydrodynamic torque and $\Mb \g$ is the net gravitational force acting on the body. Eqs.~\eqref{eq_newton_u} and~\eqref{eq_newton_w} are integrated using a forward-Euler scheme to compute $\Ur^{n+1,k+1}$, $\Wr^{n+1,k+1}$ and $\Xcom^{n+1,k+1}$. Only the vertical degree of freedom is considered free in the simulations presented in this work.

\subsubsection{Solution methodology: Projection preconditioner for the fully coupled Brinkman penalized Stokes system}
\label{sec_soultion_methodology}

To find the velocity, $\u^{n+1,k+1}$, and pressure, $p^{n+\half, k+1}$, at time step $n+1$, we solve Eqs.~\eqref{eqn_c_discrete_momentum} and \eqref{eqn_c_discrete_continuity} simultaneously. The discrete form of momentum and continuity equations in  matrix form reads as 
\begin{align}
	\M \; \x &= \b \nonumber \\
	\label{eq_stokes_system} 
\left[
	\begin{array}{cc}
	\A & \G\\
	-\vD\cdot & \mathbf{0} \\
	\end{array}
\right]
\left[
	\begin{array}{c}
  	\xu\\
  	\xp \\
	\end{array}
\right] & =
\left[
	\begin{array}{c}
 	\V{f_u}\\
	\V{0} \\
	\end{array}
\right].
\end{align}
The operator $\M$ on the left-hand side of Eq.~\eqref{eq_stokes_system} is the time-dependent, incompressible staggered Stokes operator with an additional Brinkman penalty term in the (1,1) block. We call it the Brinkman penalized Stokes operator or \emph{Stokes-BP} operator for short~\cite{Rama2023}.  Matrix $\A =  \frac{1}{\dt} \breve{\V {\rho}}^{n+1,k+1}  + \frac{1}{\kappa} \widetilde{\V{\chi}}^{n+1,k+1}  - \half \Lmu^{n+1,k+1}$ represents the discretization of the temporal, Brinkman penalty force and viscous terms, $\xu$ and $\xp$ are the velocity $\u^{n+1,k+1}$ and pressure $p^{n+1,k+1}$ degrees of freedom, and $\V{f_u}$ is the right hand side of the momentum equation
\begin{equation}
 \V{f_u} = \left(\frac{1}{\dt}  \vrho^{n} + \half \Lmu^{n}\right)\u^n + \left(\frac{\widetilde{\V{\chi}}}{\kappa} \right)^{n+1,k+1} \u_b^{n+1,k+1} - 
	\C^{n+1,k}+ \fst^{n+\half,k+1} + \V \wp^{n+1,k+1}\g.
\end{equation}
Within matrix $\A$, $\breve{\vrho}^{n+1,k+1}$ and $\widetilde{\V{\chi}}^{n+1,k+1} = \V{1} -\widetilde{\V{H}}^{\text{body}}$  are diagonal matrices of face-centered densities and body characteristic function corresponding to each velocity degree of freedom, respectively. The surface tension force $\fst$ acting on $\dOmegaF$ is modeled using the continuous surface tension formulation~\cite{brackbill1992continuum,francois2006balanced}. The continuous surface tension force reads as
\begin{equation}
\fst = \sigma \Upsilon(\phi) \grad \widetilde{B}, \label{eq_surface_tension}
\end{equation}
in which $\sigma$ is the uniform surface tension coefficient and $\Upsilon(\phi)$ is the curvature of the interface computed from the level set function $\Upsilon(\phi)^{n+\half,k+1} = -\div \left( \frac{\grad \phi}{|\grad \phi| } \right)$. 
In Eq.~\eqref{eq_surface_tension}, $ \widetilde{B}(\phi)$ represents a mollified Heaviside function that ensures that the surface tension force acts only near the two fluid interface. The immersed formulation of the level set equation implies that a fictitious surface tension also exists within the solid region. This however does not affect the momentum of the body; the fictious surface tension force is absent from the RHS of Eq.~\eqref{eq_newton_u}. 

Solving Eq.~\eqref{eq_stokes_system} iteratively becomes challenging when the body permeability $\kappa \ll 1$. This difficulty arises because a very large penalty term makes the system of equations very stiff. To address this issue, we recently proposed an effective preconditioning strategy for solving Eq.~\eqref{eq_stokes_system} using an FGMRES solver. This work adopts the same strategy. For a detailed explanation of the preconditioning method and its performance with very low values of $\kappa$, please refer to~\cite{Rama2023}.

\section{Software implementation}
\label{sec_software}
The numerical algorithms described in this work are implemented within the IBAMR library~\cite{IBAMR-web-page}, which is an open-source C++ software, enabling simulation of CFD and immersed boundary-like methods on adaptively refined grids. The code is hosted on GitHub at \url{https://github.com/IBAMR/IBAMR}. IBAMR relies on SAMRAI \cite{HornungKohn02, samrai-web-page} for Cartesian grid management and the AMR framework. Solver support in IBAMR is provided by the PETSc library~\cite{petsc-efficient, petsc-user-ref, petsc-web-page}.

\section{Results and discussion}
\label{sec_results_and_discussion}
In this section, several two-phase and three-phase flow problems are simulated to demonstrate the effectiveness of the approximate Lagrange multiplier technique. We selected the approximate Lagrange multiplier approach (henceforth called the mass loss fix) because of its simplicity, its ability to retain the signed distance property of the level set function, and its close relationship to the exact Lagrange multiplier approach, particularly if the latter method is implemented as an operator splitting technique. For three phase flows, two stringent test problems are devised to demonstrate that (i) imposing the mass conservation constraint is essential for the level set method to obtain correct dynamics; and (ii) the immersed formulation of the level set equation produces dynamics that agree very well with non-immersed and conservative methods such as moving unstructured grid-based methods, cut-cell methods, and particle-based methods.
\subsection{Two-phase flow problems}
\label{sec_2phase_cases}

\subsubsection{Vortex in a box problem}

\begin{figure}[]
   \centering
   	\includegraphics[scale= 0.45]{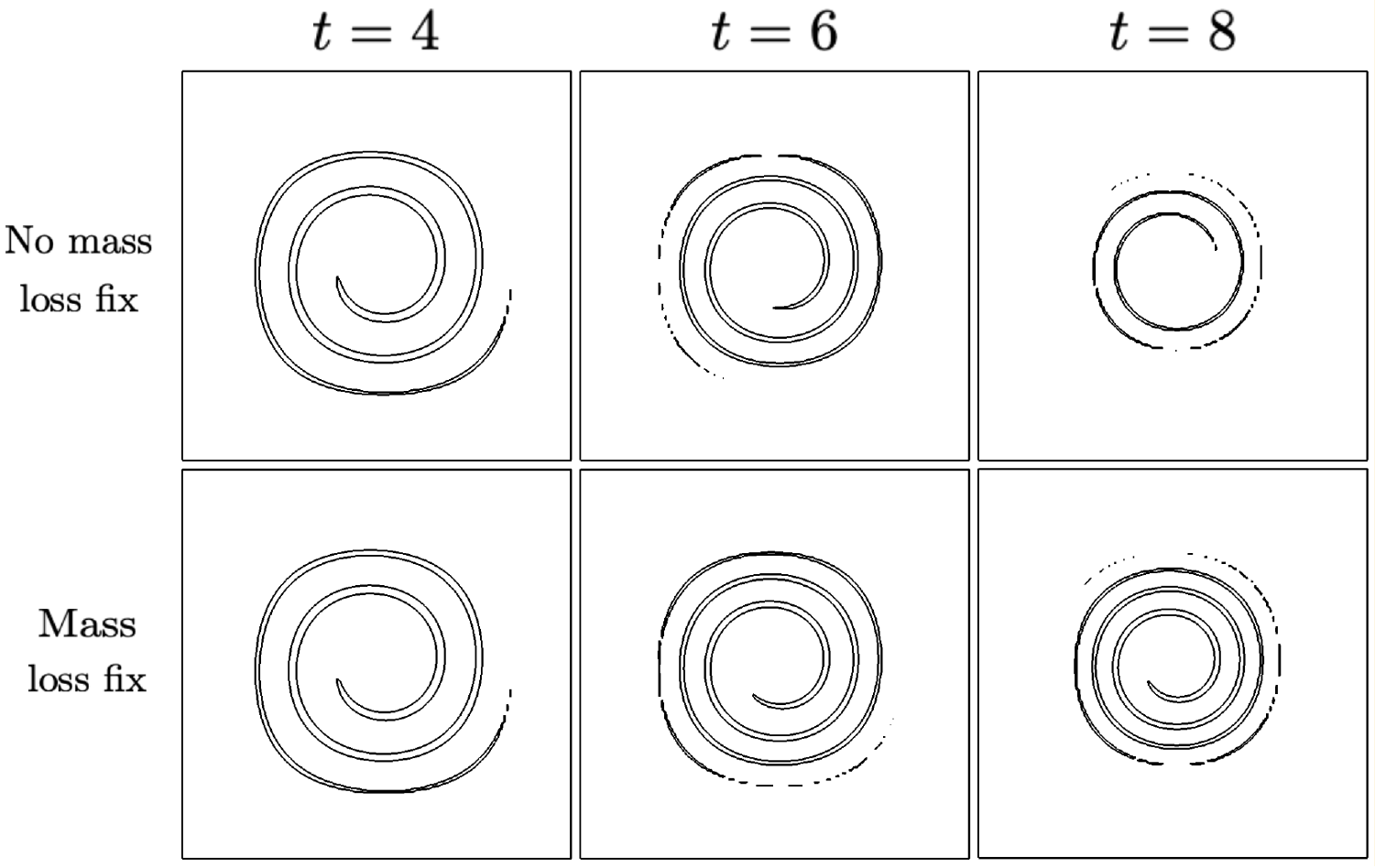}
    \caption{Time evolution of the interface for the vortex in a box problem using a 128$\times$128 uniform mesh.}
   \label{fig_vortex_128x128}
\end{figure}

Vortex in a box is a standard two-phase flow problem used in the literature~\cite{Rider1995} to test interface tracking/capturing methods' ability to resolve thin filaments. In this problem, a circular interface is stretched over time due to an imposed velocity field of the form
\begin{align*}
u &= -2\sin^2(\pi x) \sin(\pi y)\cos(\pi y), \\
v &= 2\sin(\pi x)\cos(\pi x) \sin^2(\pi y).
\end{align*}
The computational domain consists of a unit square with extents $\Omega\in [0,1]^2$. A circular interface with a radius of 0.15 has its initial centroid at (0.5, 0.75). The origin is taken to be the lower left corner. The circular interface gets stretched into a very long and thin filament due to the imposed velocity field, which over time wraps around itself. Fig.~\ref{fig_vortex_128x128} compares the interfacial dynamics with and without the mass loss fix at time instances $t$ = 4, 6, and 8. The domain is discretized into $N\times N = 128\times128$ grid cells and a uniform time-step size of $\Delta t = h/10$ is employed. As can be observed in the figure, a significant amount of mass is lost by the standard level set method compared to the mass loss fix method.

\begin{figure}[]
   \centering
     \subfigure[Normalized correction for 32$\times$32 grid]{
   	\includegraphics[scale= 0.4]{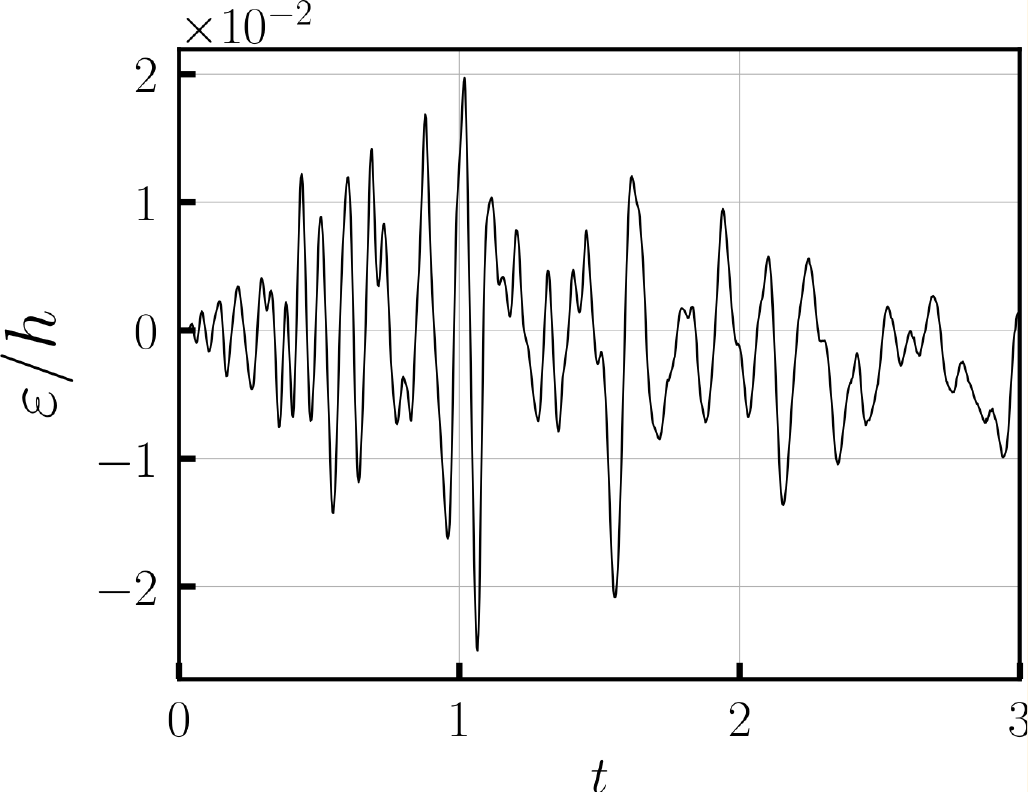}
	\label{fig_vortex_32x32_LSshift}
   } 
    \subfigure[Relative error in volume for 32$\times$32 grid]{
   	\includegraphics[scale= 0.4]{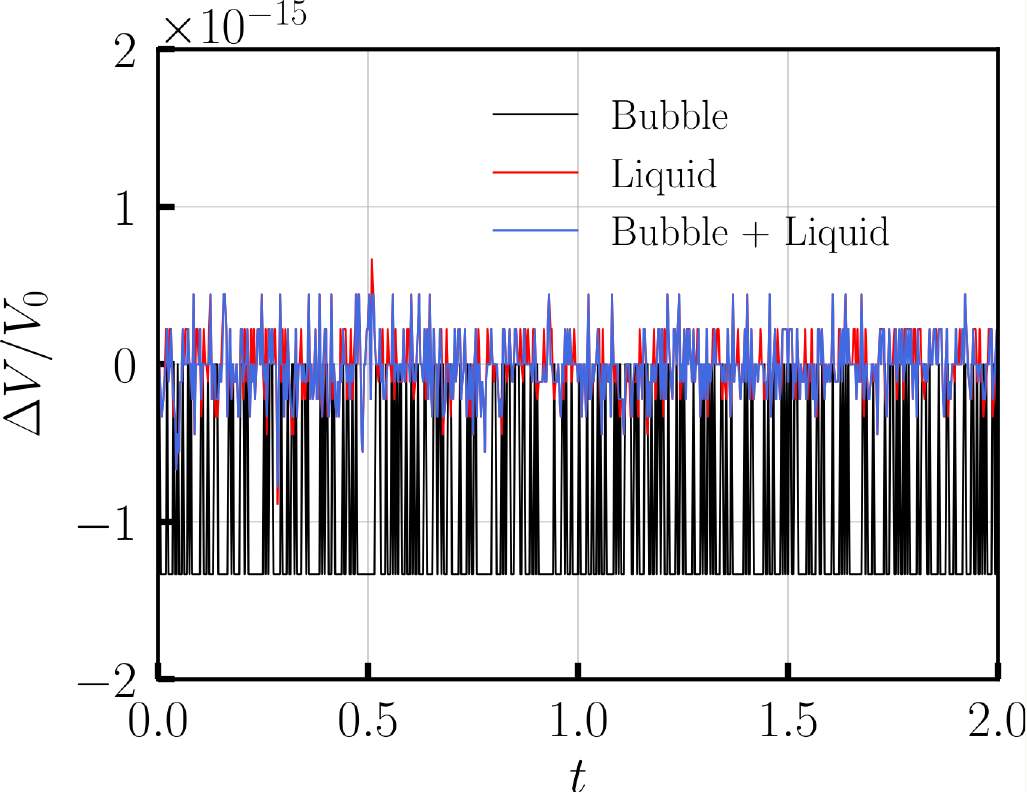}
	\label{fig_vortex_32x32_vol_error}
   }   
        \subfigure[Normalized correction for  64$\times$64 grid]{
   	\includegraphics[scale= 0.4]{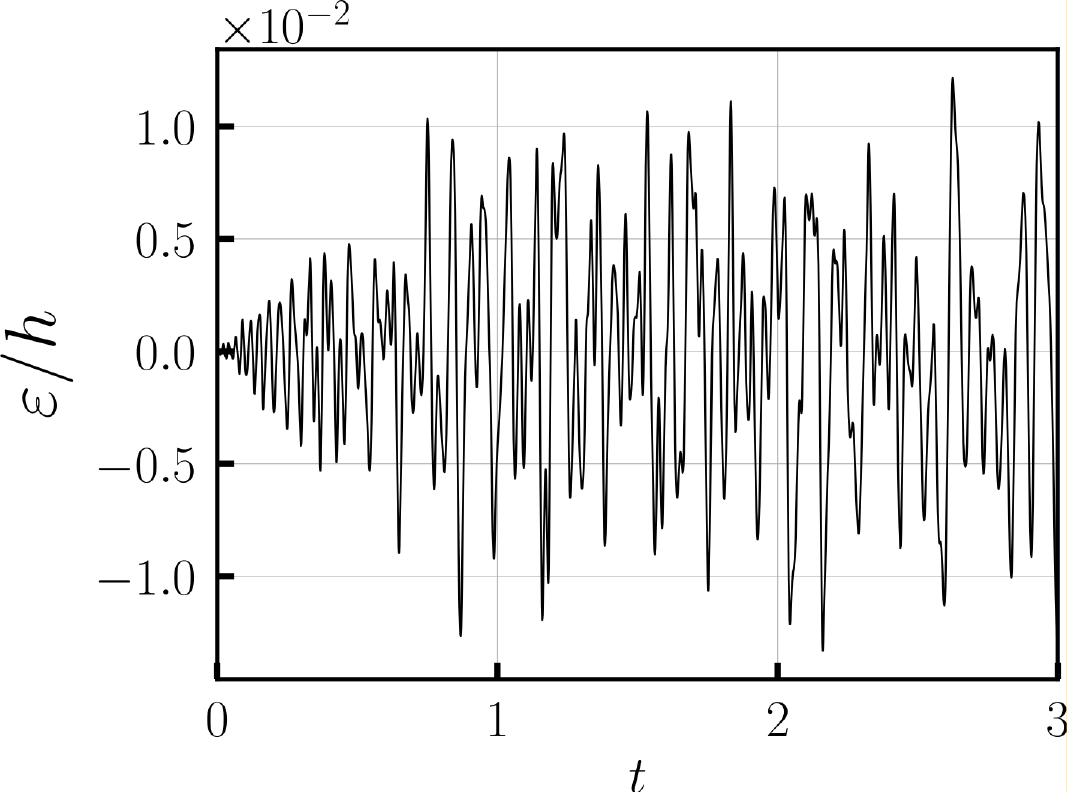}
	\label{fig_vortex_64x64_LSshift}
   } 
   \subfigure[Relative error in volume for 64$\times$64 grid]{
   	\includegraphics[scale= 0.4]{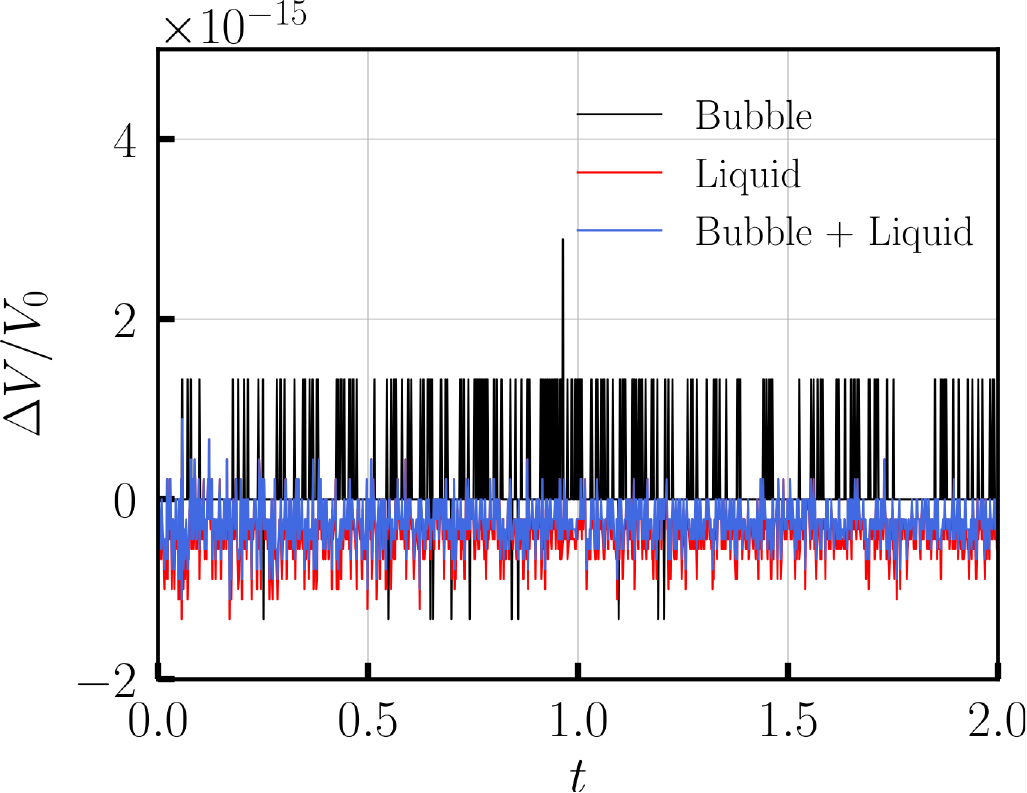}
	\label{fig_vortex_64x64_vol_error}
   }   
        \subfigure[Normalized correction for 128$\times$128 grid]{
   	\includegraphics[scale= 0.4]{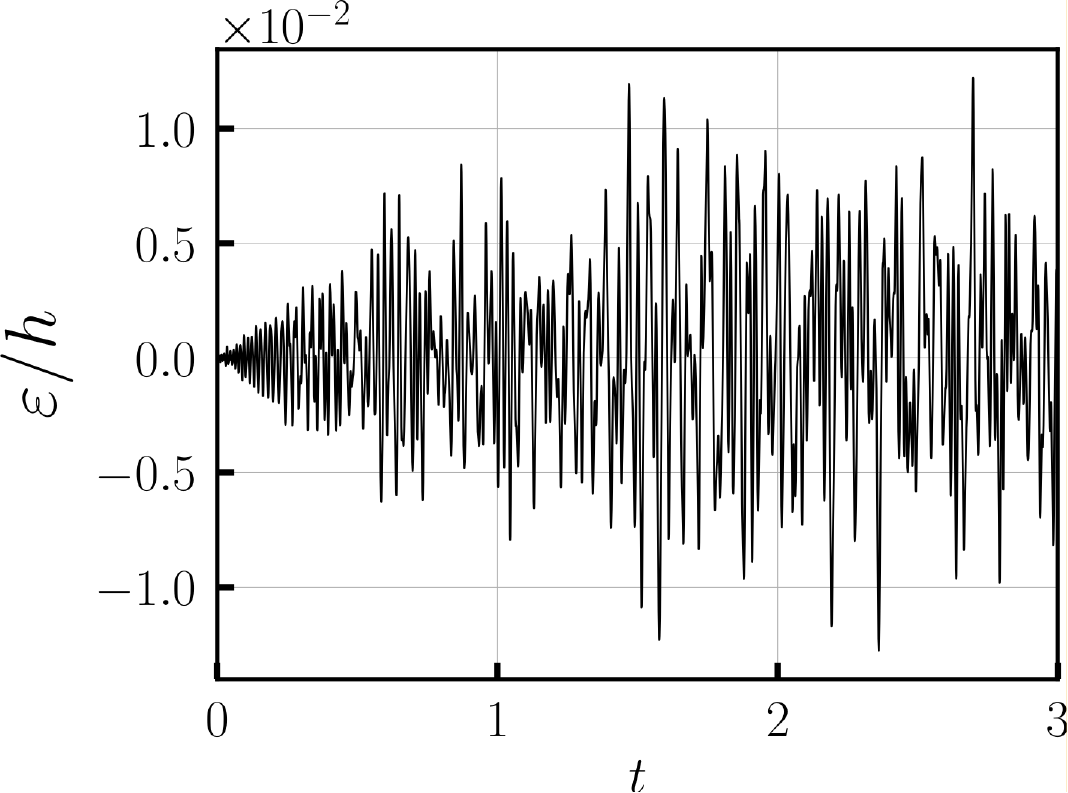}
	\label{fig_vortex_128x128_LSshift}
   } 
   \subfigure[Relative error in volume for 128$\times$128 grid]{
   	\includegraphics[scale= 0.4]{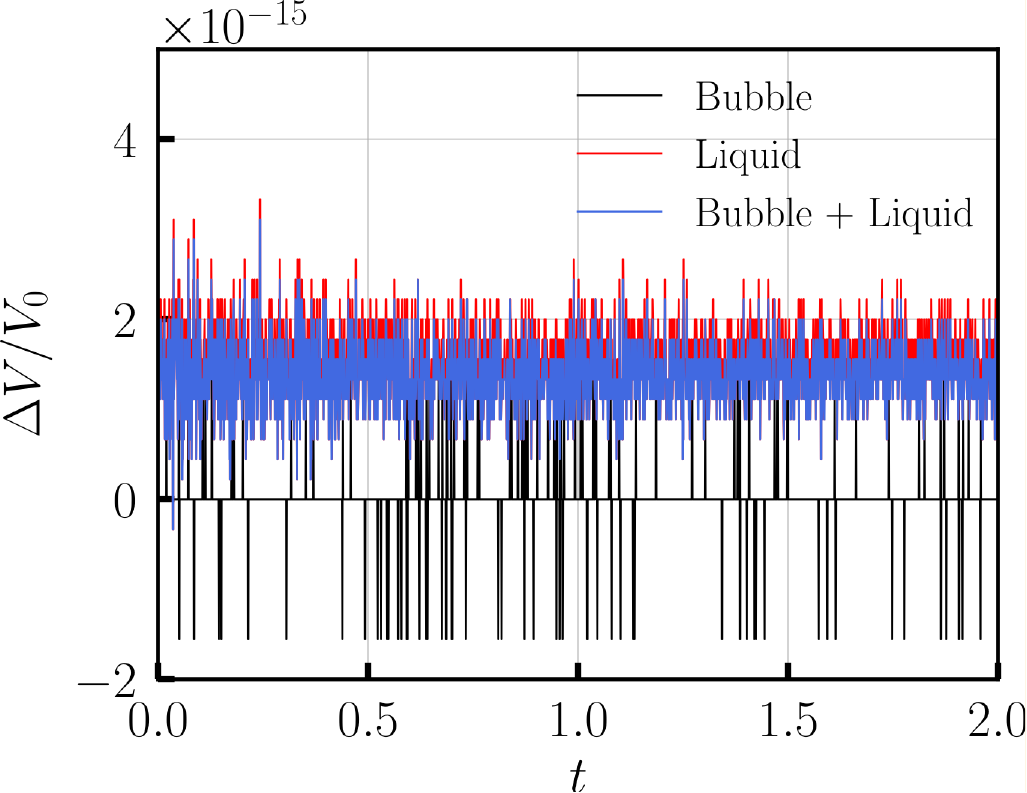}
	\label{fig_vortex_128x128_vol_error}
   }  
    \caption{Vortex in a box problem simulated using different grid sizes. The normalized correction ($\varepsilon/h$) as a function of time is shown in \subref{fig_vortex_32x32_LSshift}, \subref{fig_vortex_64x64_LSshift} and \subref{fig_vortex_128x128_LSshift}. The relative error in volume ($\Delta V/V_0$) as a function of time is shown in \subref{fig_vortex_32x32_vol_error}, \subref{fig_vortex_64x64_vol_error} and \subref{fig_vortex_128x128_vol_error}.}
   \label{fig_vortex_vol_errors}
\end{figure}

Figs.~\ref{fig_vortex_32x32_LSshift}, \ref{fig_vortex_64x64_LSshift} and \ref{fig_vortex_128x128_LSshift} show the normalized value of the uniform correction $\varepsilon/h$ as a function of time. The normalized correction values are in the order of $10^{-2}$, which means that the contours of the level set field $\widehat{\phi}$ are shifted at a sub-grid level. To quantify the amount of fluid lost/gained over time, we plot the relative change in the volume of fluid enclosed by the interface, $\Delta V/V_0 = V(t)/V_0 - 1$. Figs.~\ref{fig_vortex_32x32_vol_error}, \ref{fig_vortex_64x64_vol_error} and \ref{fig_vortex_128x128_vol_error} show the relative volume change for grid sizes 32$\times$32, 64$\times$64, and 128$\times$128. For all grids, the relative change in volume is close to machine precision. In order to solve $f(\varepsilon) = 0$ to machine precision, it usually takes 2-3 Newton iterations.   

\begin{figure}[]
   \centering
   \subfigure[$32 \times 32$]{
   	\includegraphics[scale= 0.3]{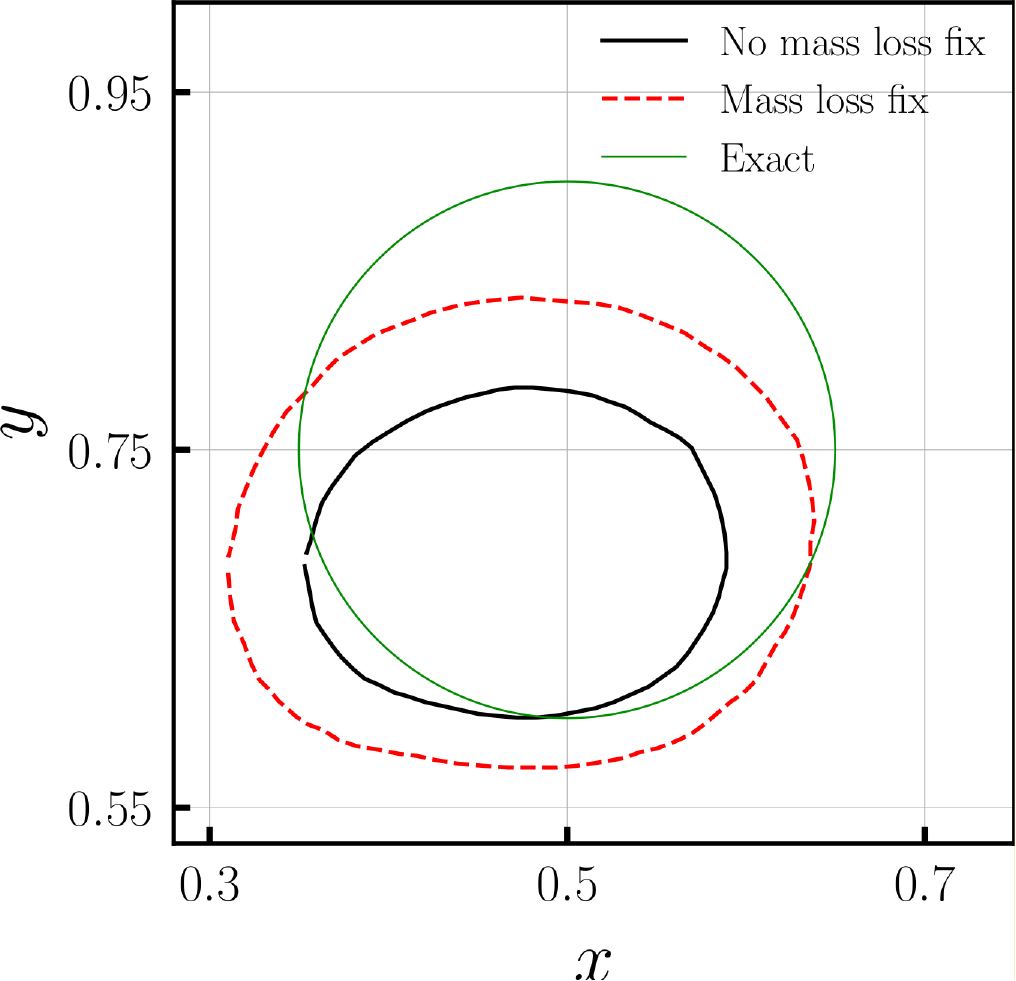}
	\label{fig_reverse_vortex_32x32}
   }
      \subfigure[$64 \times 64$]{
   	\includegraphics[scale= 0.3]{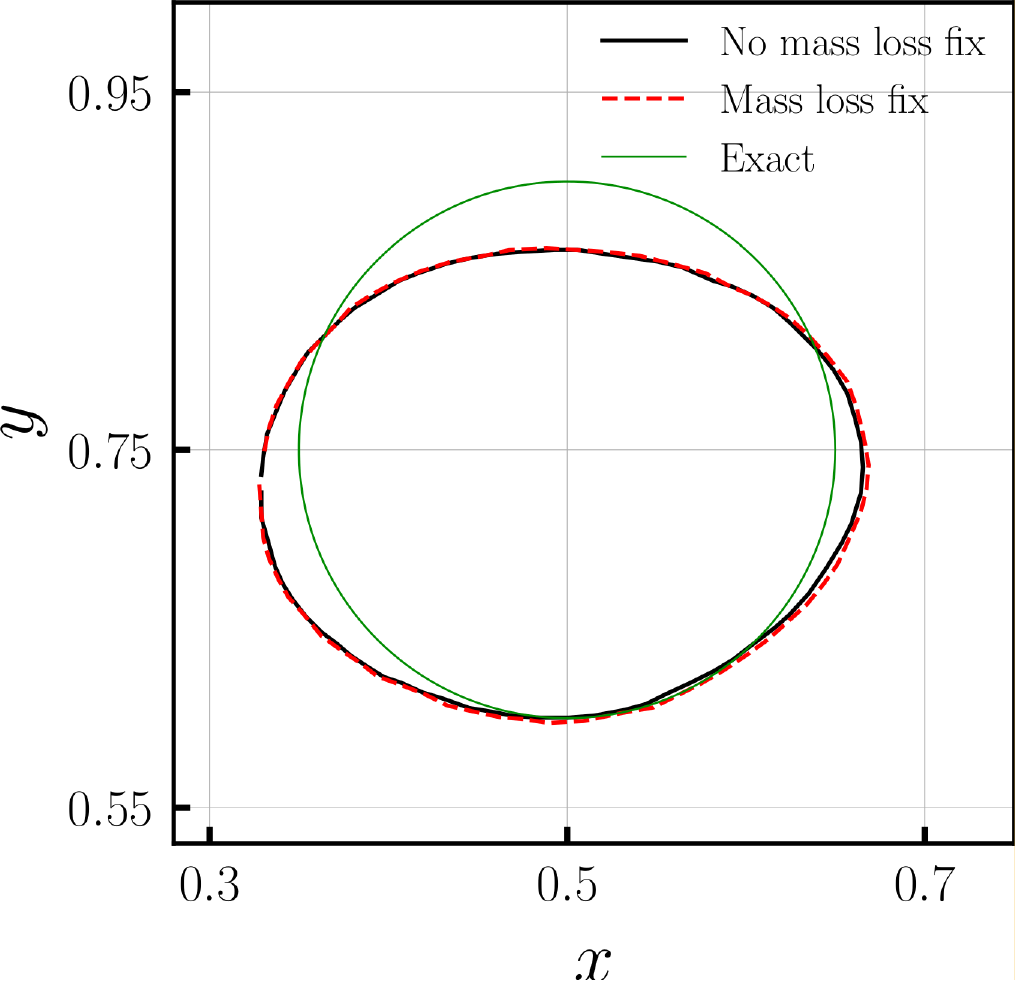}
	\label{fig_reverse_vortex_64x64}
   }     
      \subfigure[$128 \times 128$]{
   	\includegraphics[scale= 0.3]{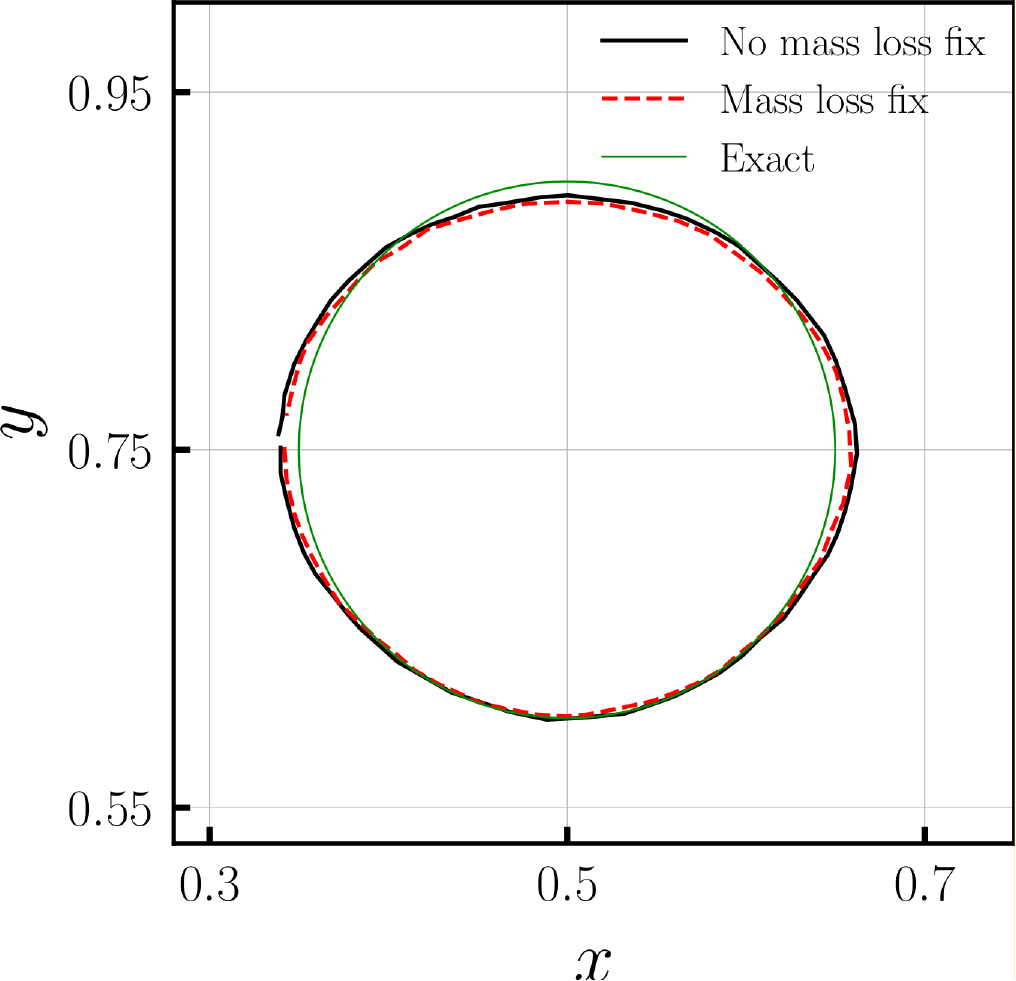}
	\label{fig_reverse_vortex_128x128}
   } 
    \caption{Results for the reverse vortex test problem. Comparison of the final interface shape with the exact solution using standard level set method and level set method with mass loss fix. Three grid sizes are considered: \ref{fig_reverse_vortex_32x32} $32 \times 32$, \ref{fig_reverse_vortex_64x64} $64 \times 64$ and \ref{fig_reverse_vortex_128x128} $128 \times 128$.}
   \label{fig_reverse_vortex_contours}
\end{figure}

Next, we will examine a somewhat different but closely related problem known as the reverse vortex test problem~\cite{Rider1995, Wang2012}. After rotating counterclockwise for $t = 4$, the initially circular interface stretches into a thin filament before rotating clockwise to return to its original shape at $t = T = 8$. In this case, velocity is a function of time, and is written as
\begin{subequations}
	\begin{align*}
	u = -2\sin^2(\pi x) \sin(\pi y)\cos(\pi y)\cos\left( \frac{\pi t}{T} \right),   \\
	v = 2\sin(\pi x)\cos(\pi x) \sin^2(\pi y)\cos\left( \frac{\pi t}{T} \right). 
	\end{align*}
\end{subequations}
Fig.~\ref{fig_reverse_vortex_contours} compares the final shape of the interface at  $t = T$ for different grid sizes, with finer grids performing better. For the coarse mesh resolution ($32 \times 32$) considered here, the mass fix method yields a ``full" circle than the standard approach, which shrinks the interface. The latter occurs because of the substantial volume loss with the standard level set method. In order to compare the two methods quantitatively, we estimate the geometric and volume errors as $E_{\rm g}$ and $E_{\rm v}$, respectively, at the end of the process:
\begin{align}
	E_{\rm g} &= \frac{\int_{\Omega} | H(\phi,t = T) - H(\phi, t = 0) |~\dOmega}{\int_{\Omega} H(\phi, t = 0)~\dOmega},  \\
	E_{\rm v} &= \frac{ | \int_{\Omega} H(\phi,t = T)~\dOmega - \int_{\Omega} H(\phi, t = 0)~\dOmega |}{\int_{\Omega} H(\phi, t = 0)~\dOmega}.  
 	\label{eq_Eg_Ev_errors}
\end{align}
Table~\ref{tab_Eg_Ev_errors} compares relative geometric and volume errors at different grids with and without mass loss fix.  The circular interface recovers well for grids $64 \times 64$ and $128 \times 128$, and geometric errors are comparable between the two approaches. Significant differences are observed in relative volume errors, however. The mass loss fix method conserves the volume enclosed by the interface to machine precision.

\begin{table}[]
\centering
\caption{Comparison of relative geometric $E_{\rm g}$ and volume $E_v$ errors at the final time instant using standard level set method and level set method with mass loss fix for the reverse vortex problem. Three grid sizes are considered.}
\begin{tabular}{|c|c|c|c|c|c|}
\hline
Errors & Method & $32 \times 32$ & $64 \times 64$ & $128 \times 128$ & $256 \times 256$  \\ \hline
\multirow{2}{*}{$E_{\rm g}$} & no mass loss fix    & $5.6274 \times 10^{-1}$      & $1.9812\times 10^{-1}$            & $6.5218 \times 10^{-2}$   &  $9.1636 \times 10^{-3}$        \\ \cline{2-6} 
                                    & mass loss fix          & $4.7861\times 10^{-1}$     & $2.0243 \times 10^{-1}$              & $6.3858 \times 10^{-2}$  & $9.6954 \times 10^{-3}$          \\ \hline
\multirow{2}{*}{$E_{\rm v}$}  & no mass loss fix     & $4.8166 \times 10^{-1}$       & $2.3117 \times 10^{-2}$              & $2.9920 \times 10^{-2}$  & $1.0634\times 10^{-3}$                  \\ \cline{2-6} 
                                    & mass loss fix          & $<\times 10^{-16}$       & $ 2 \times 10^{-16}$              & $ <\times 10^{-16}$ & $ 2 \times 10^{-16}$ \\ \hline
\end{tabular}
\label{tab_Eg_Ev_errors}
\end{table}

\begin{figure}[]
   \centering
   	\includegraphics[scale= 0.37]{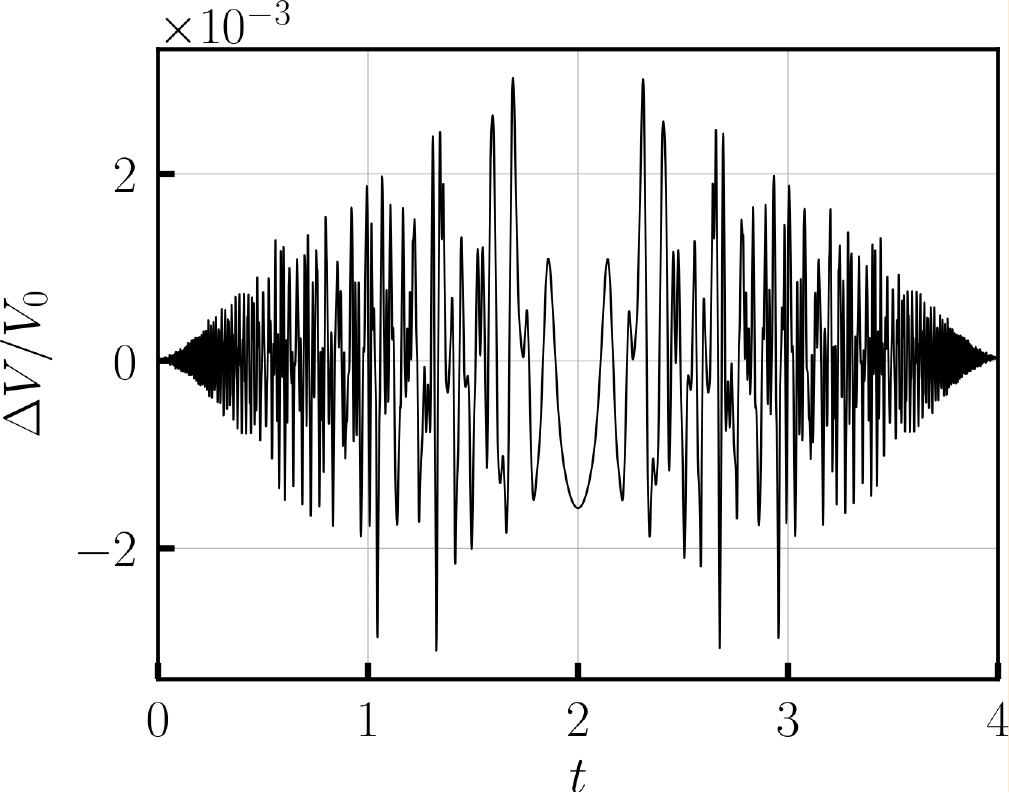}
    \caption{Relative change in volume of the fluid enclosed by the interface for the reverse vortex problem using the standard level set method. The grid size is $256 \times 256$ and the time period is $T = 4$.}
   \label{fig_reverse_vortex_Ev}
\end{figure}

Finally, we simulate the reverse vortex case using the standard level set method (without mass loss fix) for a reduced time period of $T=4$. The grid size is $256 \times 256$. The interface rotates counterclockwise until $t = 2$, at which point it reverses motion. By reducing the time period, the interface does not thin out excessively and break.  During the second half of the period, the interface follows a ``reversible" path. Fig.~\ref{fig_reverse_vortex_Ev} illustrates the relative volume change of the fluid enclosed by the interface over time. As can be seen from the figure, the level set method exhibits a very low volume error at $t = T = 4$ (specifically $\Delta V/V_0 = 2.04 \times 10^{-5}$). Based on this, we can confirm Eq.~\eqref{eq_deltaM_slsm}. It is generally not possible to find time intervals $\Delta t$ where mass loss errors are low with the standard level set method. Many practical problems have a two-phase interface that breaks, making it impossible to follow a reversible path to achieve a net zero normal displacement.     

\subsubsection{Bubble rise problem}

\begin{figure}[]
   \centering
   	\includegraphics[scale= 0.55]{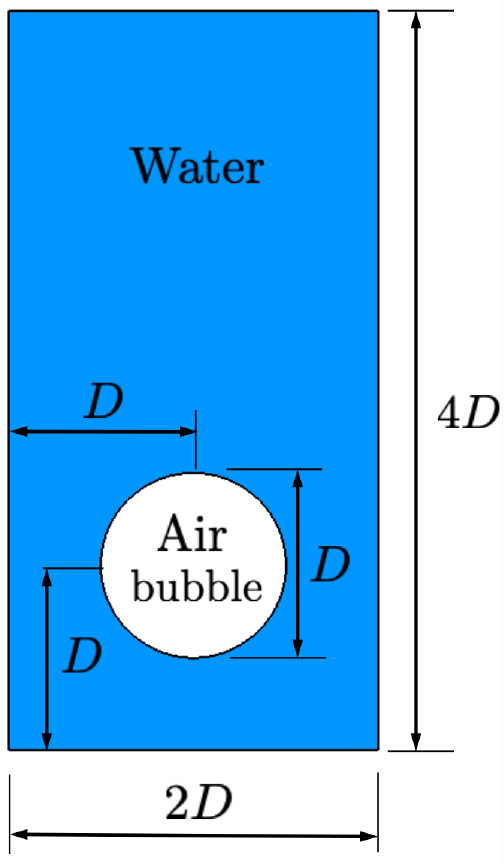}
    \caption{Schematic of the bubble rise problem.}
   \label{fig_bubble_rise_schematic}
\end{figure}

In this section, we simulate the rise of a two-dimensional bubble in water due to buoyancy, which tests the coupling between the level set method and the flow solver. The density and viscosity of air (bubble) are $\rho_\text{a}$ = 1 kg/$\text{m}^3$ and $\mu_\text{a}$ = 0.1 Pa$\cdot$s. Water density and viscosity are $\rho_\text{w}$ = 1000 kg/m$^3$ and  $\mu_\text{w}$ = 10 Pa$\cdot$s, respectively. At $t$ = 0 s, the bubble's center is located at (0.5, 0.5) m and its diameter is $D$ = 0.5 m; see Fig.~\ref{fig_bubble_rise_schematic}. There are two important non-dimensional numbers for this problem, the Reynolds number and the E\"otv\"os number, which are defined as
\begin{equation}
Re = \frac{\rho_\text{w}g^{1/2}D^{3/2}}{\mu_{\text{w}}}~~~\text{and}~~~Eo = \frac{\rho_\text{w}gD^2}{\sigma}.
\end{equation} 
Here, $g$ is the acceleration due to gravity and $\sigma$ is the surface tension coefficient. The specific values of the non-dimensional numbers for the simulated case are $Re = 35$ and $Eo = 125$. The computational domain is discretized into uniform cells using a 
grid size of $N_x \times N_y = 256\times512$. A uniform time-step size of $\Delta t$ = $10^{-3}$ s is used. As the bubble rises its $y$-center of mass position is obtained as
\begin{equation}
\text{y}_\text{COM} = \frac{\int_\Omega y H(-\phi)~\dOmega}{\int_\Omega H(-\phi)~\dOmega}.
\label{eq_ycom_calculation}
\end{equation}

\begin{figure}[]
   \centering
   	\includegraphics[scale= 0.55]{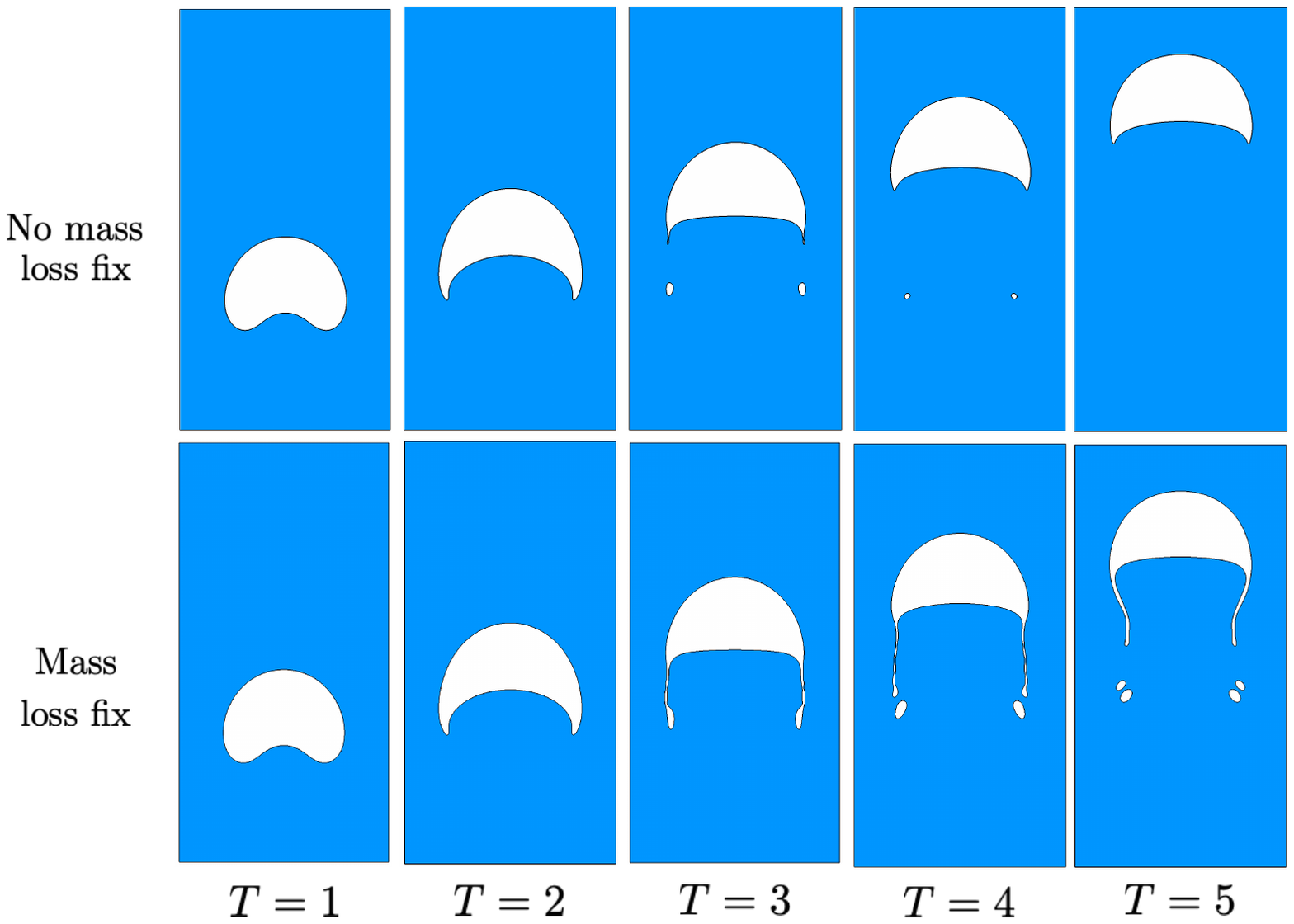}
    \caption{Evolution of the bubble interface with and without the mass loss fix technique at non-dimensional time instances $T$ = 1, 2, 3, 4, and 5.}
   \label{fig_bubble_rise_contours}
\end{figure}

Fig.~\ref{fig_bubble_rise_contours} compares the evolution of the bubble interface with and without the mass loss fix technique at non-dimensional time instances $T$ = 1, 2, 3, 4, and 5. Time is non-dimensionalized using $t_{\rm scale} = D/U_{\rm g}$, in which $U_{\rm g} = \sqrt{gD}$ represents the velocity scale. The bubble starts to rise significantly around $T$ = 1 and forms rounded lower ends. Due to the large deformation of the bubble, the rounded ends break into smaller bubbles. In the absence of the mass loss fix, small bubbles quickly disappear. Mass is lost as a result. The pinched bubbles stay in the domain longer when the mass loss fix is applied.  

\begin{figure}[]
   \centering
   	\includegraphics[scale= 0.37]{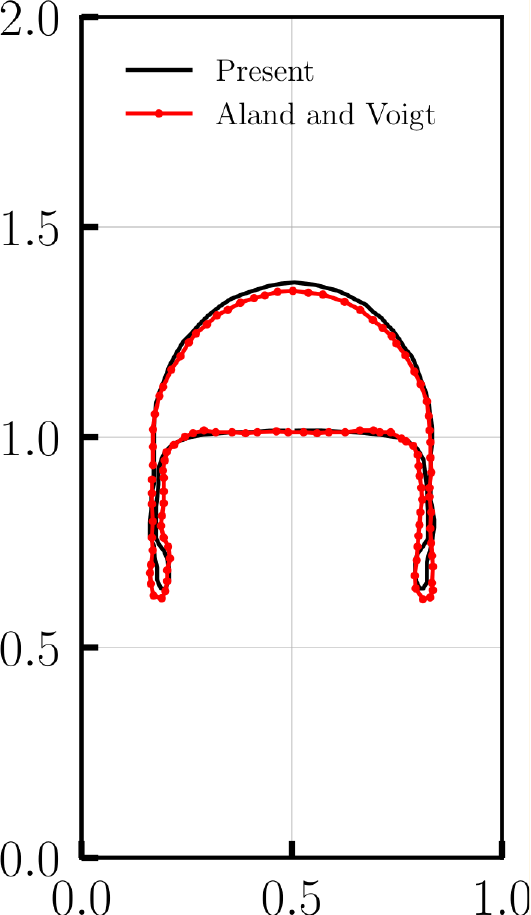}
    \caption{Comparison of the bubble shape using the mass loss fix method with the  benchmarking numerical solution of Aland and Voigt~\cite{Aland2012} at $T = 3$.}
   \label{fig_bubble_rise_benchmark_comparison}
\end{figure}

\begin{figure}[h!]
   \centering
   	\includegraphics[scale= 0.42]{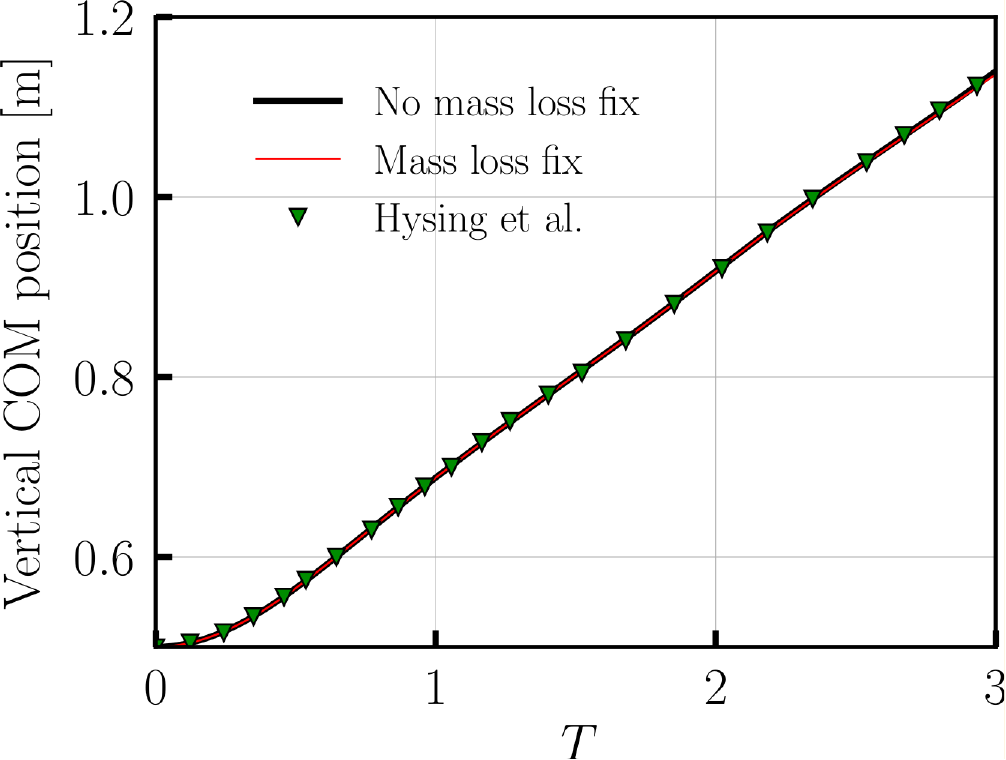}
    \caption{Temporal evolution of the vertical coordinate of the bubble center of mass.}
   \label{fig_ycom_bubble_rise}
\end{figure}

\begin{figure}[]
   \centering
   \subfigure[Normalized correction]{
   	\includegraphics[scale= 0.37]{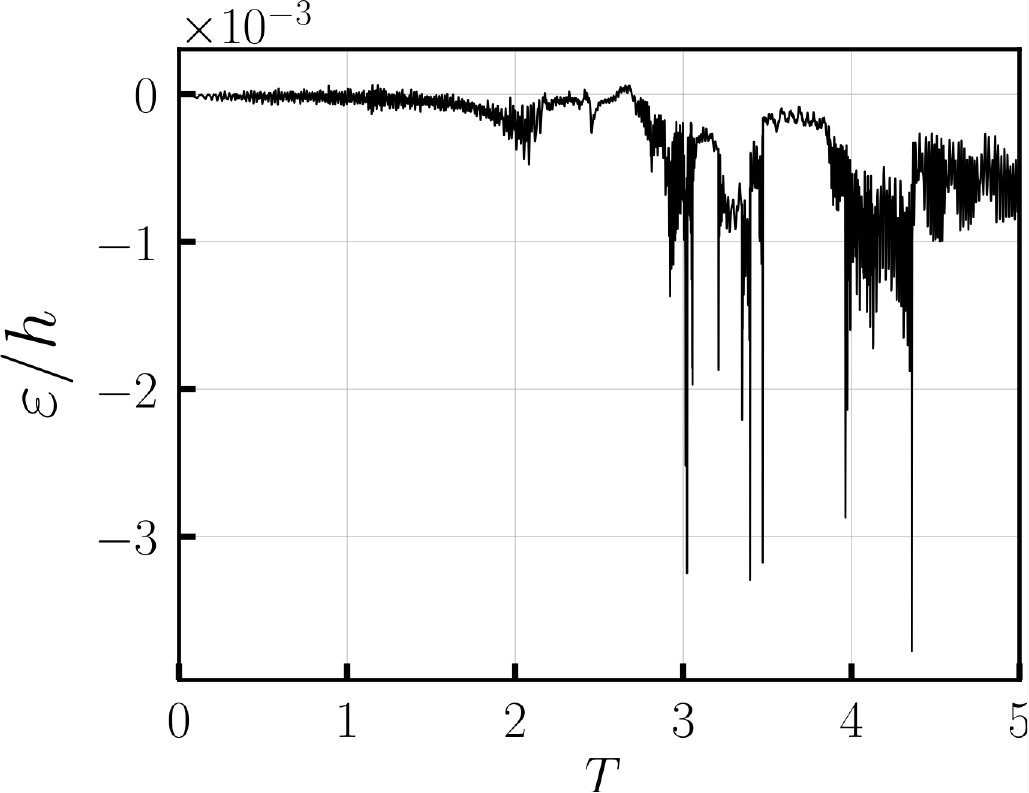}
	\label{fig_bubble_LSshift}
   }
      \subfigure[Relative change in volume]{
   	\includegraphics[scale= 0.37]{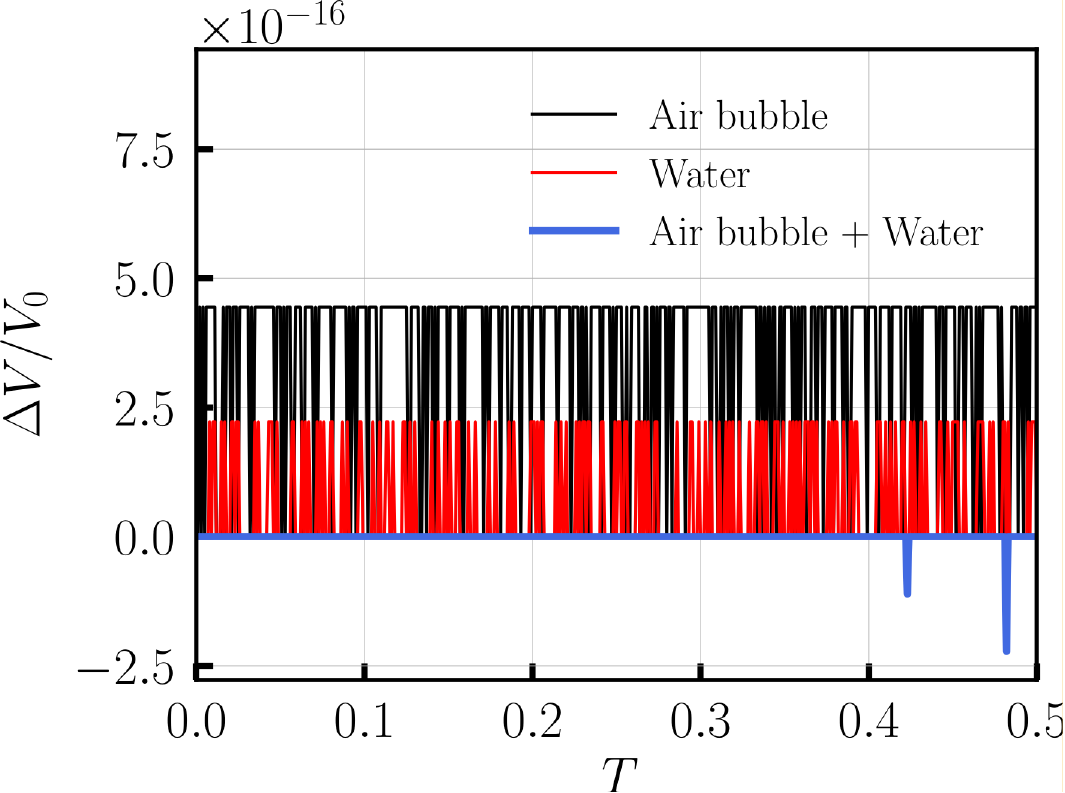}
	\label{fig_relative_volume_error_bubble_rise}
   }  
    \caption{\subref{fig_bubble_LSshift} Temporal evolution of the normalized shift in the level set contours and \subref{fig_relative_volume_error_bubble_rise} relative change in volumes of air and water phases.}
   \label{fig_bubble_rise_ycom_vol_error}
\end{figure}

Fig.~\ref{fig_bubble_rise_benchmark_comparison} compares the bubble shape achieved with the mass loss fix with the benchmarking numerical solution of Aland and Voigt~\cite{Aland2012} at $T = 3$, who used Cahn-Hilliard phase field method for capturing the two-phase interface. The two are in excellent agreement. Fig.~\ref{fig_ycom_bubble_rise} shows the evolution of the vertical coordinate of the bubble center of mass over time. The results of $y_\text{COM}$ obtained with and without the mass fix approach are compared with those of Hysing et al.~\cite{Hysing2009}. Results are in agreement. It also shows that level set correction $\varepsilon$ does not affect the momentum of the system. This can be understood from Fig.~\ref{fig_bubble_LSshift} which shows the normalized value of $\varepsilon$ over time. The values are in the order of $10^{-3}$. As a result, the level set contours are shifted at a sub-grid level per time step. Performing mass corrections frequently is crucial to achieving sub-grid level changes in $\widehat{\phi}$.  Otherwise, an abrupt shift in the contours can change the system's momentum. Additionally, Fig.~\ref{fig_relative_volume_error_bubble_rise} illustrates the relative change in bubble and water volumes. Volume changes are close to machine accuracy. In summary, this example demonstrates that the approximate Lagrange multiplier method conserves mass to machine precision without affecting momentum.  

\subsection{Fluid-structure interaction cases}
\label{sec_FSI_cases}
We test the performance of the approximate Lagrange multiplier method on problems involving rigid bodies interacting with two-fluid interfaces in this section. The FSI problems discussed here are relevant to ocean engineering. To accurately account for buoyancy forces in these FSI problems, we conserve the heavier fluid volume. The volume of the union of complementary phases (light fluid and solid) is also conserved. 

\subsubsection{Floating rectangular block problem}
\label{subsubsec_floating_rect}

\begin{figure}[]
   \centering
   	\includegraphics[scale= 0.35]{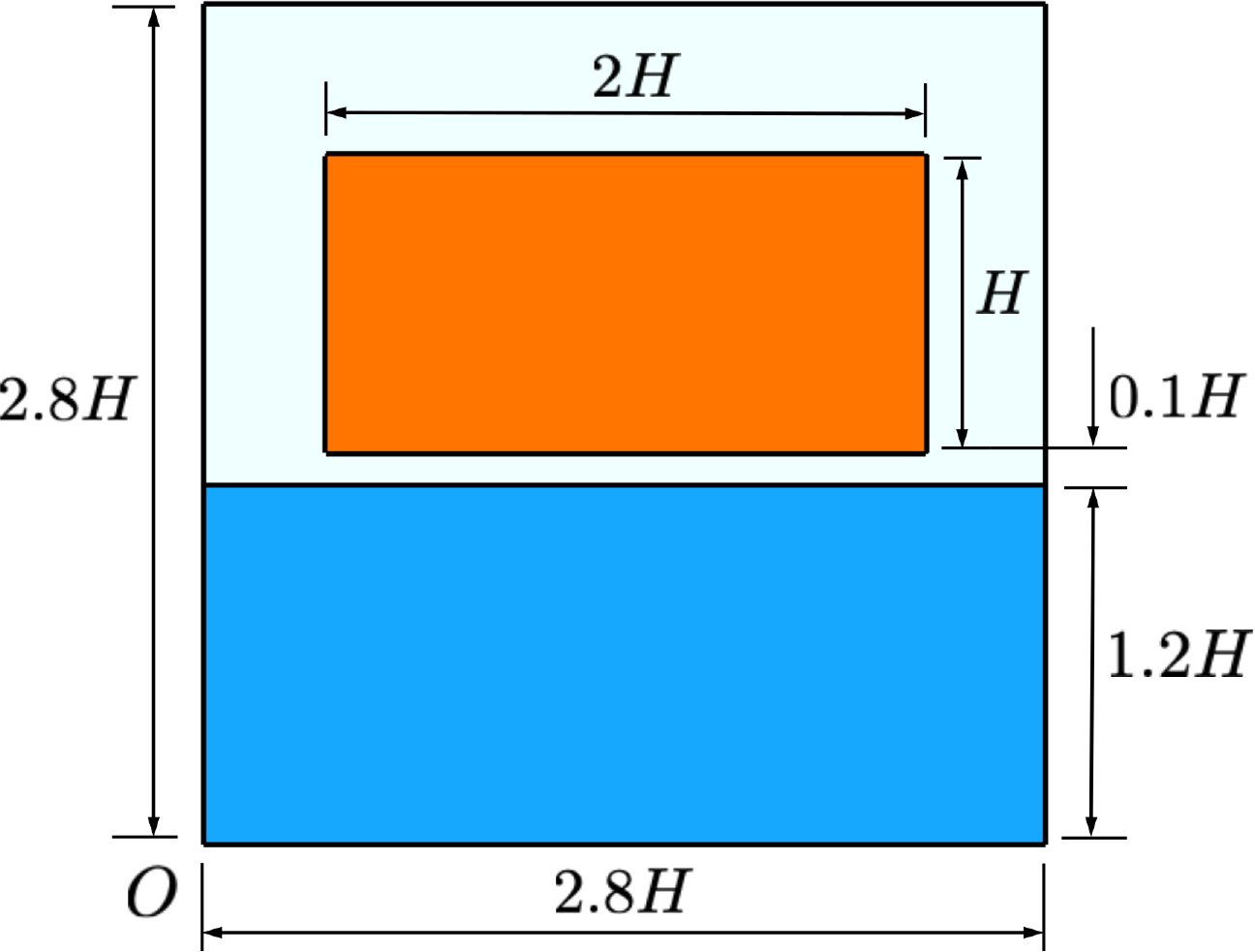}
    \caption{Schematic of the floating rectangular block problem.}
   \label{fig_rect_schematic}
\end{figure}

Ocean engineering problems in the literature typically consider wider domains when studying fluid-structure interactions~\cite{Nangia2019, Calderer2014, Dafnakis2020, Bhalla2020}. There is little change in the mean water level when a solid body interacts with the air-water interface in these large domains. Thus, spurious mass/volume losses do not significantly affect FSI dynamics. To exaggerate the effects of mass loss errors, we consider a relatively large rectangular block interacting with water inside a relatively small tank. The solid's density is half that of water and it is released from a small distance above the air-water interface. Fig.~\ref{fig_rect_schematic} shows the problem setup schematic. Since the domain is narrow, the water level rises appreciably at equilibrium. The rectangle will be half submerged at equilibrium if the domain is very wide. With narrow domains, this is not the case. Using conservation of mass and Archimedes' principles, the exact equilibrium position of the rectangle and air-water interface can be found analytically. Appendix Sec.~\ref{sec_analytical_rectangle_calculations} provides the derivation.  

The height and width of the rectangular block are $H$ = 0.75 m and $W = 2H$, and its initial centroid is located at ($1.4H$, $1.8H$) within a closed square domain of extents $2.8H\times2.8H$. The origin is considered to be at the lower left corner. Initially, the water level is set at $y = 1.2H$. The density and viscosity of water are $\rho_\text{w} = 1000$ kg/$\text{m}^3$ and $\mu_\text{w} = 10^{-3}$ Pa$\cdot$s. For air, these values are $\rho_\text{a} = 1$ $\text{m}^3$ and $\mu_\text{a} = 1.8\times10^{-5}$ Pa$\cdot$s. The solid density is $\rho_\text{s} = 500$ kg/$\text{m}^3$ and its fictitious viscosity\footnote{The fictitious viscosity in the solid domain does not affect the FSI dynamics.} is the same as water, i.e., $\mu_\text{s} = \mu_\text{w}$. The block is allowed to heave freely on the air-water interface and its remaining degrees of freedom are locked. The equilibrium positions of the block and air-water interface can be analyzed analytically, however transient dynamics have not been investigated in the literature.  To compare our transient FD/BP results, we examined the same problem using other numerical codes. These include: (i) ANSYS Fluent~\cite{ANSYS_Fluent} that uses the geometric volume of fluid (VOF) method with moving unstructured grids; (ii) CONVERGE CFD~\cite{ConvergeCFD3pt1} that uses geometric VOF with Cartesian cut-cells; and (iii) DualSPHysics~\cite{Crespo2015} that employs smooth particle hydrodynamics. The latter is a mesh-free Lagrangian method, while the other two are Eulerian methods. Our fictitious domain approach differs from the other numerical techniques in that we allow the air-water interface to pass through the immersed solid, whereas the others do not. Mass and volume conservation are not critical in these methods, since they are inherently mass-conserving. However, the FD/BP approach in conjunction with the approximate Lagrange multiplier approach is much easier to implement than Cartesian cut-cells and/or moving meshes. All four numerical codes do not consider the contact angle condition at the material triple points.     
 
\begin{figure}[]
   \centering
   \subfigure[\ADDITION{Grid convergence using FD/BP method}]{
   	\includegraphics[scale= 0.33]{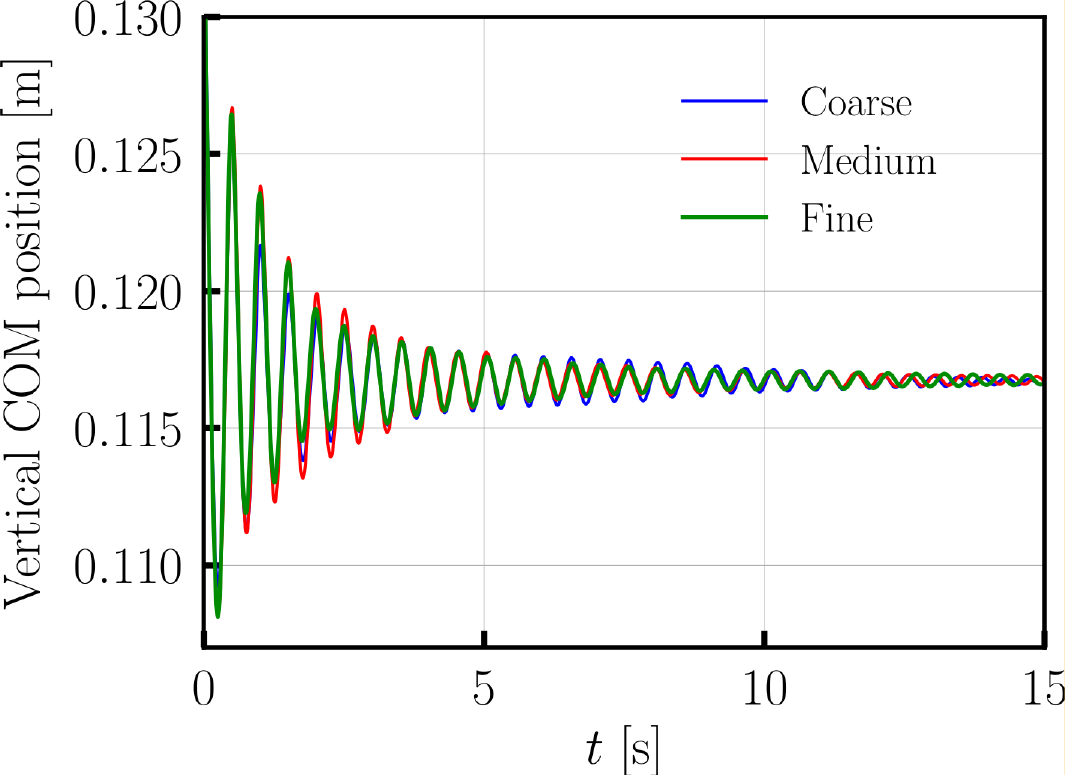}
	\label{fig_rect_grid_convergence_ibamr}
   }
    \subfigure[Grid convergence using CONVERGE CFD]{
   	\includegraphics[scale= 0.33]{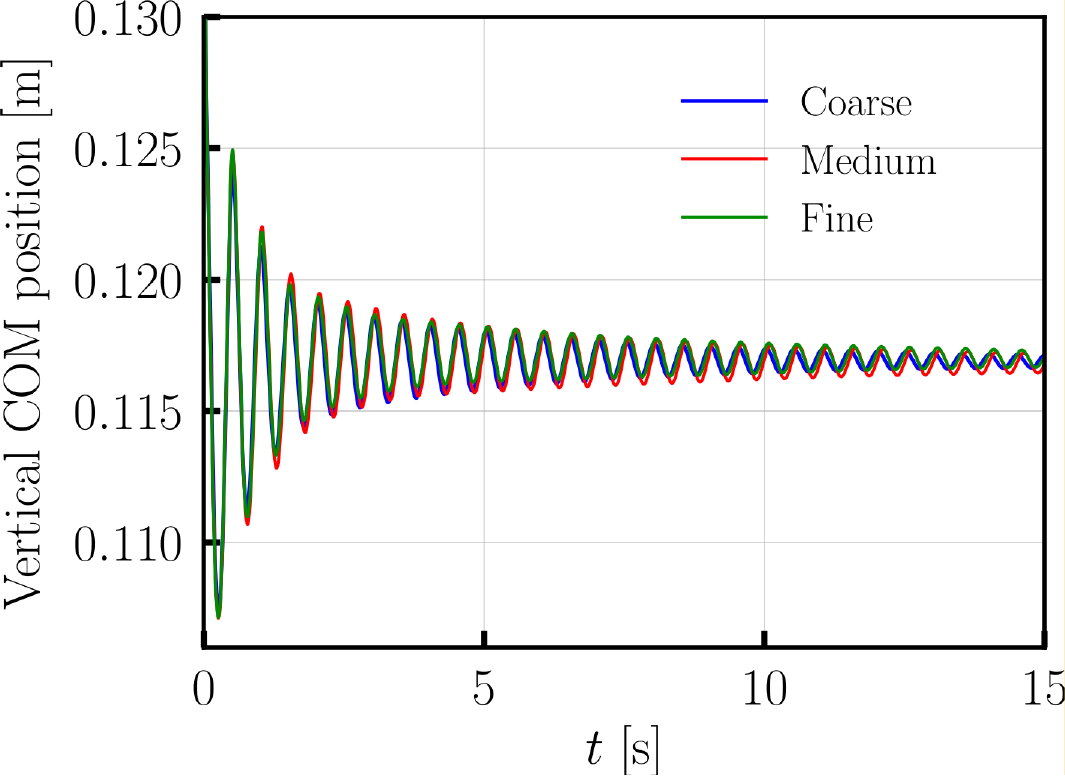}
	\label{fig_rect_grid_convergence_convergecfd}
   }     
    \subfigure[Grid convergence using ANSYS Fluent]{
   	\includegraphics[scale= 0.33]{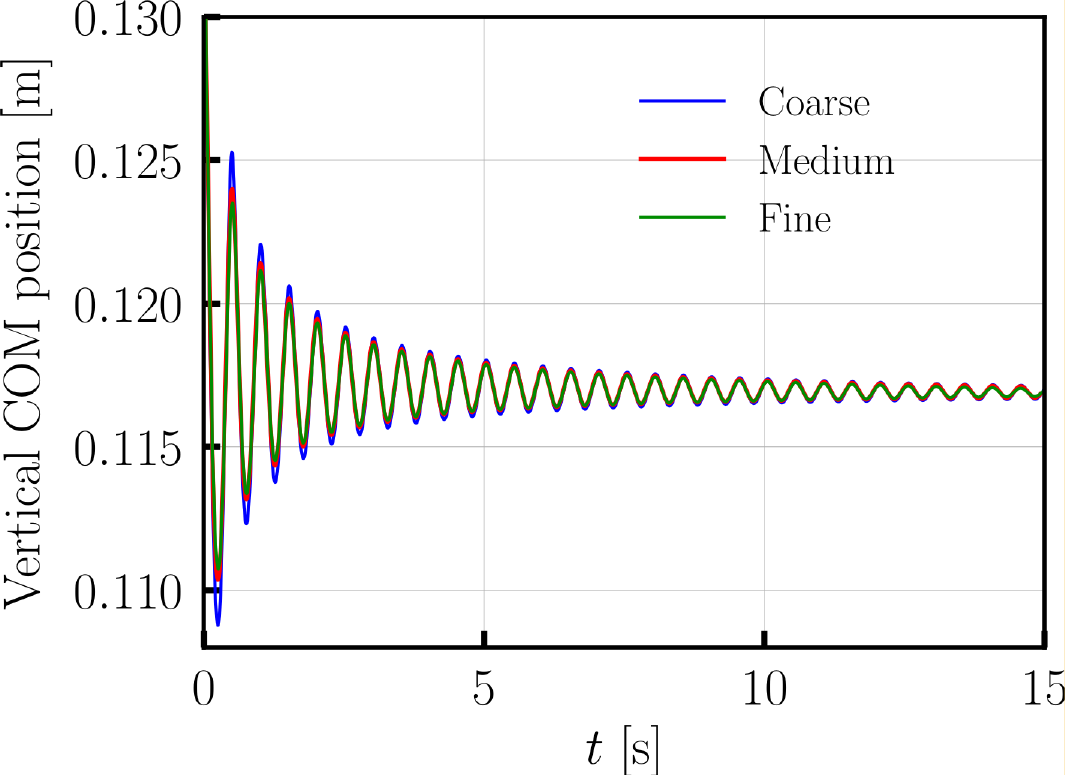}
	\label{fig_rect_grid_convergence_ansys}
   }   
    \subfigure[Grid convergence using DualSPHysics]{
   	\includegraphics[scale= 0.33]{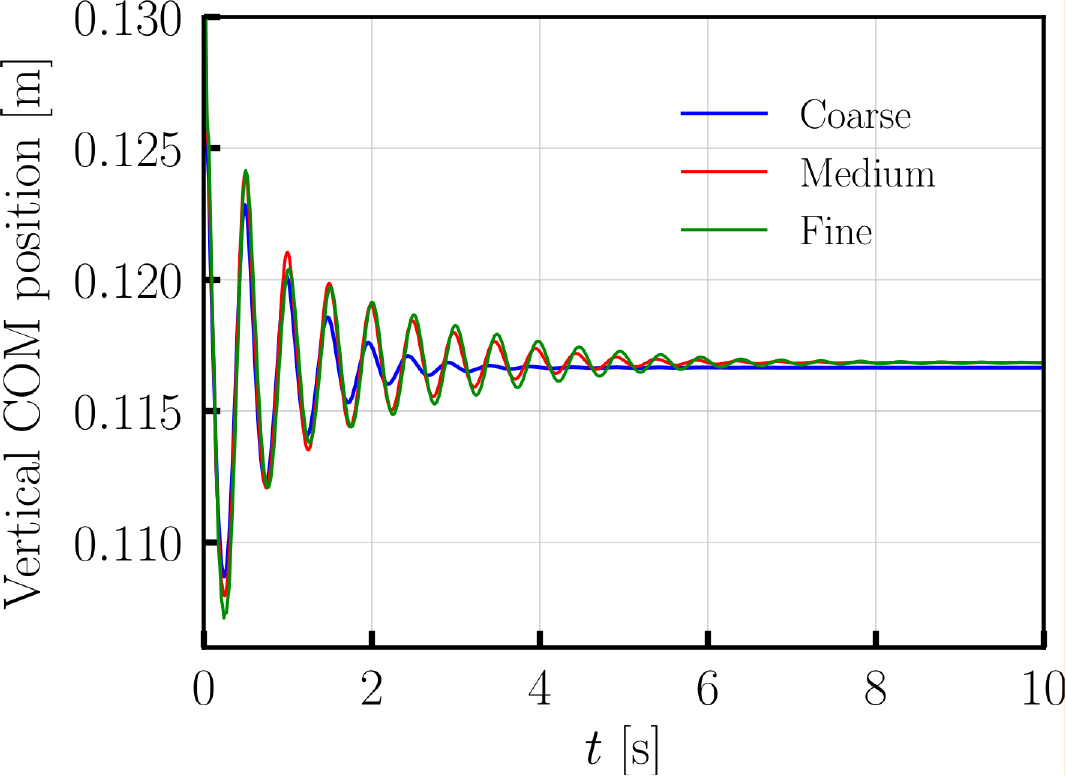}
	\label{fig_rect_grid_convergence_sph}
   }
    \caption{Time evolution of the vertical center of mass position of the block at various grid sizes using \subref{fig_rect_grid_convergence_ibamr} the FD/BP method (present work), \subref{fig_rect_grid_convergence_convergecfd} CONVERGE CFD, \subref{fig_rect_grid_convergence_ansys} ANSYS Fluent, and \subref{fig_rect_grid_convergence_sph} DualSPHysics.}
   \label{fig_grid_convergence_rect_softwares}
\end{figure}

\begin{figure}[]
   \centering
   	\includegraphics[scale= 0.35]{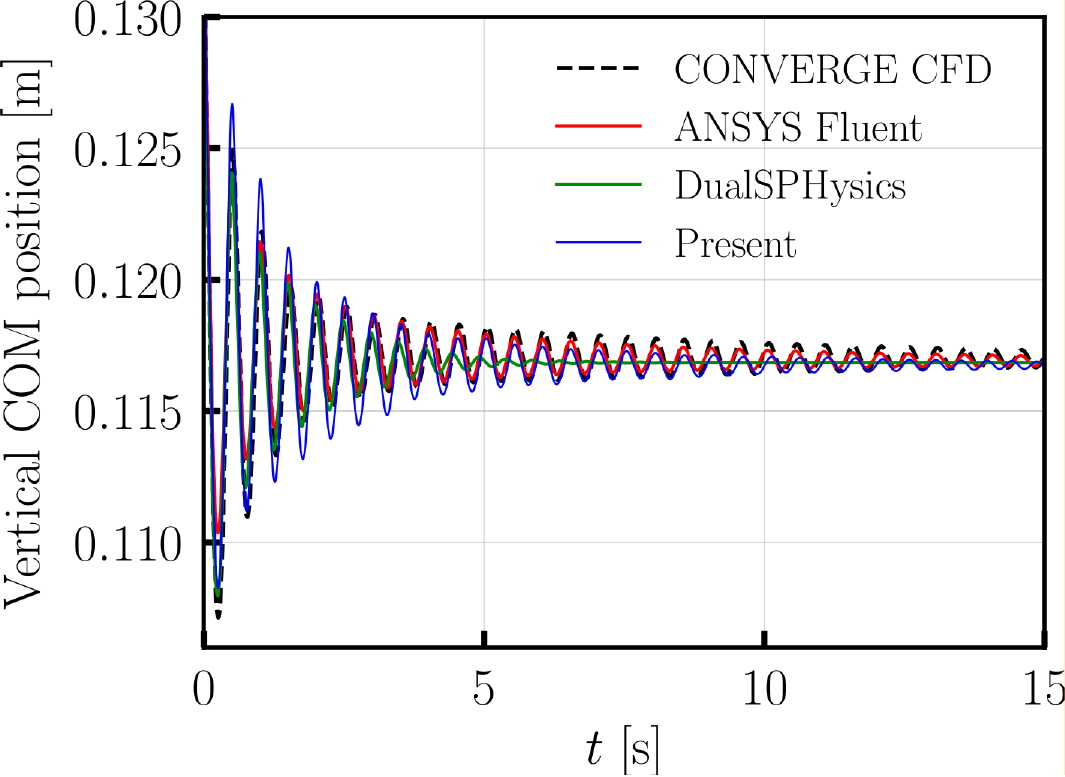}
    \caption{\ADDITION{Comparison of the converged dynamics of the floating rectangular block using different numerical codes.}}
   \label{fig_rect_softwares_compared}
\end{figure}

To obtain respective converged FSI solutions, a grid convergence study is conducted with each of the four numerical codes. In all four codes, adaptive time-stepping is used. Table~\ref{tab_grid_convergence_rect_ibamr} lists the grid cell size $h$ and the maximum time step size $\Delta t_\text{max}$. The time-step size of the simulation is adjusted as the rectangular block hits the water surface.  Fig.~\ref{fig_grid_convergence_rect_softwares} shows the block's center of mass position over time as a function of grid size. Fig.~\ref{fig_rect_softwares_compared} shows a comparison of the converged results from the four solvers. The FD/BP method produces very similar heaving dynamics as the other three solvers. An overly damped motion is predicted by DualSPHysics at a later time, which is different from the predictions from the other three methods. Table~\ref{fig_grid_convergence_rect_equil_positions} compares the equilibrium vertical center of mass position of the block with the analytical solution. As can be observed, all codes predict the block's equilibrium position correctly.

\begin{figure}[]
   \centering
   \subfigure[$t = 4.5$ s]{
   	\includegraphics[scale= 0.32]{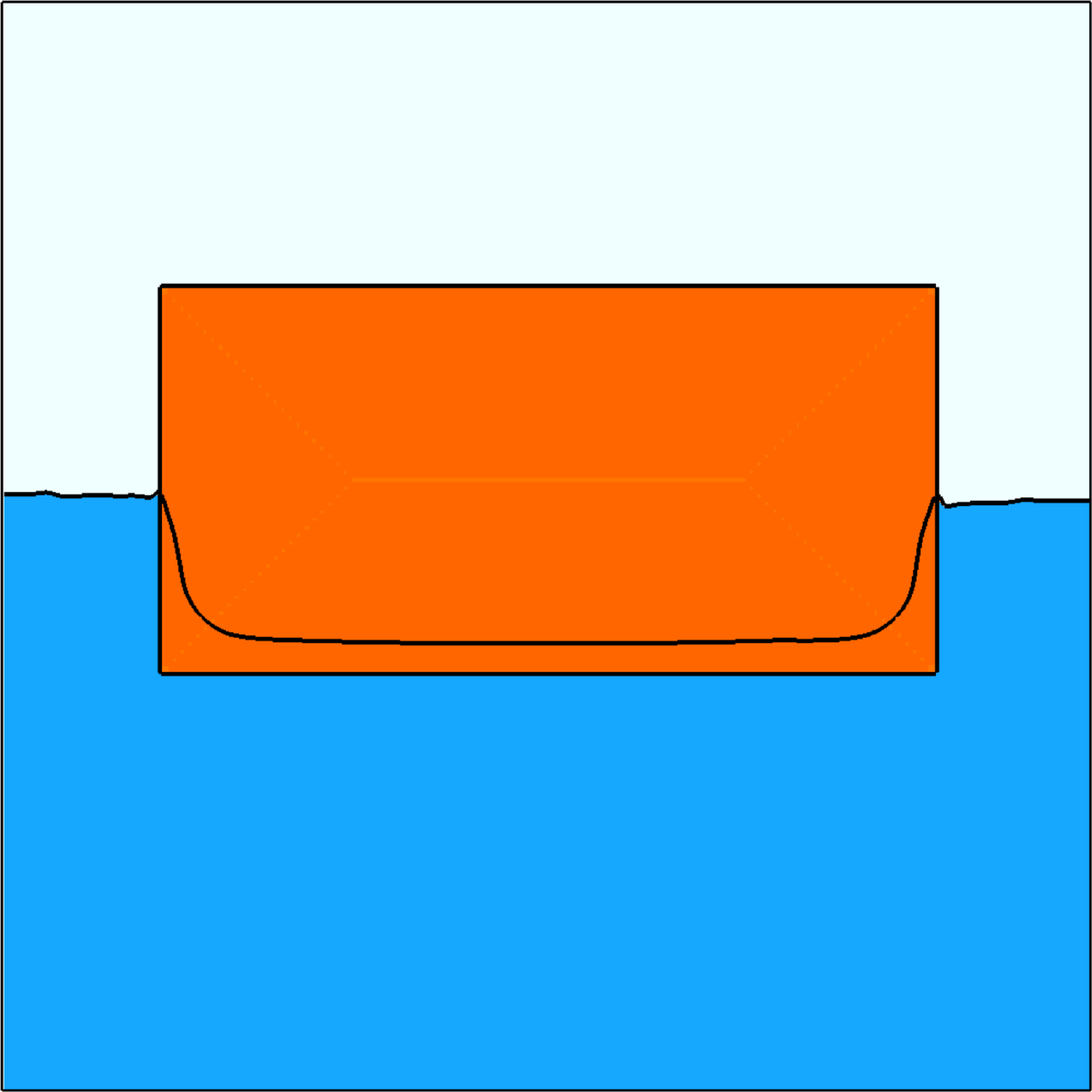}
	\label{rect_kn1_t4pt5s}
   }
    \subfigure[$t = 25$ s]{
   	\includegraphics[scale= 0.32]{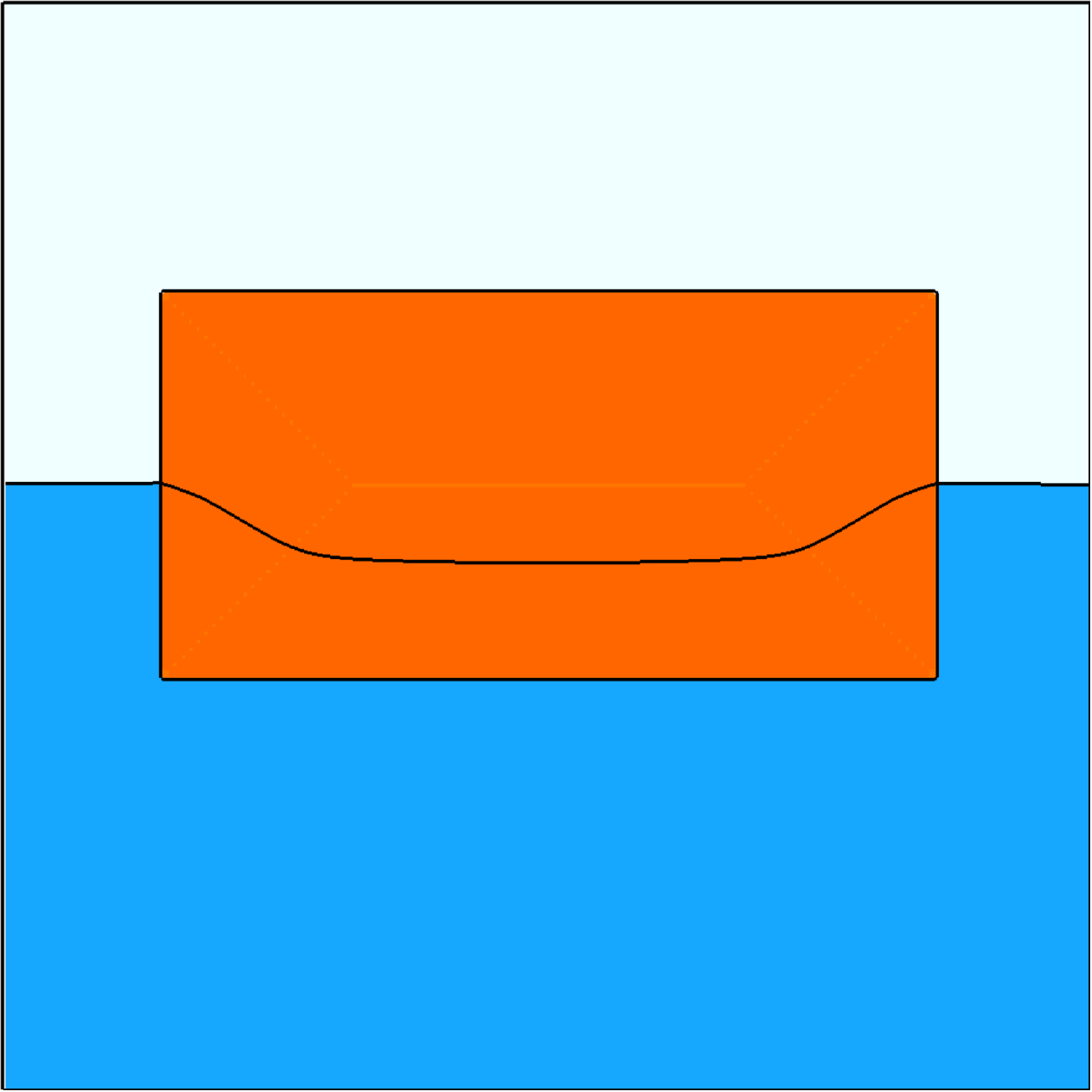}
	\label{rect_kn1_t25s}
   }     
    \caption{Air-water interface passing through the floating rectangular block at \subref{rect_kn1_t4pt5s} $t = 4.5$ s, and \subref{rect_kn1_t25s} $t = 25$ s using the FD/BP method.}
   \label{fig_rect_kn1_FSI}
\end{figure}

\begin{figure}[]
   \centering
   	\includegraphics[scale= 0.7]{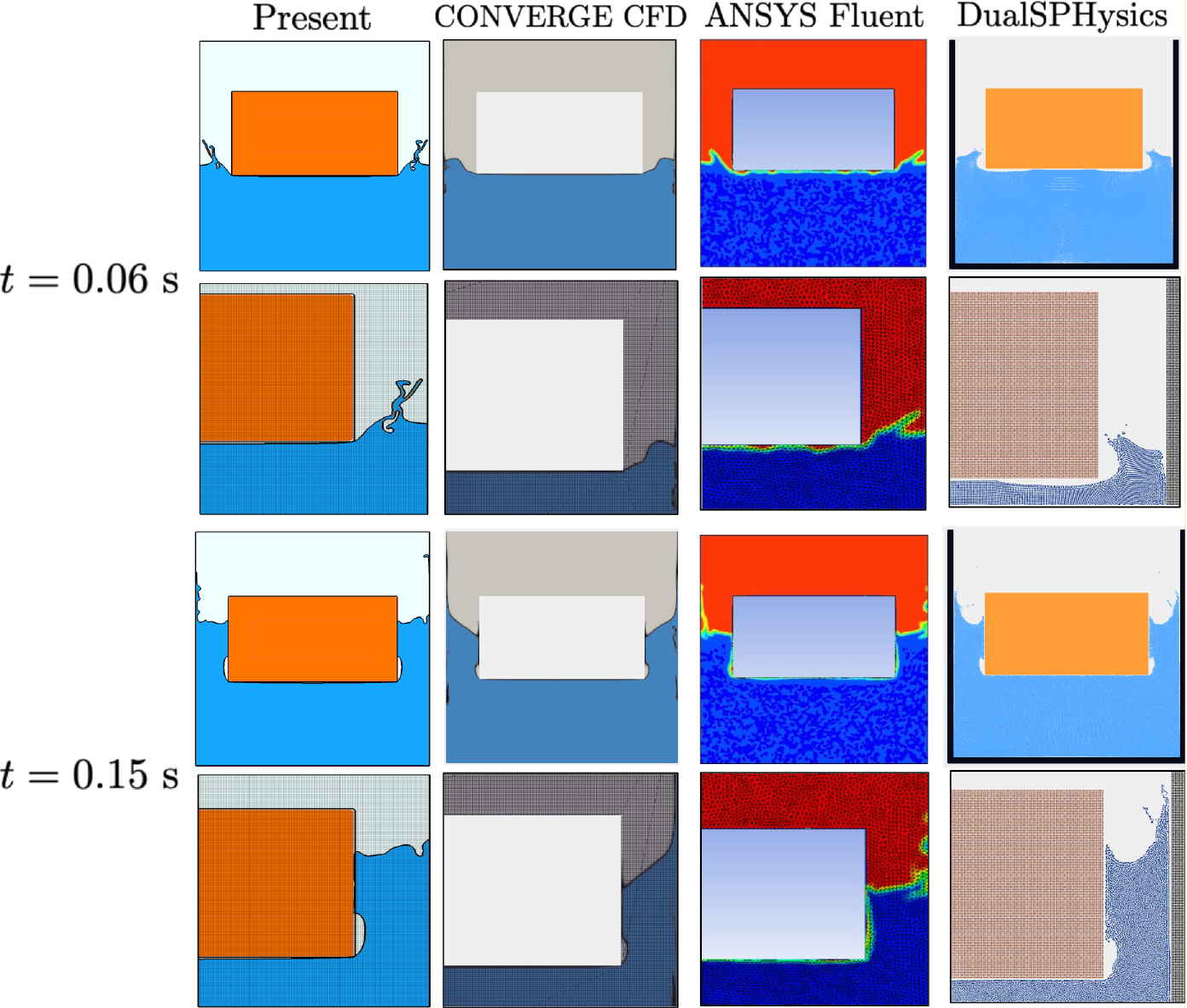}
    \caption{Fluid-structure interaction of a rectangular block impacting the air-water interface at $t = 0.06$ s and $t = 0.15$ s using the FD/BP method with mass loss fix, CONVERGE CFD, ANSYS Fluent and DualSPHysics.}
   \label{fig_rect_FSI}
\end{figure}

Fig.~\ref{fig_rect_FSI} illustrates the converged fluid-structure interaction and mesh around the bottom right corner of the block at $t = 0.06$ s and $t = 0.15$ s with four numerical techniques. In both time instances, the three Eulerian codes produce qualitatively the same wave dynamics. In contrast to other predictions, DualSPHysics predicts an ``inverted" wave near the block at $t =0.06$ s. This potentially unphysical behavior might be attributed to artificial gaps arising from SPH kernel interactions between the water and solid material points of the rectangular block and tank walls (as illustrated in Fig.~\ref{fig_rect_FSI}). The DualSPHysics simulation was configured based on accompanying online examples and guidelines provided in the software's user guide for ocean engineering problems.  Further, Fig.~\ref{fig_rect_kn1_FSI} illustrates the fictitious air-water interface passing through the body using the present solver. The air-water interface begins to penetrate the solid body around $t = 4.5$ s and completely penetrates it around $t = 25$ s. Fig.~\ref{fig_rect_kn1_FSI} contrasts sharply with Fig.~\ref{fig_rect_kn50}, which prevents the interface from entering the body. $\kappa \approx \Delta t/\rho_\text{s}$, and $\mathcal{K}_\text{n} = \mathcal{K}_\text{t} = 1$ are used to obtain physically correct dynamics. We have also proposed these penalty parameters in our previous works~\cite{Bhalla2020,Khedkar2021,Khedkar2022, Rama2023}.

\begin{table}[]
\centering
 \caption{Grid convergence test parameters $h$ (m), and $\dt_\text{max}$ (s) for the floating rectangular block problem.}
\begin{tabular}{|c|cc|cc|cc|cc|}
\hline
\multirow{2}{*}{Grid} & \multicolumn{2}{c|}{Present}                             & \multicolumn{2}{c|}{CONVERGE CFD}                            & \multicolumn{2}{c|}{ANSYS Fluent}                           & \multicolumn{2}{c|}{DualSPHysics} \\ \cline{2-9} 
                      & \multicolumn{1}{c|}{$h$} & $\dt_\text{max}$ & \multicolumn{1}{c|}{$h$} & $\dt_\text{max}$ & \multicolumn{1}{c|}{$h$} & $\dt_\text{max}$ & \multicolumn{1}{c|}{$h$}  & $\dt_\text{max}$  \\ \hline
Coarse          & \multicolumn{1}{c|}{7.5$\times 10^{-4}$}  &  7.5$\times 10^{-5}$ & \multicolumn{1}{c|}{1$\times 10^{-3}$}   &  1$\times 10^{-4}$ & \multicolumn{1}{c|}{1$\times 10^{-3}$}   &  1$\times 10^{-3}$   &   \multicolumn{1}{c|}{2$\times 10^{-3}$} & 1$\times 10^{-2}$   \\ \hline
Medium         & \multicolumn{1}{c|}{5$\times 10^{-4}$} & 5$\times 10^{-5}$   & \multicolumn{1}{c|}{7.5$\times 10^{-4}$}      &   1$\times 10^{-5}$   & \multicolumn{1}{c|}{7.5$\times 10^{-4}$}      &   1$\times 10^{-2}$   & \multicolumn{1}{c|}{1$\times 10^{-3}$}      &    1$\times 10^{-2}$  \\ \hline
Fine               & \multicolumn{1}{c|}{3.75$\times 10^{-4}$}   &  3.75$\times 10^{-5}$  & \multicolumn{1}{c|}{5$\times 10^{-4}$}      &  1$\times 10^{-4}$     & \multicolumn{1}{c|}{5$\times 10^{-4}$}      &   7.5$\times 10^{-3}$   & \multicolumn{1}{c|}{5$\times 10^{-4}$}      &    1$\times 10^{-2}$  \\ \hline
\end{tabular}
 \label{tab_grid_convergence_rect_ibamr}
\end{table}

\begin{table}[]
\centering
 \caption{Vertical center of mass position of the rectangle at equilibrium using FD/BP method (present work), CONVERGE CFD, ANSYS Fluent, DualSPHysics, and analysis.}
\begin{tabular}{|c|c|c|c|c|c|}
\hline
Grid   & Present  & CONVERGE CFD & ANSYS Fluent & DualSPHysics & Analytical   \\ \hline
Coarse & 0.1166 & 0.1168       & 0.1169       & 0.1208 &\multirow{3}{*}{0.1167} \\ \cline{1-5}
Medium & 0.1167 & 0.1167       & 0.1168       &     0.1199      &              \\ \cline{1-5}
Fine   & 0.1167 & 0.1168       & 0.1169       &          0.1184      &         \\ \hline
\end{tabular}
   \label{fig_grid_convergence_rect_equil_positions}
\end{table}

\begin{figure}[]
   \centering
   	\includegraphics[scale= 0.38]{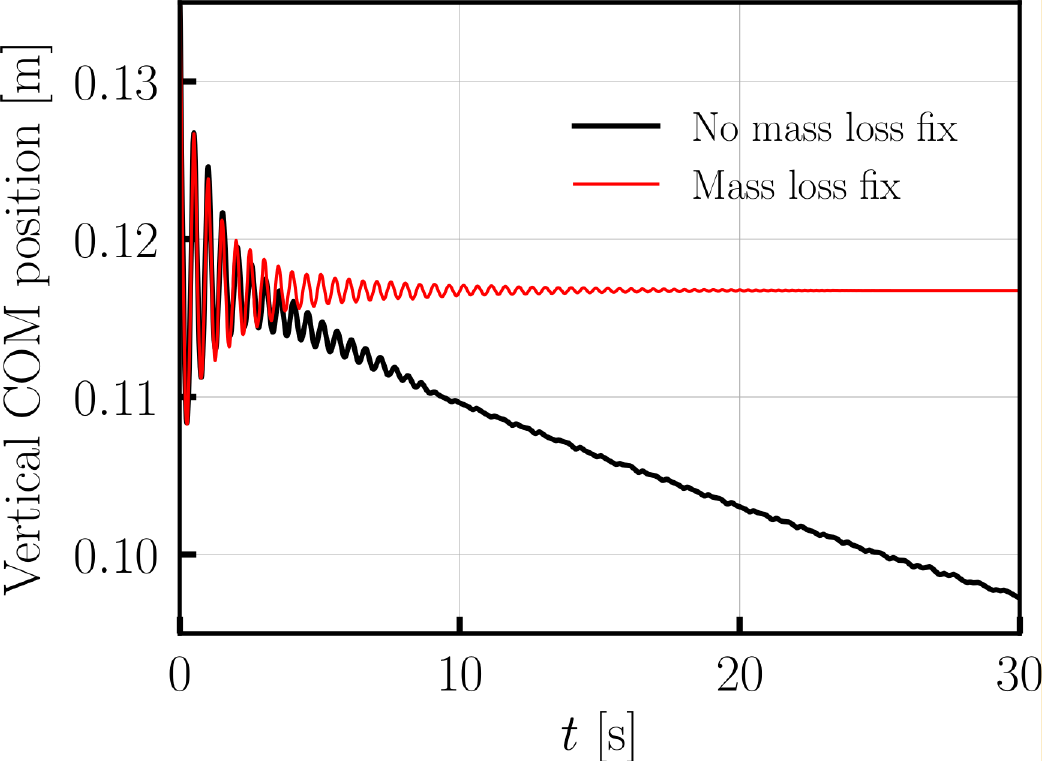}
    \caption{\ADDITION{Temporal evolution of the vertical center of mass of the rectangular block with and without the mass loss fix.}}
   \label{fig_rect_mass_fix_comparison}
\end{figure}

\begin{figure}[h]
   \centering
   \subfigure[\ADDITION{Normalized correction}]{
   	\includegraphics[scale= 0.37]{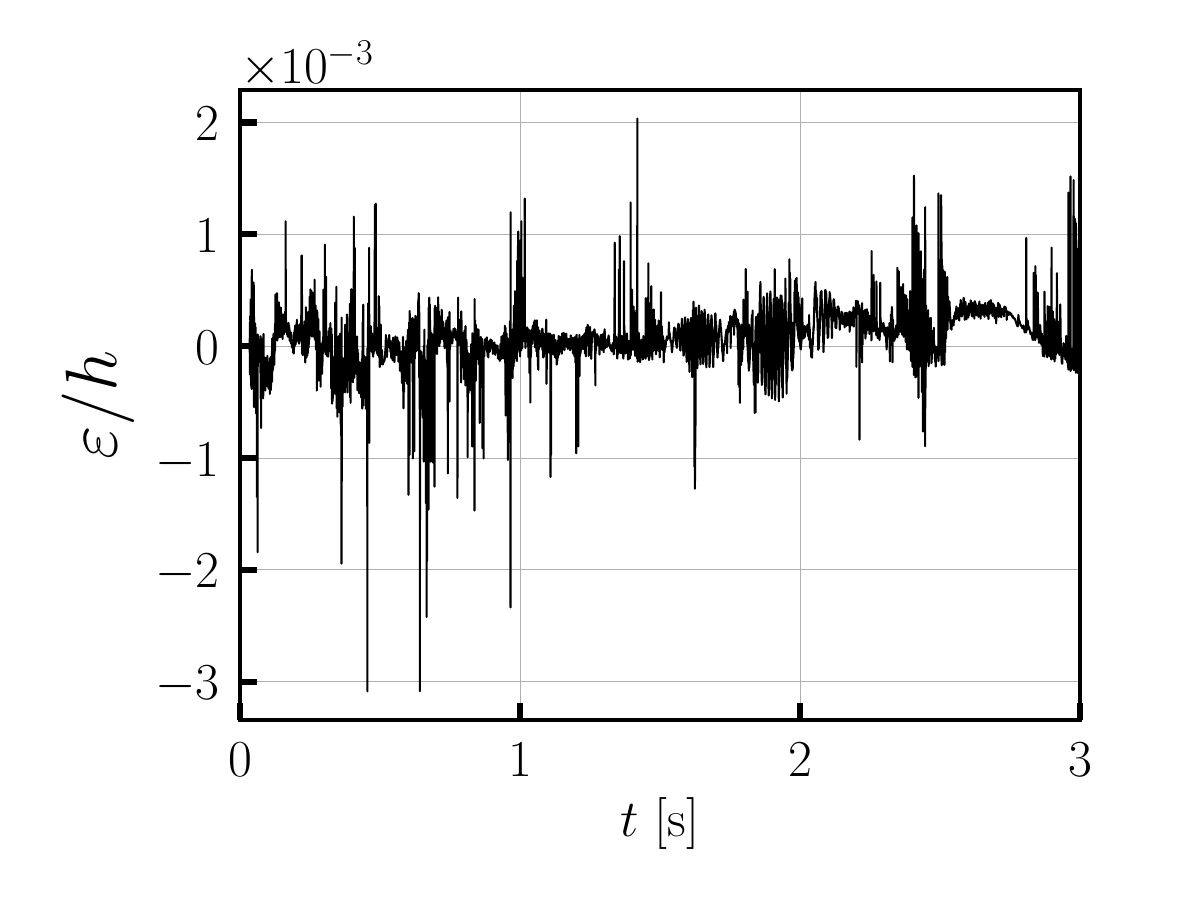}
	\label{fig_rect_LSshift}
   }
      \subfigure[\ADDITION{Relative change in volume}]{
   	\includegraphics[scale= 0.37]{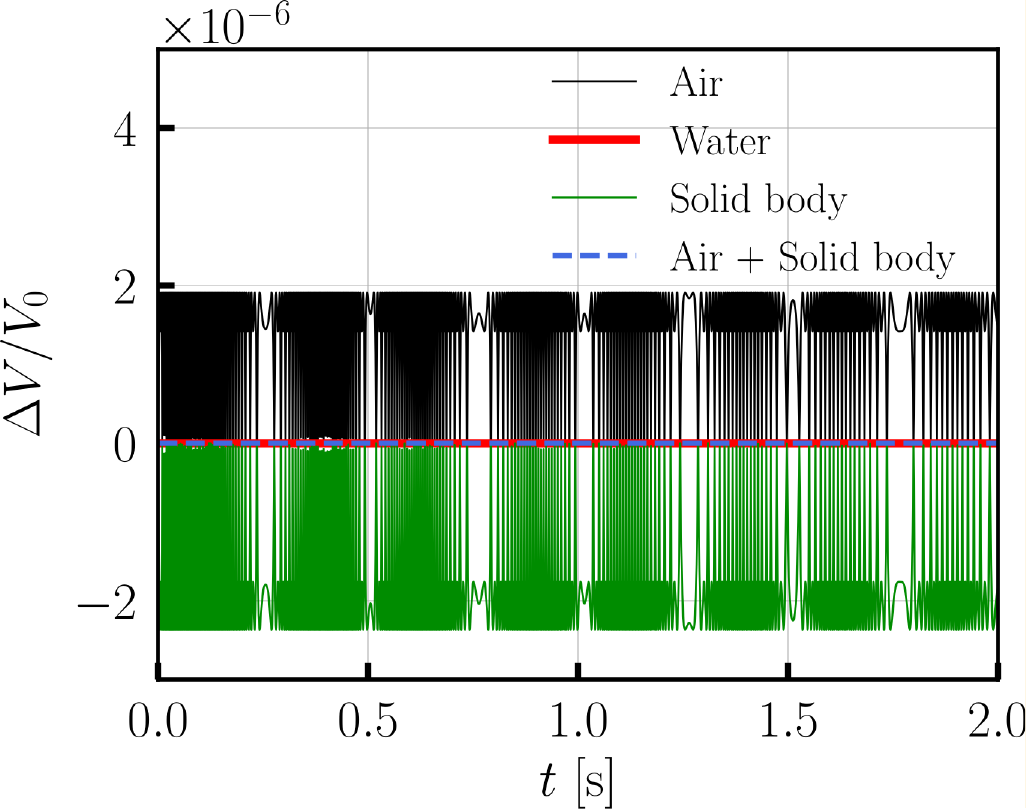}
	\label{fig_relative_volume_error_rect}
   }  
    \caption{\ADDITION{Time evolution of \subref{fig_rect_LSshift} the normalized correction to the level set function, and \subref{fig_relative_volume_error_rect} the relative volume change for air, water and solid phases for the floating rectangular block problem. A medium grid resolution is used here.}}
   \label{fig_rect_ycom_vol_error}
\end{figure}

Next, we examine the importance of conserving water mass/volume for this problem using the FD/BP method. Fig.~\ref{fig_rect_mass_fix_comparison} shows how the block heaves  with and without the mass loss fix (using a medium grid resolution). Without the mass loss fix method, the block settles at a different location. This corresponds to the initial water level position $y = 1.2 H$. This could be explained as follows. The original level set method without the mass loss fix conserves (under grid refinement) the sum total of water volume outside and inside the solid body. Without a volume conservation constraint, there is no distinction between real and fictitious fluids. By imposing a volume conservation constraint on the actual fluid, the approximate Lagrange multiplier method corrects the level set field at each time step. This can be observed from Fig.~\ref{fig_rect_LSshift}, which shows the normalized value of the correction as a function of time. Normalized correction values are in the order of $10^{-3}$, which is considerably smaller than the cell size $h = 0.0005$ m. FSI dynamics are not affected by this subgrid level shift. 

Fig.~\ref{fig_relative_volume_error_rect} shows the relative volume change for air, water and solid phases over time when the mass loss fix method is applied. The target fluid volume, in this case water, is conserved to machine precision. In addition, the sum of volumes of the remaining phases, i.e., air and block, is also conserved to machine accuracy.  

\subsubsection{Floating cylinder problem}
\label{subsec_floating_cylinder}

In this section, we simulate a case similar to the previous Sec.~\ref{subsubsec_floating_rect}, but with a different solid geometry, a cylinder of radius $R$ = 0.06 m. In comparison to the rectangular geometry, the cylindrical geometry leads to a nontrivial relationship between the submerged volume and the water level rise at equilibrium conditions. The steady state version of the problem is solved analytically in the appendix Sec.~\ref{sec_analytical_cylinder_calculations}. This example tests the accuracy of the approximate Lagrange multiplier method when dealing with complex geometries. 

\begin{figure}[]
   \centering
   	\includegraphics[scale= 0.38]{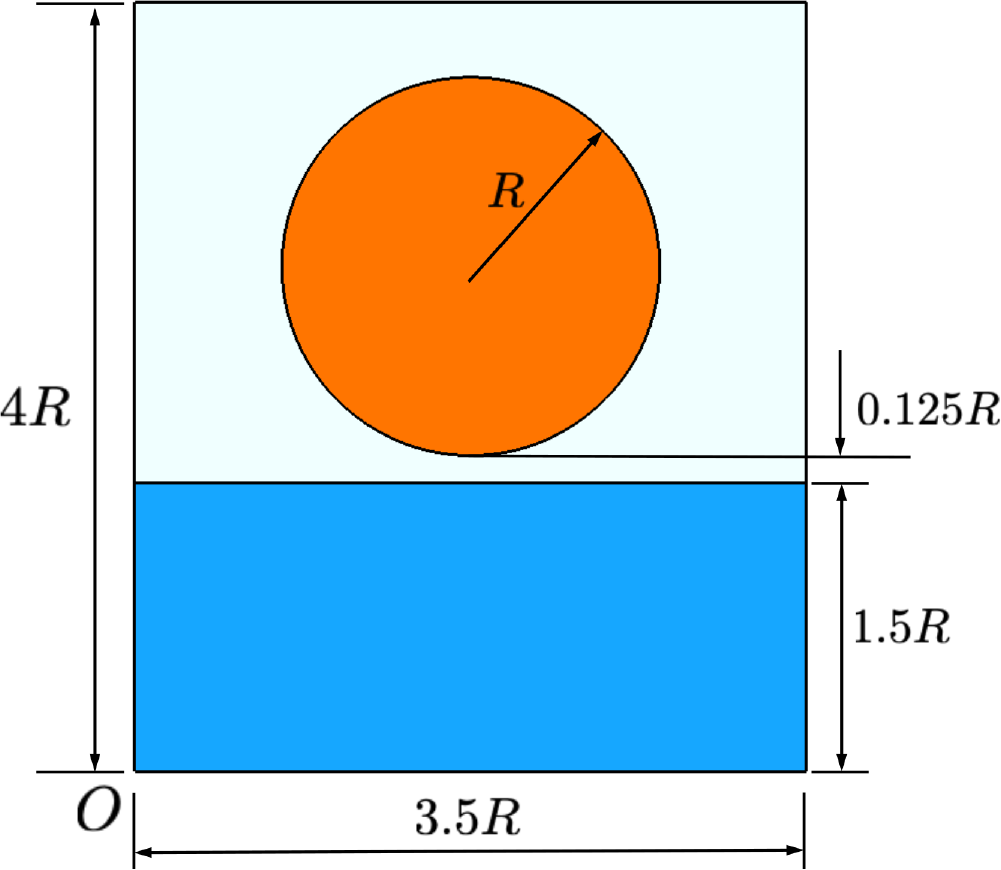}
    \caption{Schematic of the floating cylinder problem.}
   \label{fig_cyl_schematic}
\end{figure}

\begin{figure}[]
   \centering
   \subfigure[\ADDITION{Grid convergence study}]{
   	\includegraphics[scale= 0.37]{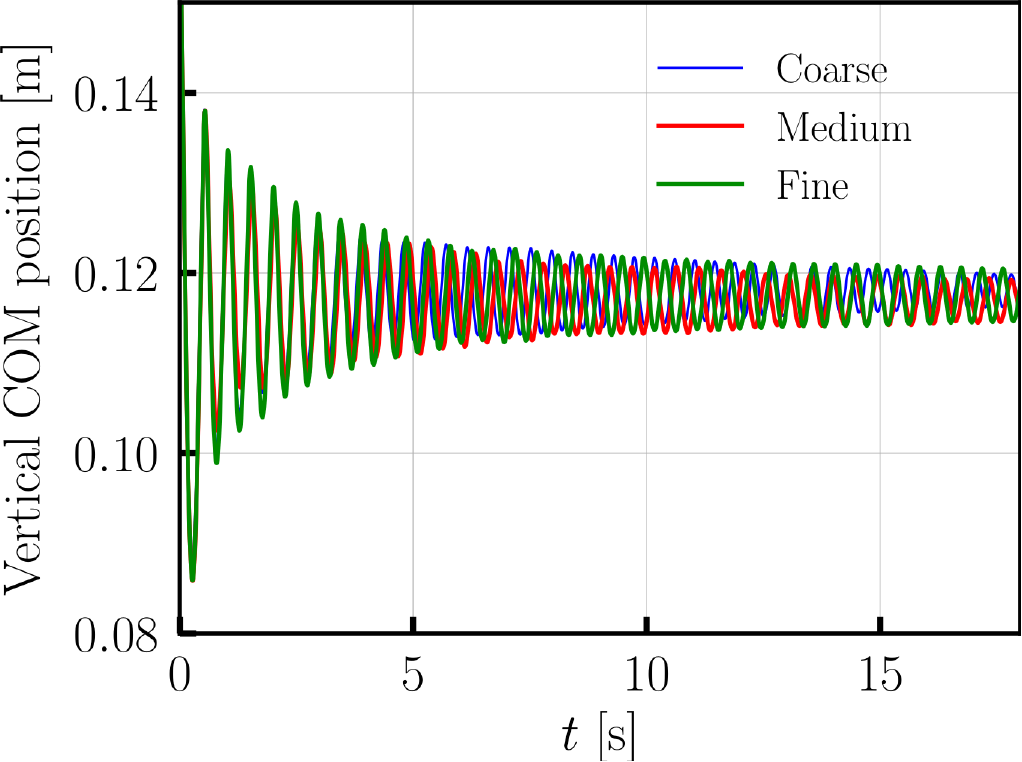}
	\label{fig_cyl_ibamr_grid_convergence}
   }
      \subfigure[\ADDITION{Heave dynamics}]{
   	\includegraphics[scale= 0.37]{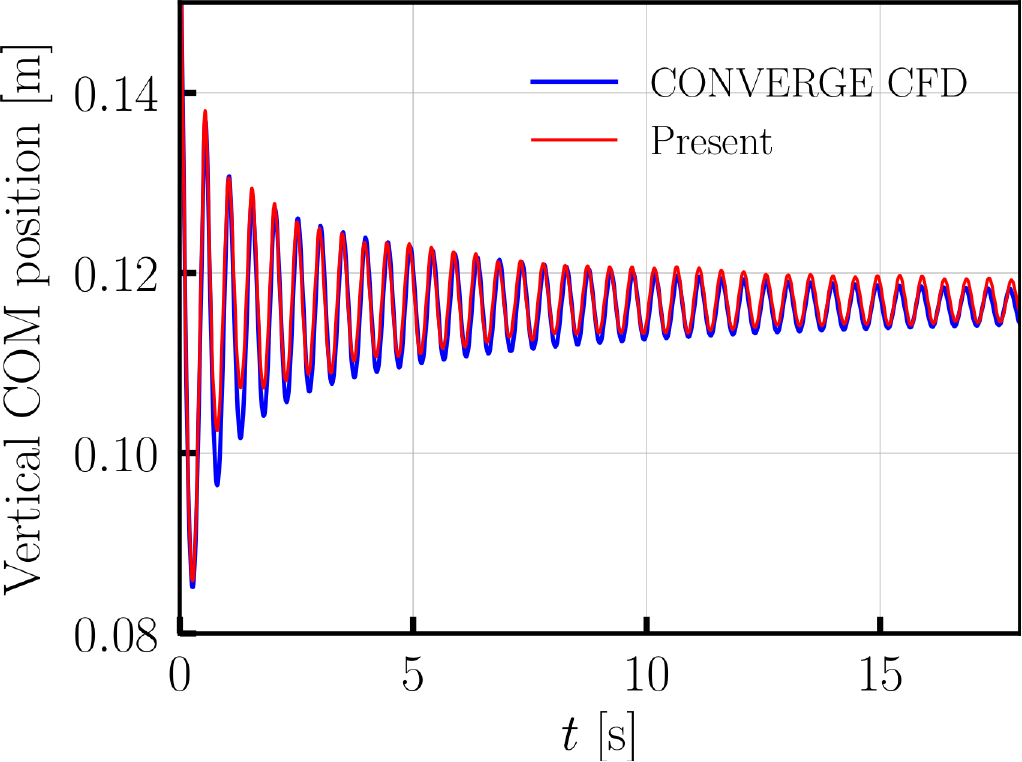}
	\label{fig_cyl_ibamr_converge_comparison}
   }  
    \caption{\ADDITION{\subref{fig_cyl_ibamr_grid_convergence} Temporal evolution of the cylinder's vertical center of mass with three grids. \subref{fig_cyl_ibamr_converge_comparison} Comparison of the heaving dynamics of the cylinder using CONVERGE CFD software and the FD/BP method (present work).}}
   \label{fig_rect_results}
\end{figure}

\begin{figure}[]
   \centering
   	\includegraphics[scale= 0.4]{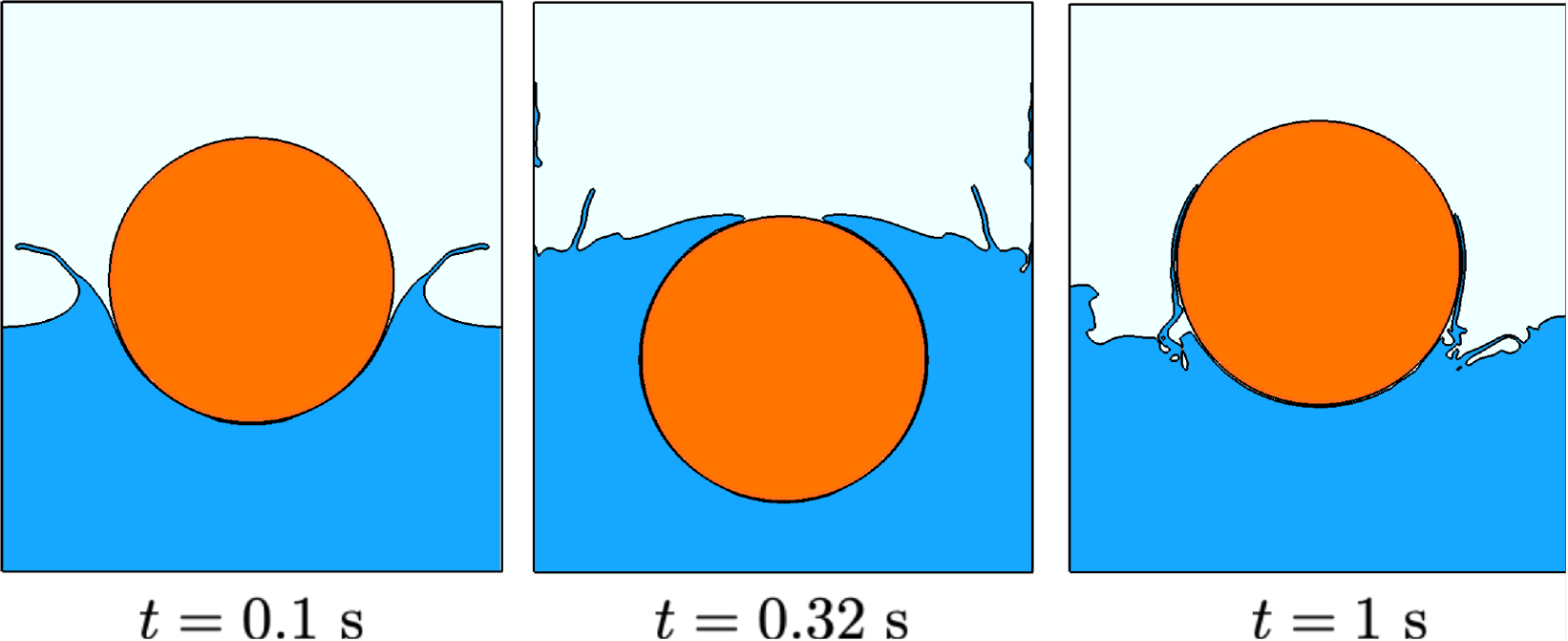}
    \caption{Fluid-structure interaction of the floating cylinder at different time instants using the FD/BP method (present work).}
   \label{fig_cyl_FSI}
\end{figure}

The schematic of the problem setup is shown in Fig.~\ref{fig_cyl_schematic}. Initially, the cylinder center is located at $(1.75R, 2R)$ inside a closed box of size $(3.5R,4R)$. $y = 1.5R$ is the initial water level. In this case, the cylinder can heave only vertically while the other degrees of freedom are locked. A grid convergence study is conducted on three uniform grid sizes of $280 \times 320$ (coarse), $420 \times 480$ (medium), and $560 \times 640$ (fine), corresponding to cell sizes of $h = 10^{-3}$ m, $7.5\times10^{-4}$ m and $5\times10^{-4}$ m, respectively. $t_\text{max}$ for these grids is set to be the same as in the floating rectangular block problem (see Table~\ref{fig_grid_convergence_rect_equil_positions}). The mass loss fix method is used for all simulations. Fig.~\ref{fig_cyl_ibamr_grid_convergence} compares the temporal evolution of the vertical center of mass position of the cylinder for the three grids. Considering our results, we conclude that medium grid resolution is sufficient. In order to validate the transient dynamics of the heaving cylinder, CONVERGE CFD software is used to simulate the same problem. Fig.~\ref{fig_cyl_ibamr_converge_comparison} presents a comparison of the two numerical codes' heaving dynamics. The CONVERGE CFD simulation is conducted on a mesh with cell size of $h = 0.00075$ m and $t_\text{max} = 10^{-5}$ s. Table~\ref{tab_grid_convergence_cyl_equil_positions} shows the final equilibrium position of the cylinder's center of mass and compares it with the analytical solution. Under grid refinement, the present solver's final equilibrium position approaches the analytical value. CONVERGE CFD also predicts the cylinder's final equilibrium position correctly.
Fig.~\ref{fig_cyl_FSI} illustrates the wave dynamics due to cylinder impact at $t = 0.01$, $t = 0.32$, and $t = 1$ s using the FD/BP method.

\begin{table}[]
\centering
 \caption{Vertical center of mass position of the cylinder at equilibrium using the FD/BP method (present work), CONVERGE CFD, and analysis.}
\begin{tabular}{|c|c|c|c|}
\hline
Grid      & Present  & CONVERGE CFD & Analytical              \\ \hline
Coarse & 0.1179   & -                       & \multirow{3}{*}{0.1169} \\ \cline{1-3}
Medium & 0.1178 & 0.1162              &                         \\ \cline{1-3}
Fine       & 0.1175 & -                       &                         \\ \hline
\end{tabular}
\label{tab_grid_convergence_cyl_equil_positions}
\end{table}

\begin{figure}[]
   \centering
   	\includegraphics[scale= 0.4]{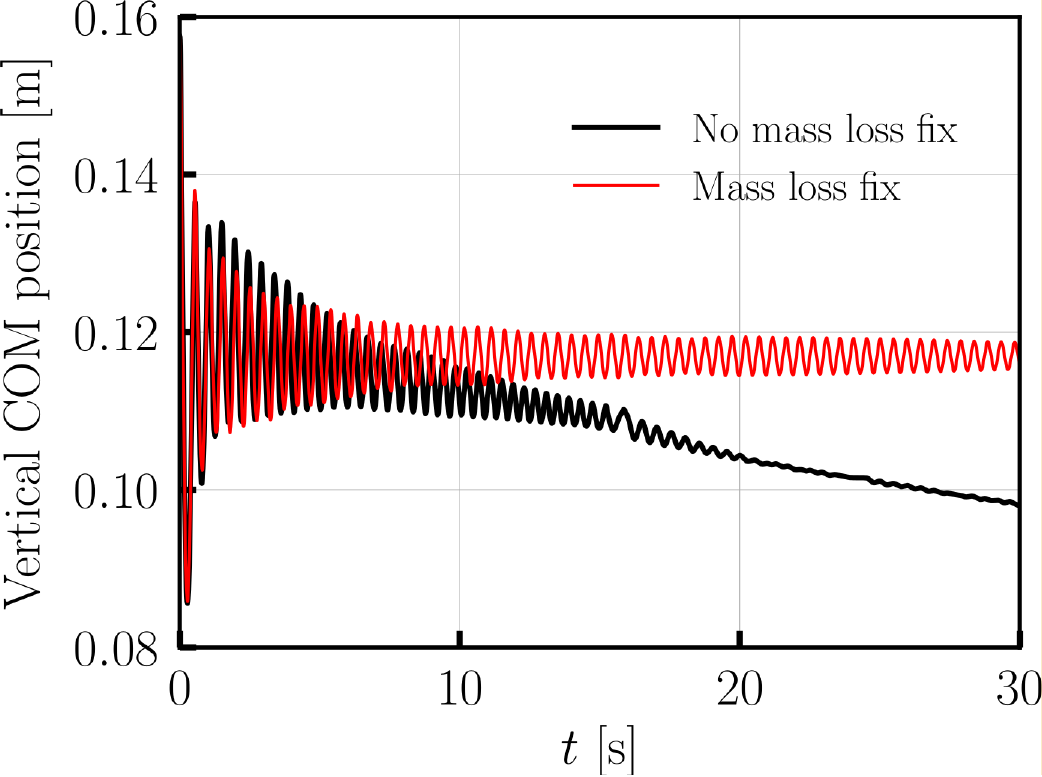}
    \caption{\ADDITION{Temporal evolution of the cylinder's vertical center of mass with and without mass loss fix.}}
   \label{fig_cyl_mass_fix_comparison}
\end{figure}

\begin{figure}[h]
   \centering
   \subfigure[\ADDITION{Normalized correction}]{
   	\includegraphics[scale= 0.37]{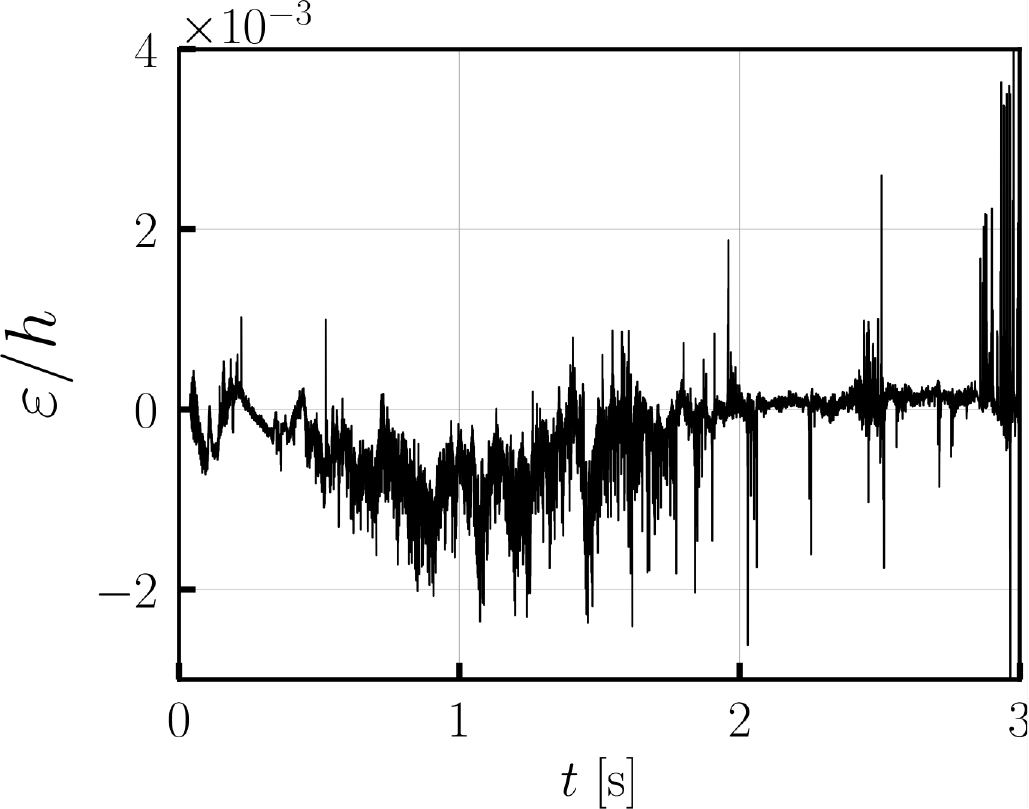}
	\label{fig_cyl_LSshift}
   }
      \subfigure[\ADDITION{Relative change in volume}]{
   	\includegraphics[scale= 0.37]{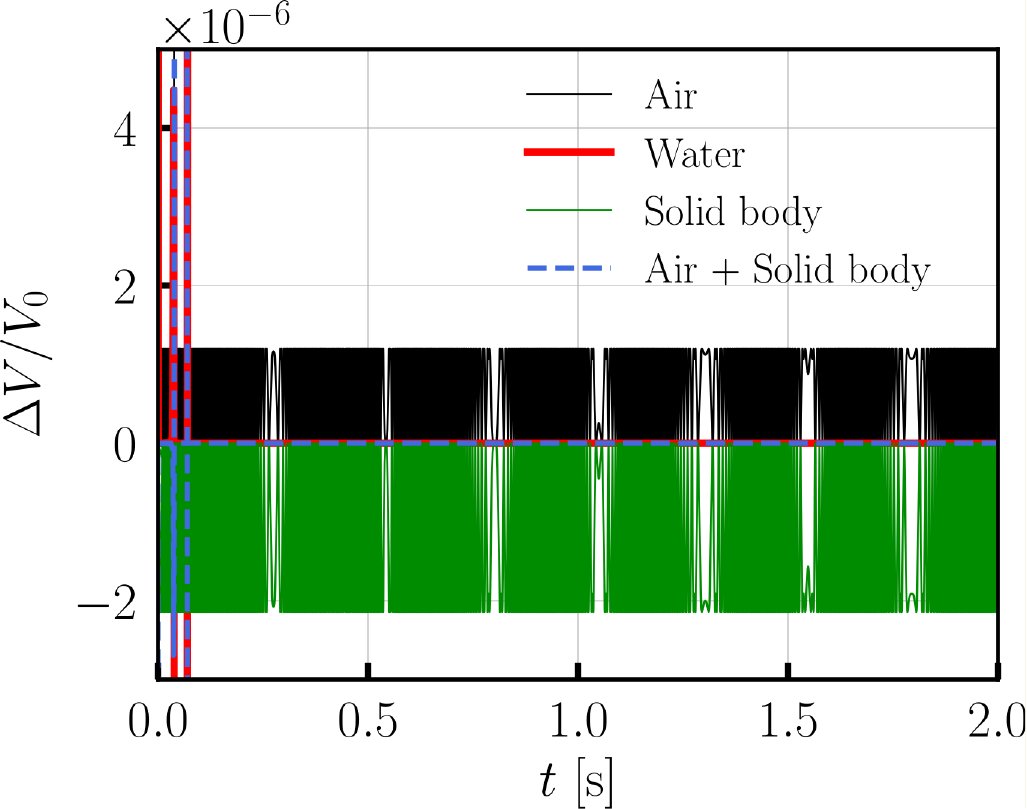}
	\label{fig_relative_volume_error_cyl}
   }  
    \caption{\ADDITION{Time evolution of \subref{fig_cyl_LSshift} the normalized correction to the level set function, and \subref{fig_relative_volume_error_cyl} the relative volume change for air, water and solid phases for the floating cylinder problem.}}
   \label{fig_cyl_vol_error}
\end{figure}

In addition, we compare the dynamics of the cylinder with and without the mass loss fix. The results are shown in Fig.~\ref{fig_cyl_mass_fix_comparison}. In the absence of the mass loss fix method, the cylinder exhibits highly irregular heaving dynamics and settles in the wrong equilibrium position (which is close to the initial water level). Fig.~\ref{fig_cyl_LSshift} shows the normalized value of the correction applied to the level set function in the mass fix approach. Subgrid level corrections are observed. Finally, Fig.~\ref{fig_relative_volume_error_cyl} shows that water phase volume is conserved to machine precision for this problem as well. 

\subsubsection{Free fall of a 2D Wedge}
\label{subsec_2dwedge}

\begin{figure}[]
   \centering
   	\includegraphics[scale= 0.5]{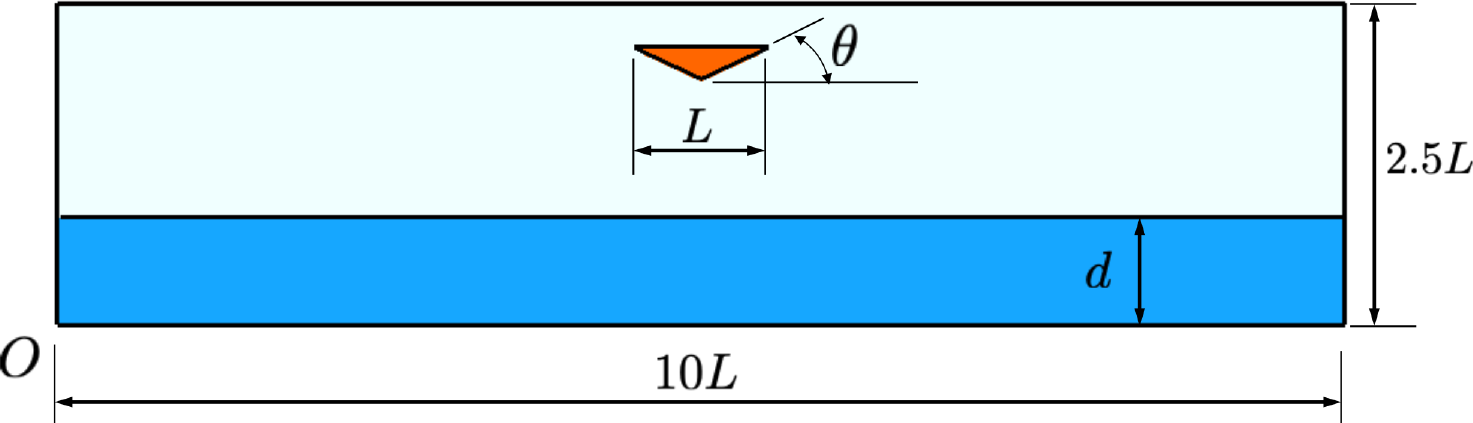}
    \caption{Schematic of the 2D wedge problem.}
   \label{fig_2d_wedge_schematic}
\end{figure}

Here we present a numerical investigation of the standard benchmark problem in the ocean engineering literature: the free fall of a 2D wedge impacting the air-water interface in a wide water tank, as described in~\cite{Yettou2006, Pathak2016, Nangia2019}. The schematic of the problem setup is illustrated in Fig.~\ref{fig_2d_wedge_schematic}. The computational domain's extents are $\Omega = [0, 10L]\times[0, 2.5L]$, in which $L = 1.2$ m represents the largest side of the wedge. The wedge has an angle of $\theta = 25^\circ$ and is initially positioned with its lowermost vertex at $(L/2, 2.3)$ m within the domain. The initial water depth is $d = 1$ m. Water density is $\rho_\text{w}$ 1000 kg/$\text{m}^3$ and viscosity is $\mu_\text{w} = 10^{-3}$ Pa$\cdot$s. For air, these values are $\rho_\text{a} = 1.2$ kg/$\text{m}^3$ and $\mu_\text{a} = 1.8\times10^{-5}$ Pa$\cdot$s. The wedge has a density of $\rho_\text{s}$ = 466.6 kg/m$^3$ and its fictitious viscosity is set equal to that of water. The wedge is free to heave vertically, while all other degrees of freedom are constrained.

\begin{figure}[]
   \centering
   \subfigure[Vertical displacement]{
   	\includegraphics[scale= 0.38]{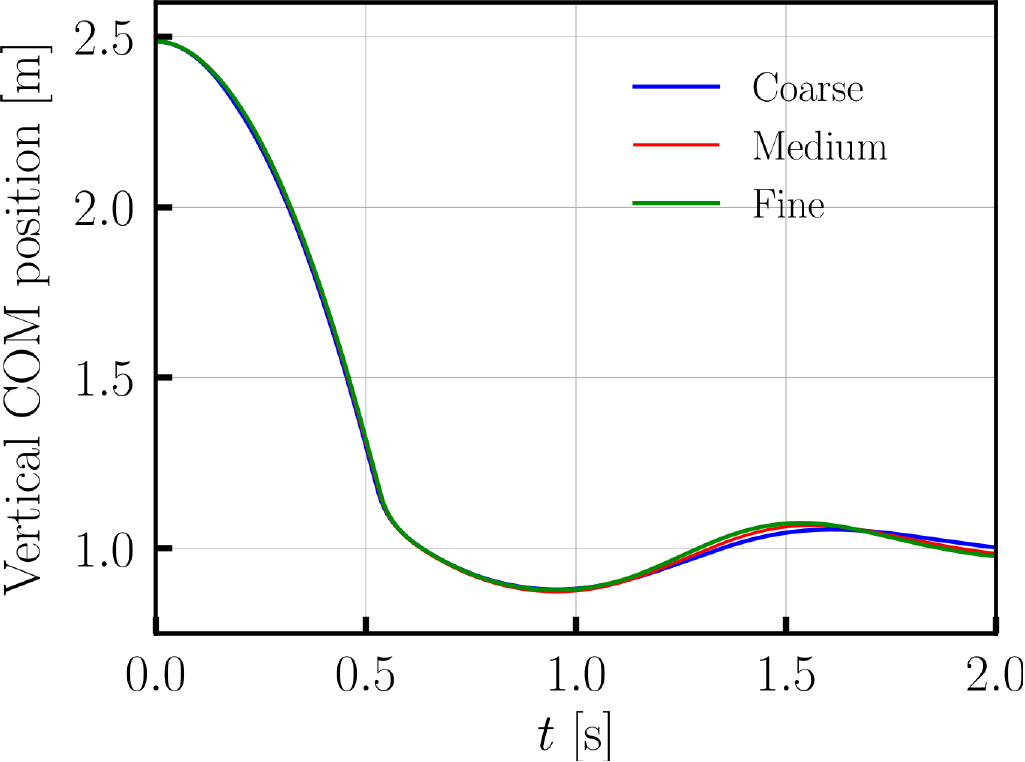}
	\label{fig_displacement_2dwedge_grid}
   }
      \subfigure[Vertical velocity]{
   	\includegraphics[scale= 0.38]{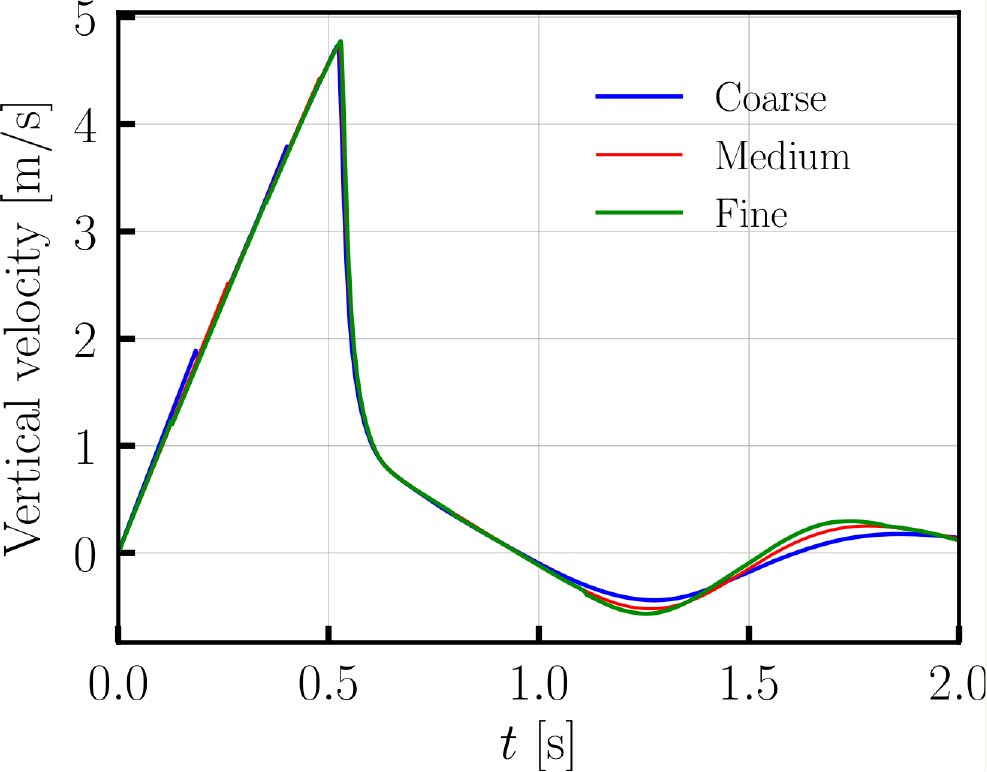}
	\label{fig_velocity_2dwedge_grid}
   }  
    \caption{Time evolution of \subref{fig_displacement_2dwedge_grid} vertical center of mass position and \subref{fig_velocity_2dwedge_grid} vertical velocity of a freely falling 2D wedge at different mesh resolutions.}
   \label{fig_2dwedge_convergence}
\end{figure}

\begin{figure}[]
   \centering
   \subfigure[Vertical displacement]{
   	\includegraphics[scale= 0.38]{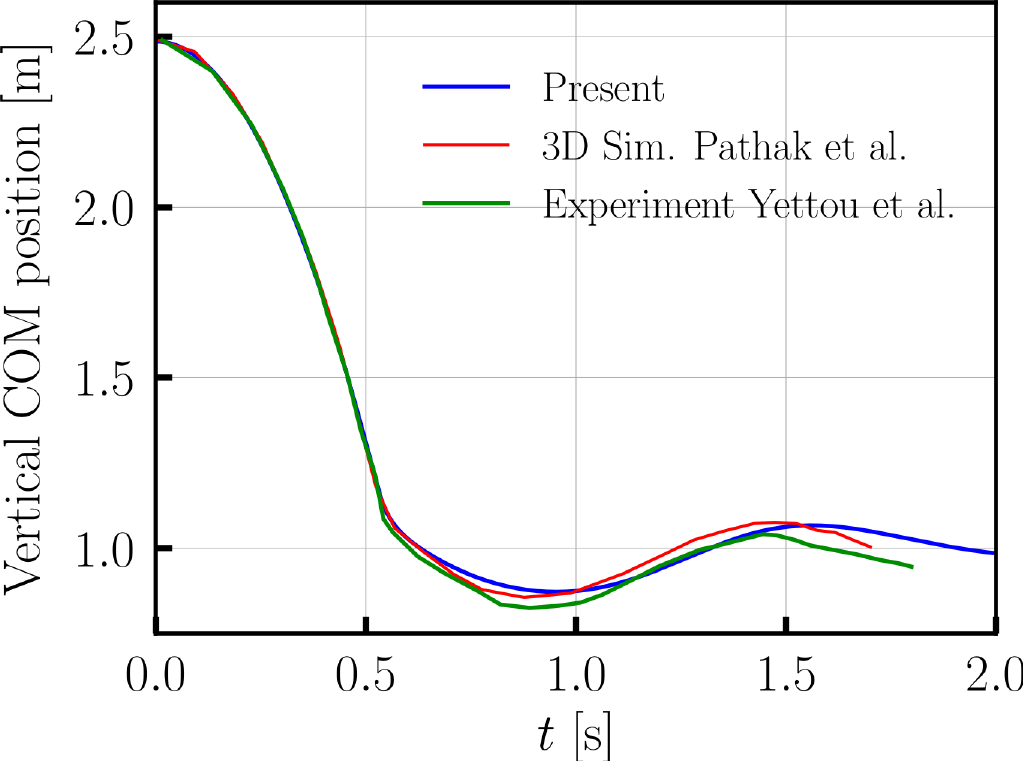}
	\label{fig_displacement_2dwedge_comparison}
   }
      \subfigure[Vertical velocity]{
   	\includegraphics[scale= 0.38]{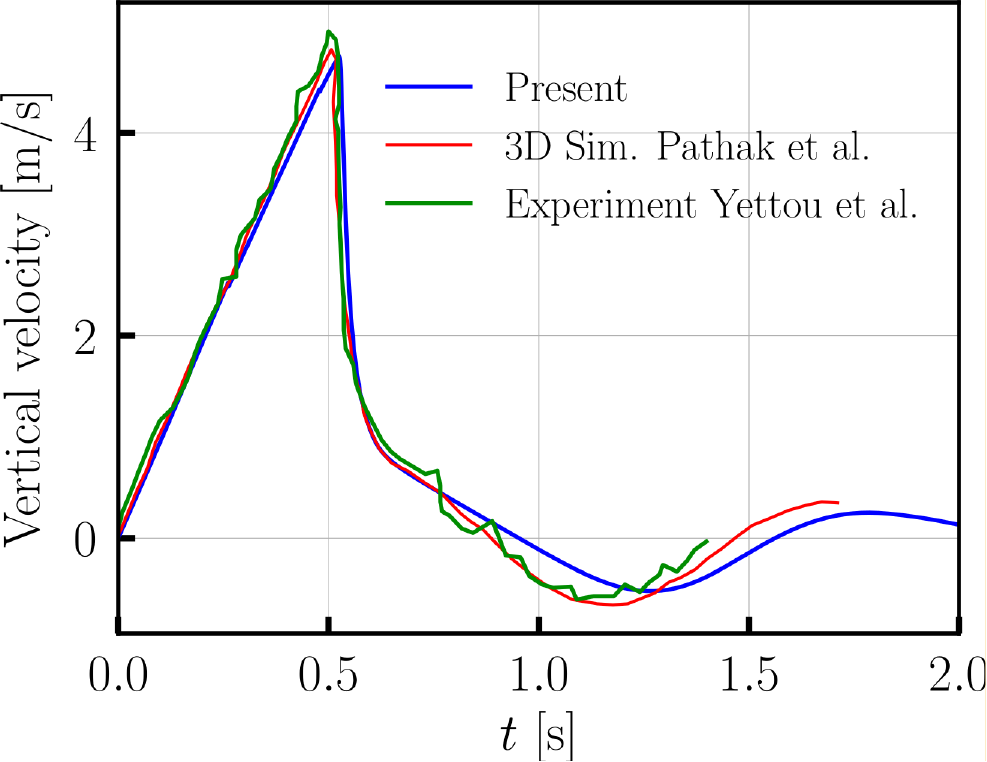}
	\label{fig_velocity_2dwedge_comparison}
   }  
    \caption{Comparison of \subref{fig_displacement_2dwedge_comparison} vertical center of mass position and \subref{fig_velocity_2dwedge_comparison} vertical velocity with 3D volume of fluid simulations of Pathak et al.~\cite{Pathak2016} and experiments of Yettou et al.~\cite{Yettou2006}.}
   \label{fig_2dwedge_comparison}
\end{figure}

\begin{figure}[]
   \centering
   \subfigure[$t = 0.8125$ s]{
   	\includegraphics[scale= 0.25]{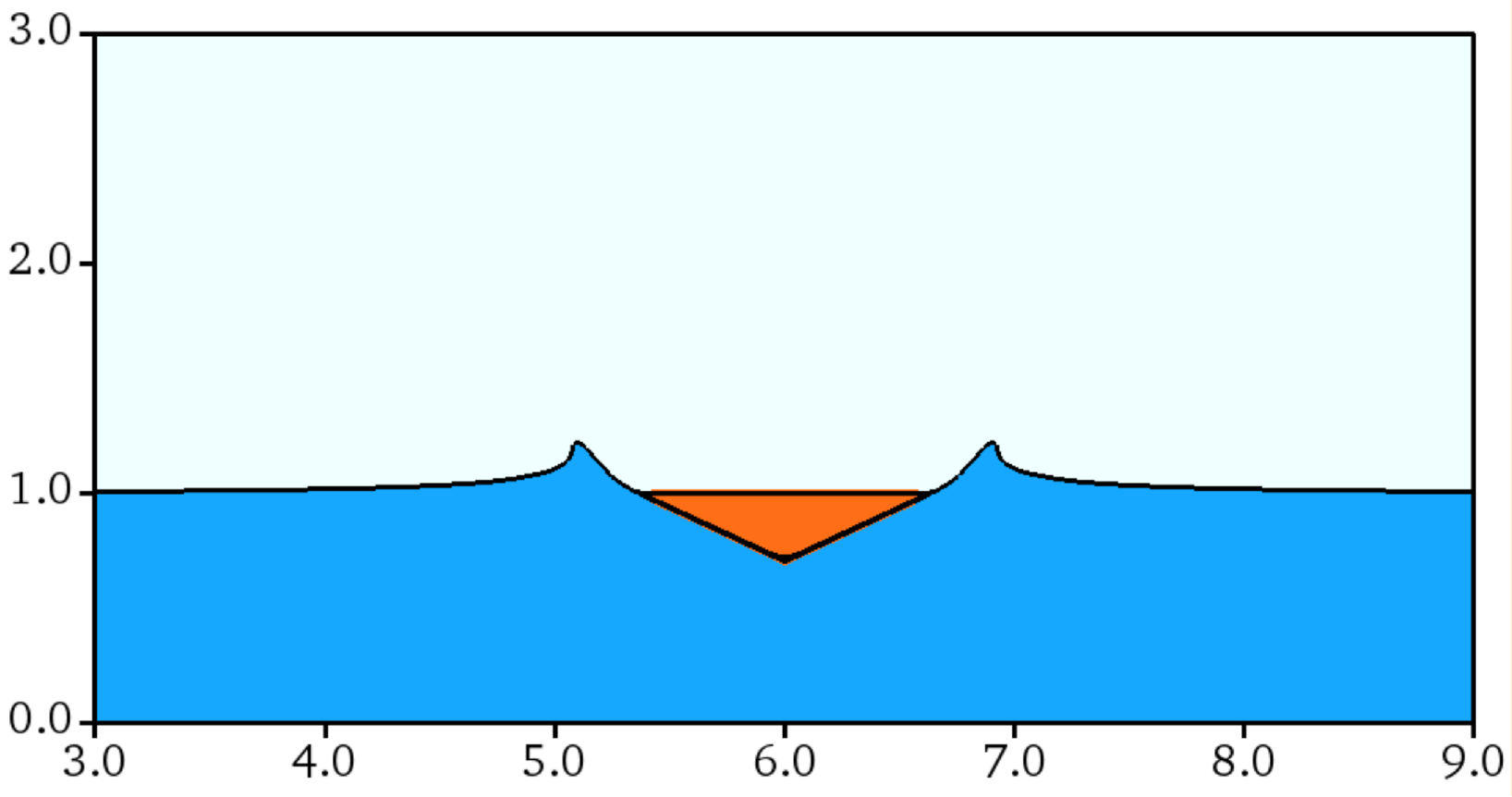}
	\label{fig_2dwedge_rho_t1}
   }
      \subfigure[$t = 0.8125$ s]{
   	\includegraphics[scale= 0.25]{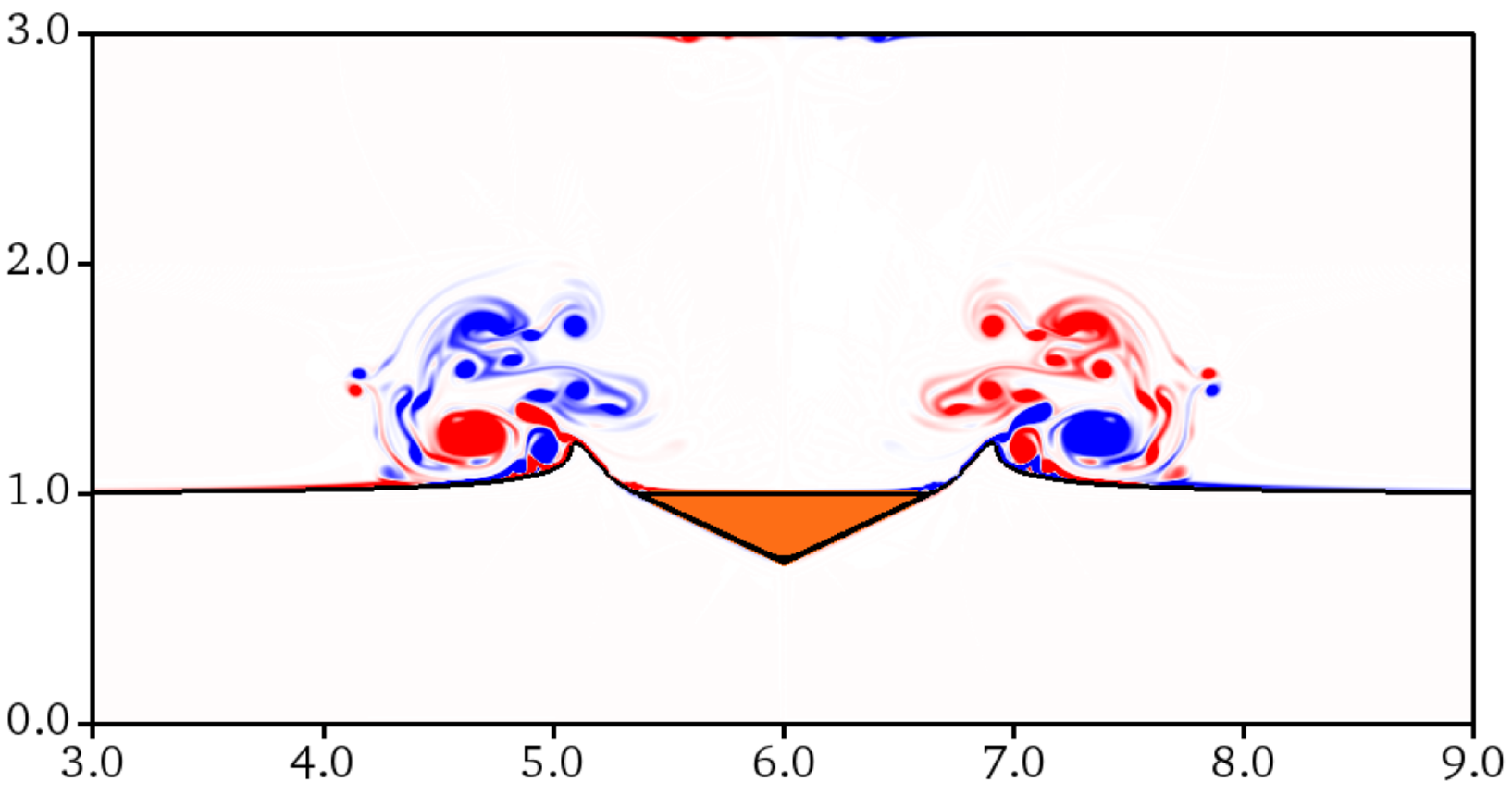}
	\label{fig_2dwedge_w_t1}
   }
      \subfigure[$t = 1$ s]{
   	\includegraphics[scale= 0.25]{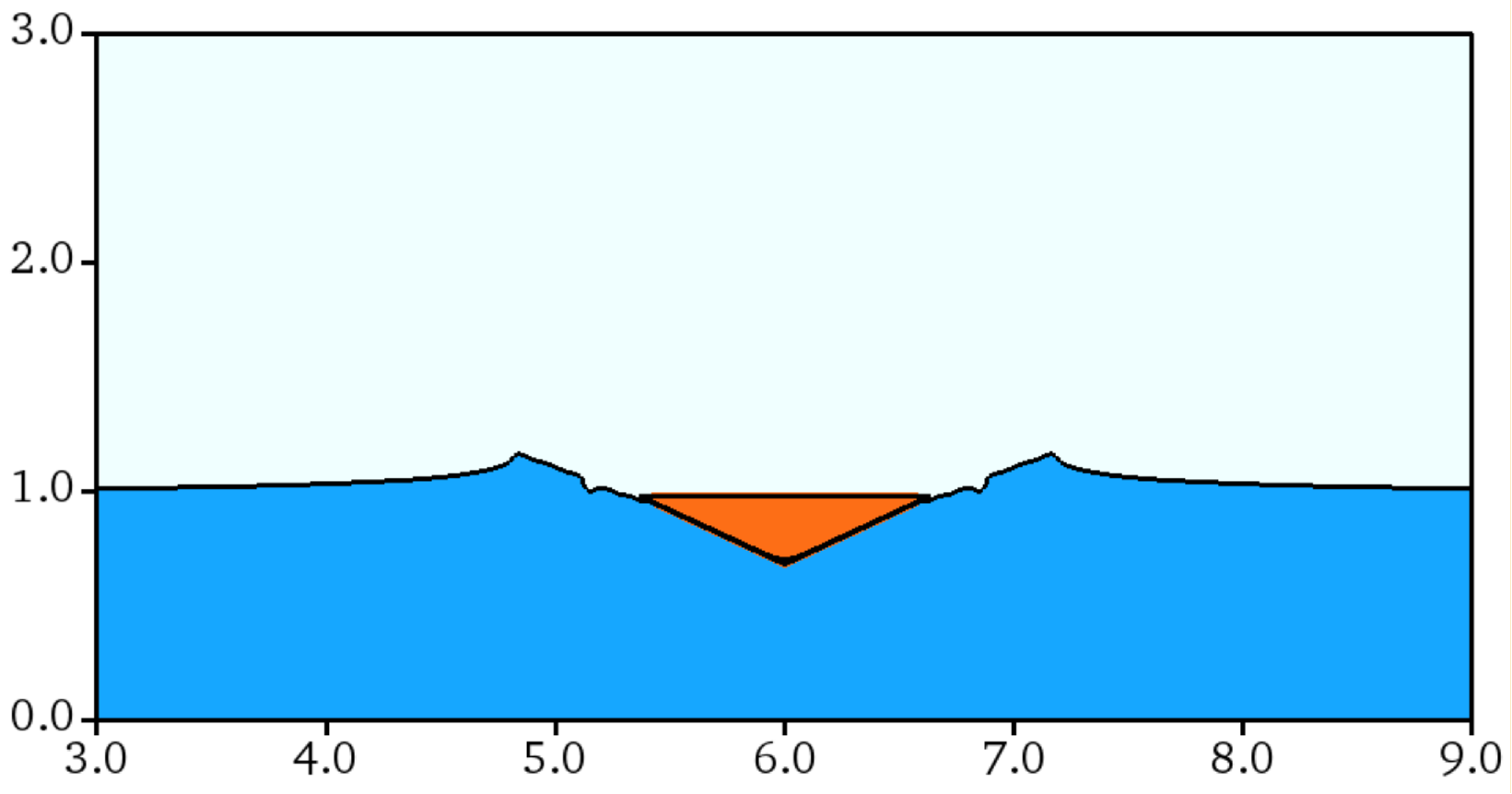}
	\label{fig_2dwedge_rho_t2}
   }  
     \subfigure[$t = 1$ s]{
   	\includegraphics[scale= 0.25]{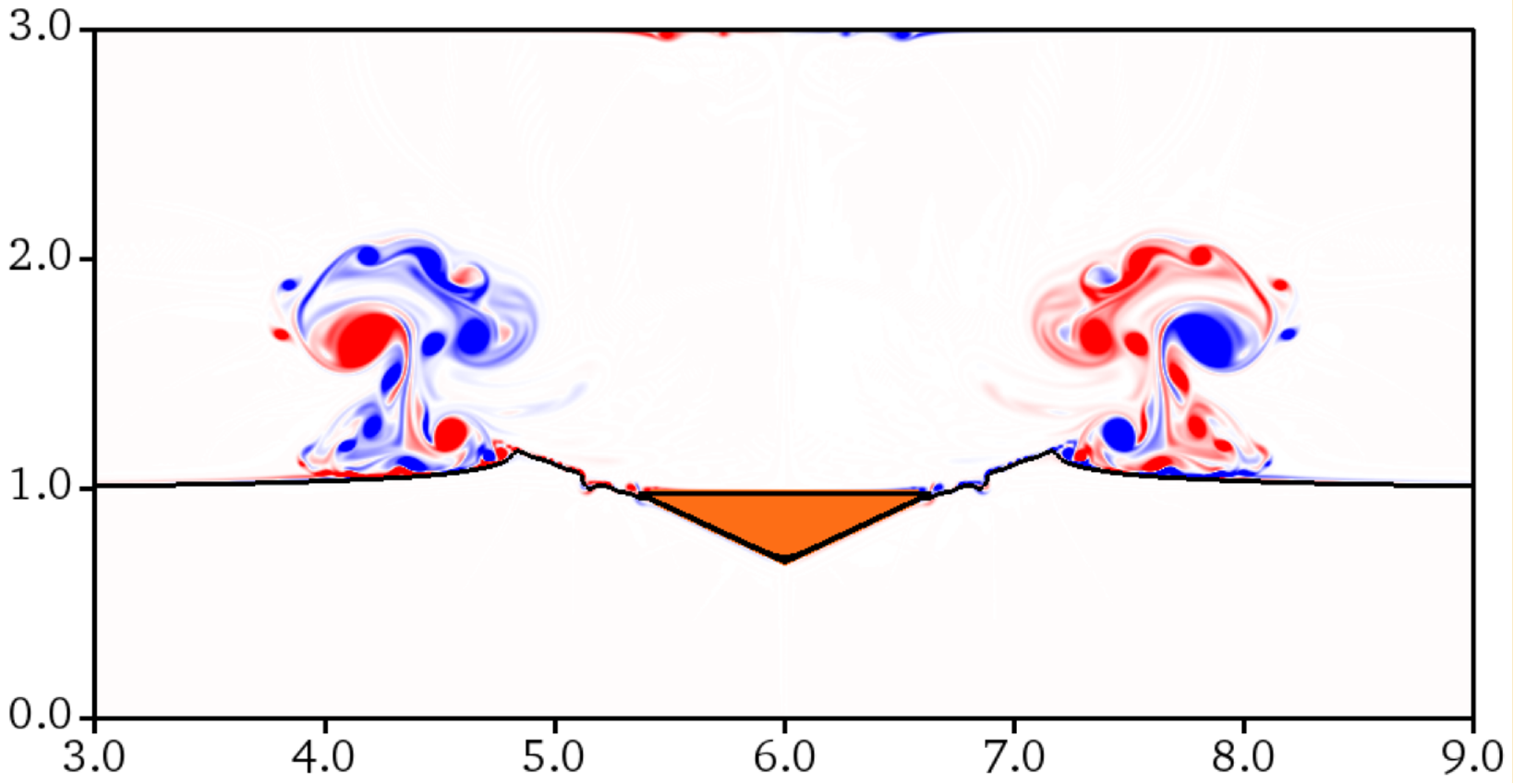}
	\label{fig_2dwedge_w_t2}
   }  
    \caption{FSI dynamics of a free falling 2D wedge impacting the air-water interface: (left) density and (right) vorticity (in the range -100 to 100 $\text{s}^{-1}$) plots.}
   \label{fig_2dwedge_FSI}
\end{figure}

A grid convergence study is conducted for this problem using three uniform grid sizes listed in Table~\ref{tab_grid_convergence_2dwedge}. Figs.~\ref{fig_displacement_2dwedge_grid} and~\ref{fig_velocity_2dwedge_grid} compare the vertical center of mass position and velocity at different mesh resolutions. From the plots, it can be seen that FSI dynamics can be resolved with a medium grid resolution. We compare the converged results with those from the prior numerical~\cite{Pathak2016} and experimental~\cite{Yettou2006} studies in Fig.~\ref{fig_2dwedge_comparison}. There is good agreement between the dynamics. Additionally, Fig.~\ref{fig_2dwedge_FSI} illustrates the wave and vortex dynamics that occur as a consequence of the FSI.

\begin{table}[]
\centering
 \caption{Grid convergence test parameters for the free falling problem.}
\begin{tabular}{|c|c|c|}
\hline
Grid   & $h$ (m)     & $\Delta t_\text{max}$ (s)   \\ \hline
Coarse & 0.01   & 6.25$\times10^{-5}$   \\ \hline
Medium & 0.005  & 3.125$\times10^{-5}$  \\ \hline
Fine   & 0.0025 & 1.5652$\times10^{-5}$ \\ \hline
\end{tabular}
\label{tab_grid_convergence_2dwedge}
\end{table}

\begin{figure}[]
   \centering
   \subfigure[Vertical displacement]{
   	\includegraphics[scale= 0.38]{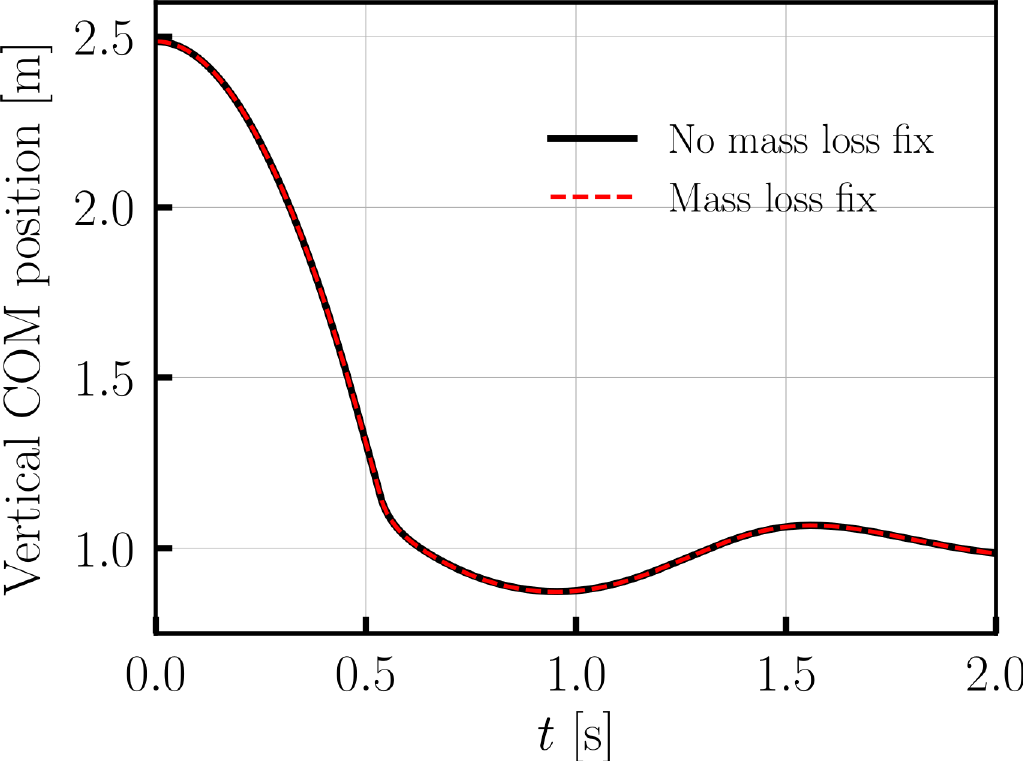}
	\label{fig_displacement_2dwedge_mass_comparison}
   }
      \subfigure[Vertical velocity]{
   	\includegraphics[scale= 0.38]{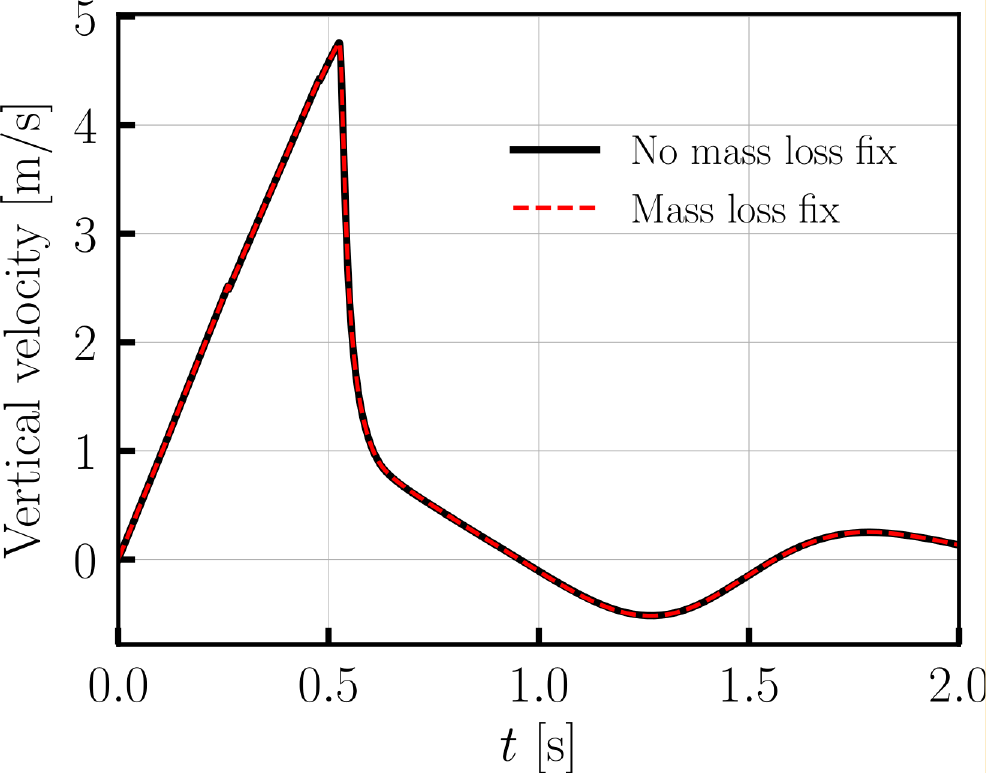}
	\label{fig_velocity_2dwedge_mass_comparison}
   }  
    \caption{Comparison of \subref{fig_displacement_2dwedge_mass_comparison} vertical center of mass position and \subref{fig_velocity_2dwedge_mass_comparison} velocity of the falling wedge with and without the mass loss fix.}
   \label{fig_2dwedge_mass_comparison}
\end{figure}

\begin{figure}[]
   \centering
   \subfigure[Normalized correction]{
   	\includegraphics[scale= 0.38]{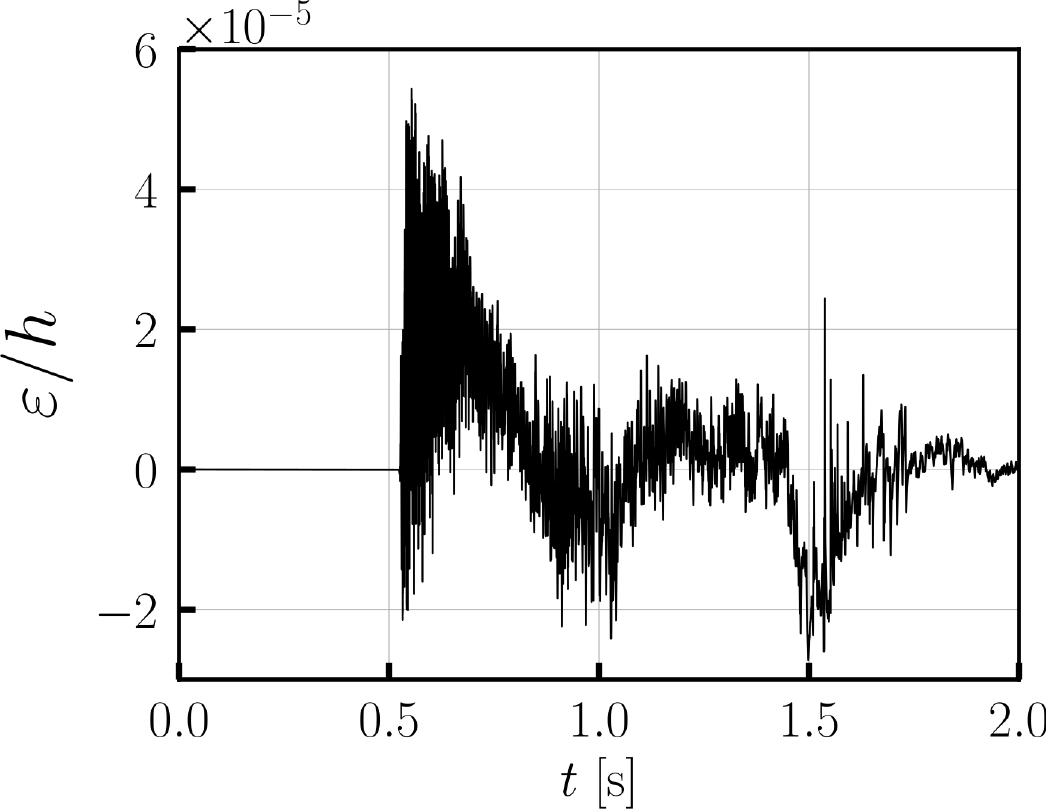}
	\label{fig_2dwedge_LSshift}
   }
      \subfigure[Relative change in volumes]{
   	\includegraphics[scale= 0.38]{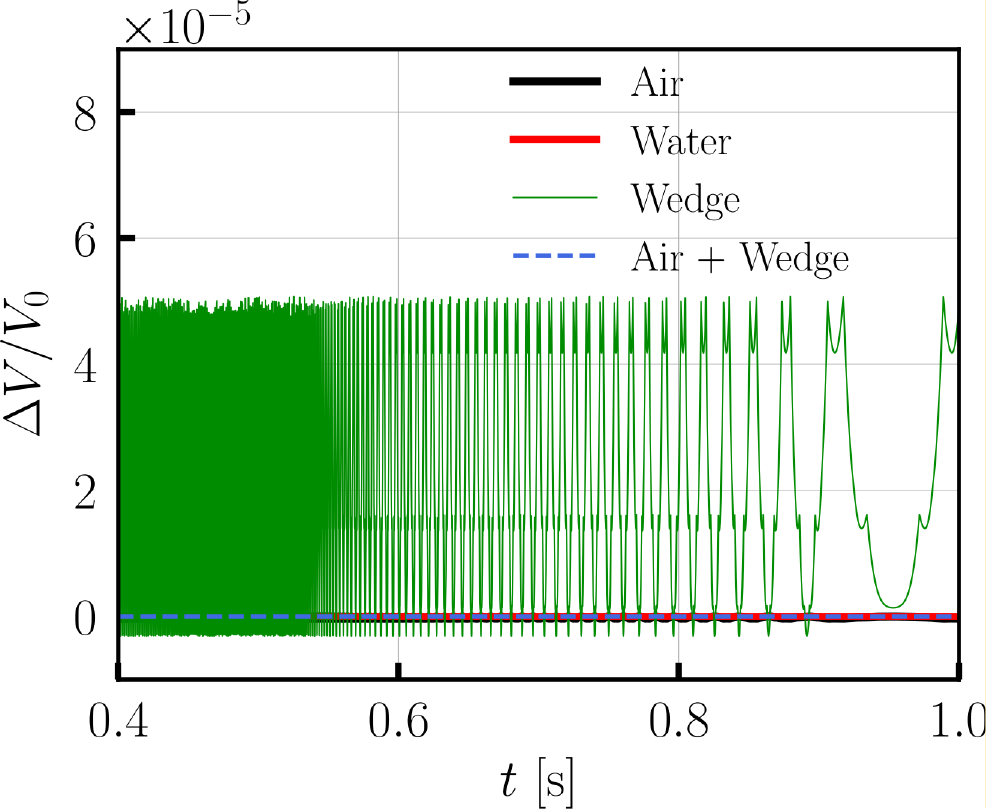}
	\label{fig_vol_error_2dwedge}
   }  
    \caption{Time evolution of \subref{fig_2dwedge_LSshift} the normalized correction to the level set function, and \subref{fig_vol_error_2dwedge} the relative volume change for air, water and solid phases for the falling wedge problem.}
   \label{fig_2dwedge_vol_error_LSshift}
\end{figure}

Given the large computational domain used for this benchmark problem, we hypothesize that maintaining a constant water volume is not essential for capturing the correct fluid-structure interaction dynamics of the wedge impact. This is because the water level rise is expected to be negligible, even at equilibrium. To verify this, we compare the wedge's heave motion with and without the mass loss fix method. As shown in Fig.~\ref{fig_2dwedge_mass_comparison}, the dynamics remain identical. Consequently, mass-conservative schemes, such as cut-cell methods or geometric volume of fluid (which are also more challenging to implement), offer less advantage for these types of problems. Their strengths are more evident in problems like those discussed in Secs.~\ref{subsubsec_floating_rect} and \ref{subsec_floating_cylinder}. Finally, Figs.~\ref{fig_2dwedge_LSshift} and~\ref{fig_vol_error_2dwedge} illustrate the normalized correction and relative volume change for different phases, respectively, when the mass-loss fix is applied. 

\subsubsection{Wave energy converter (WEC) problem}
\label{subsec_vcyl}

\begin{figure}[]
   \centering
   	\includegraphics[scale= 0.5]{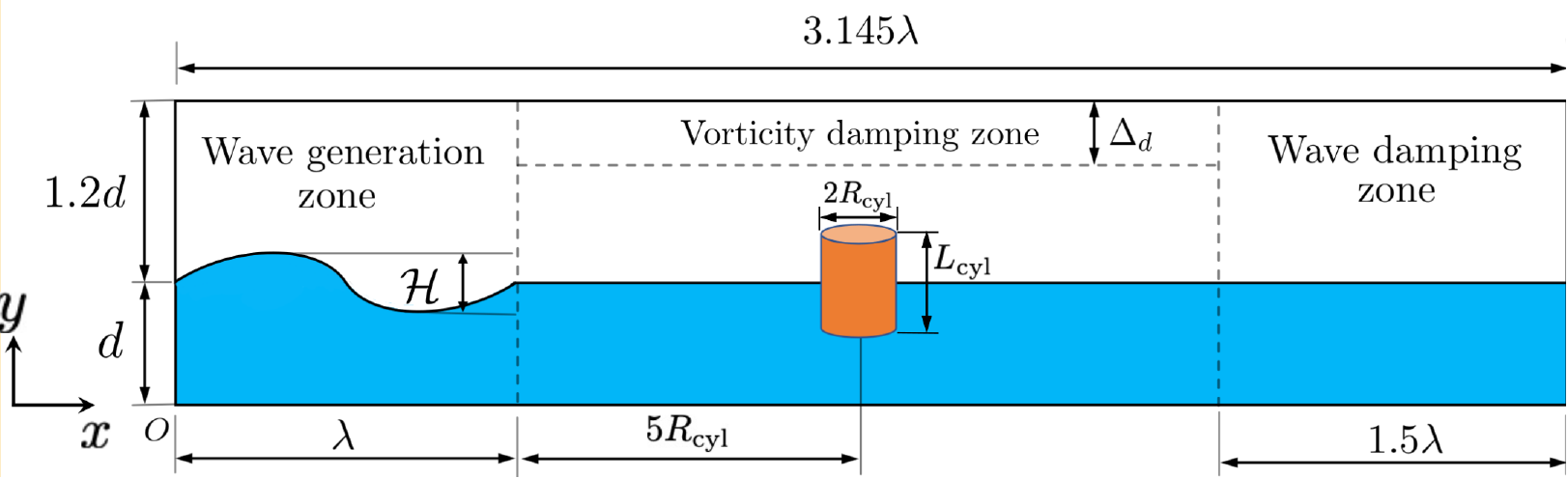}
    \caption{Schematic of the 3D vertical cylinder heaving at air-water interface.}
   \label{fig_3d_vcyl_schematic}
\end{figure}

This section presents a numerical investigation of wave-structure interaction (WSI) for a three-dimensional wave energy converter (WEC). We focus on the dynamics of a 1:20 scaled down model of a converter studied for optimal control in Cretel et al.~\cite{Cretel2010}. Unlike Cretel et al., who employed the boundary element method for WSI resolution, we simulate the converter's heaving response due to incoming waves in a (long) numerical wave tank (NWT). All boundaries of the tank are modeled as stationary walls, except for the top boundary which is considered to be open (zero pressure boundary condition). The schematic of the wave tank with the converter is depicted in Fig.~\ref{fig_3d_vcyl_schematic}.

The computational domain is of extents $\Omega = [0, 3.145\lambda]\times[0, 2.2d]\times[0, 12R_\text{cyl}]$ m, in which $\lambda$ is the wavelength of the incoming waves, $d = 2$ m is the mean depth of water in the tank  and $R_\text{cyl} = 0.25$ m is the radius of the cylinder. The cylinder has a length of $L_\text{cyl} = 0.8$ m and its density is half that of water. The domain origin is located at the lower left corner. First-order Stokes waves (regular waves) with height $\mathcal{H} = 0.1$ m and period $\mathcal{T}_p = 1.5652$ s are generated in the wave generation zone. These waves have a wavelength to water depth ratio that classifies them as deep water waves. The Stokes waves satisfy the dispersion relation, which reads as
\begin{equation}
	\omega^2 = g\kappa\tanh(\kappa d)
	\label{eq_dispersion_relation}
\end{equation}
in which, $\omega = 2\pi/\mathcal{T}_p$ is the wave frequency, $g = 9.81$ m/s$^2$ is the acceleration due to gravity, and $\kappa = 2\pi/\lambda$ is the wave number. The water waves propagate in the positive $x$ direction, interact with the device, and encounter a wave damping zone on the right side of the tank (Fig.~\ref{fig_3d_vcyl_schematic}). The wave generation zone prevents waves reflected from the device from interfering with the left boundary. The wave damping zone on the right of the domain prevents waves reflected from the right boundary from traveling into the tank and affecting the device dynamics. The wave generation zone width is $\lambda$ and that of the wave damping zone is $1.5\lambda$. These dimensions are selected based on authors' experience in modeling wave energy converter devices in NWTs; see~\cite{Dafnakis2020, Khedkar2021, Khedkar2022}. 
The wave generation and damping zones utilize the relaxation method~\cite{Jacobsen2012} to smoothly generate and absorb waves, respectively. Similarly, a vorticity damping zone is implemented at the top boundary using an additional force term in the momentum equation to suppress vortices generated by the device interaction.  For more details on the numerical wave tank  implementation see our previous works~\cite{Khedkar2021, Khedkar2022}.

\begin{figure}[]
   \centering
   \subfigure[Locally refined Cartesian grid]{
   	\includegraphics[scale= 0.28]{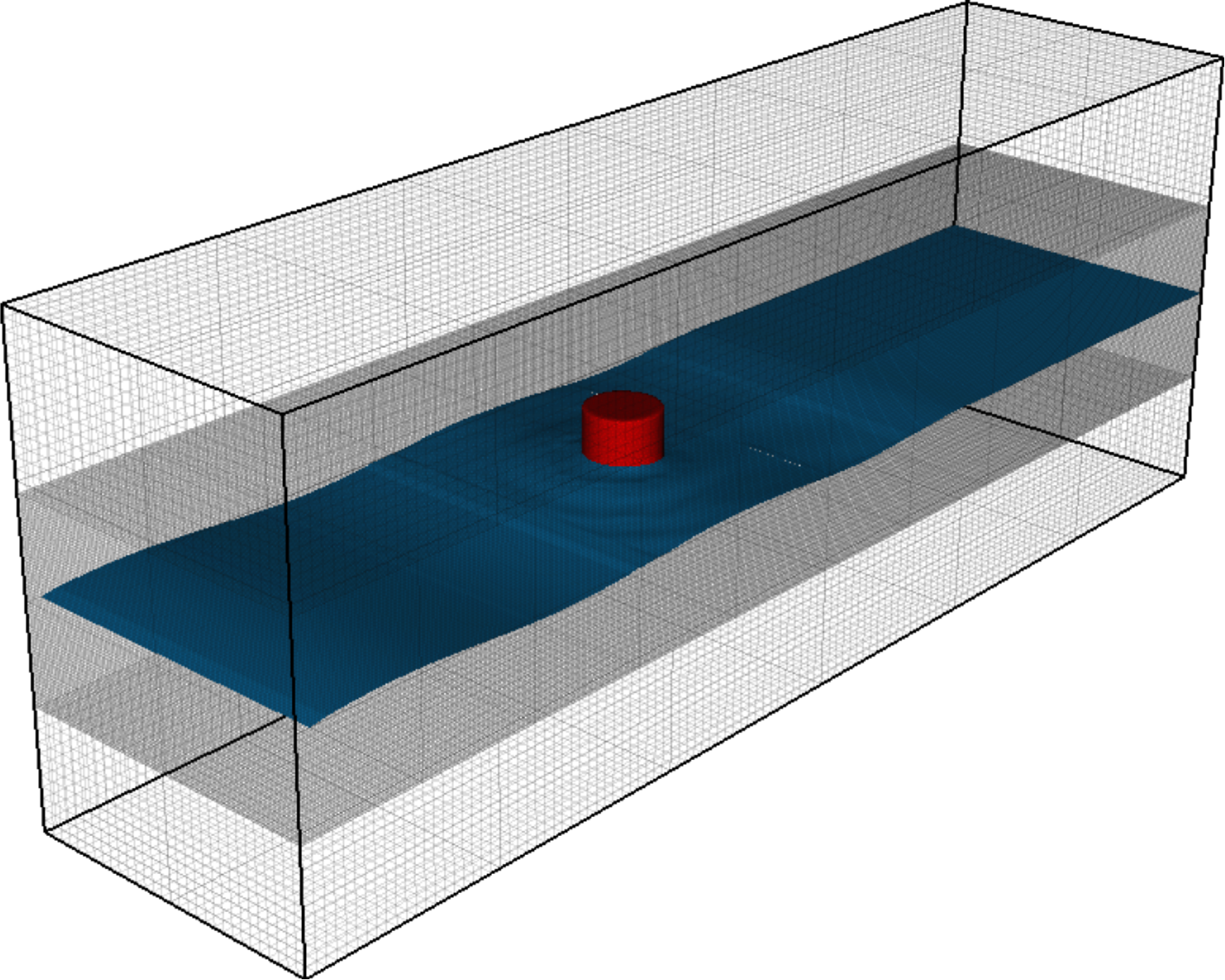}
	\label{fig_vcyl_locally_refined_grid}
   }
      \subfigure[A WEC heaving under the action of regular waves]{
   	\includegraphics[scale= 0.28]{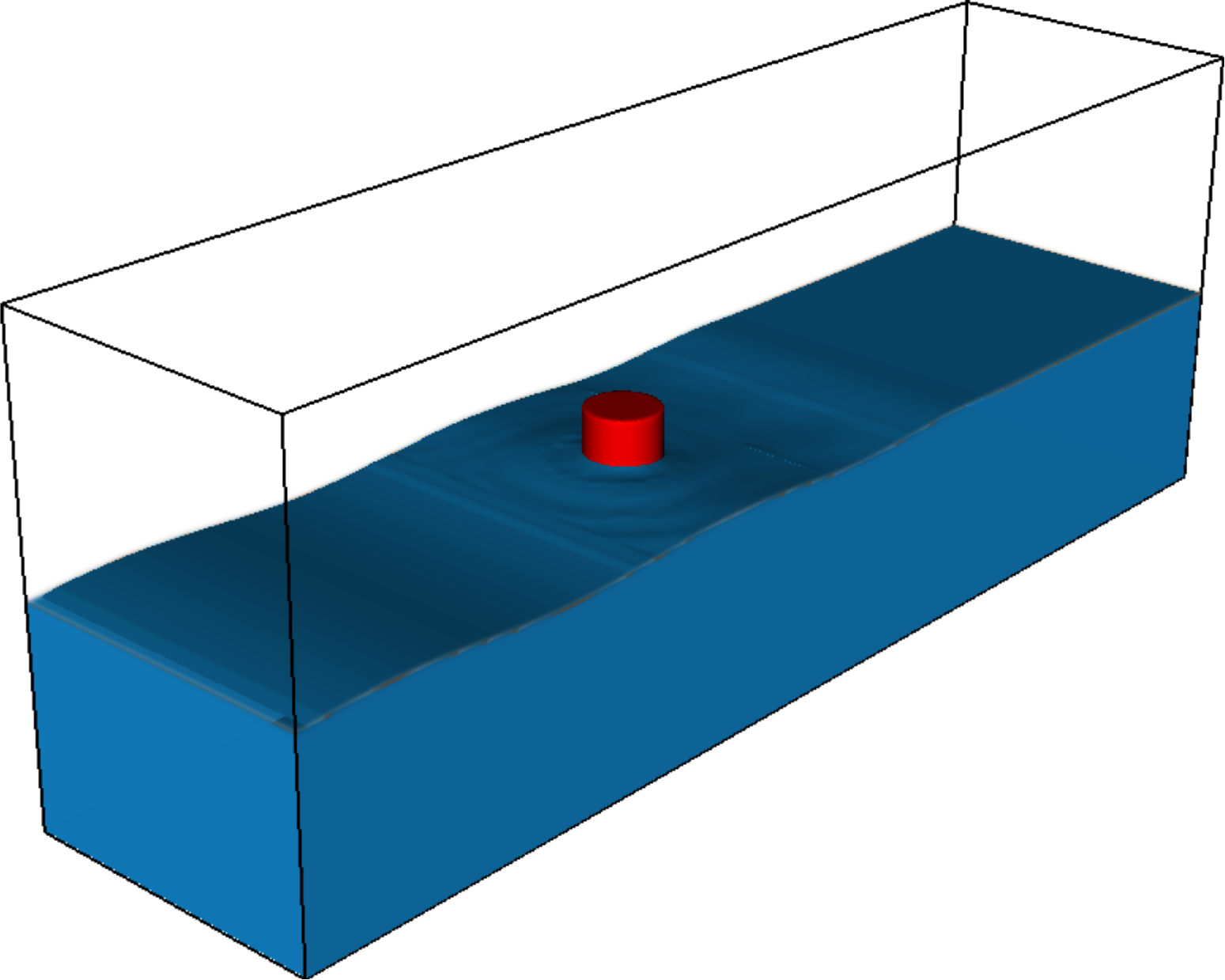}
	\label{fig_vcyl_on_reg_waves}
   }  
    \caption{\subref{fig_vcyl_locally_refined_grid} Locally refined Cartesian grid and \subref{fig_vcyl_on_reg_waves} WSI of a 3D WEC heaving due to regular water waves in the NWT at $t = 12.9$ s.}
   \label{fig_vcyl_refined_grid_WSI}
\end{figure}

\begin{figure}[]
   \centering
   \subfigure[Vertical dispacement vs time]{
   	\includegraphics[scale= 0.38]{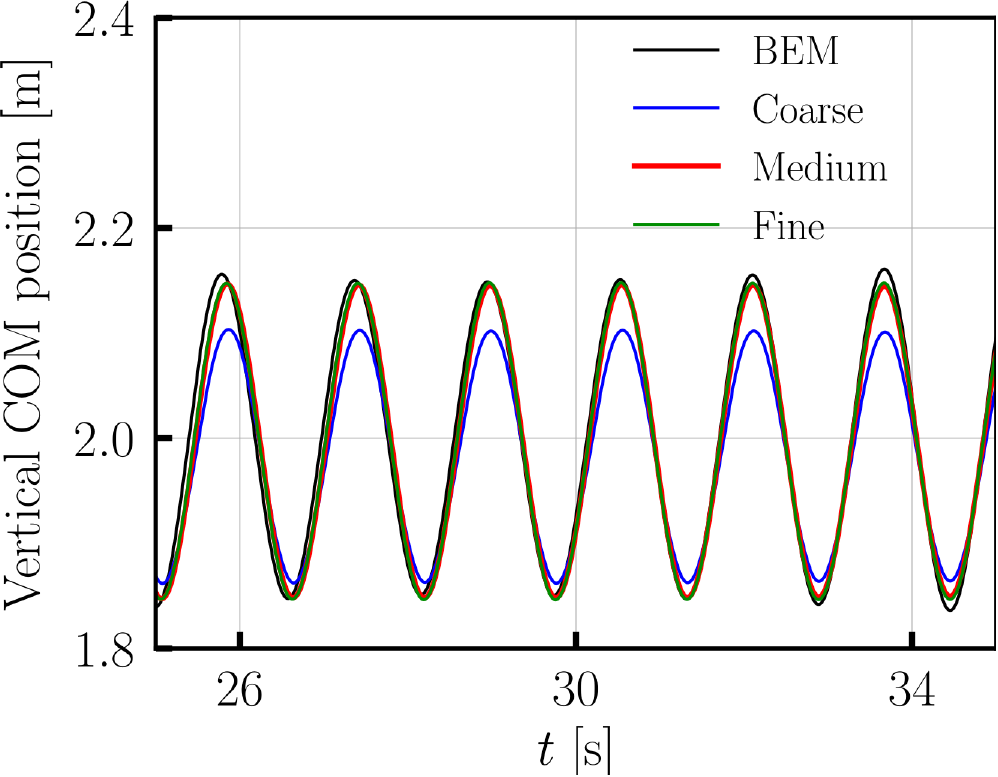}
	\label{fig_vcyl_displacement_grid_refinement}
   }
      \subfigure[Vertical velocity vs time]{
   	\includegraphics[scale= 0.38]{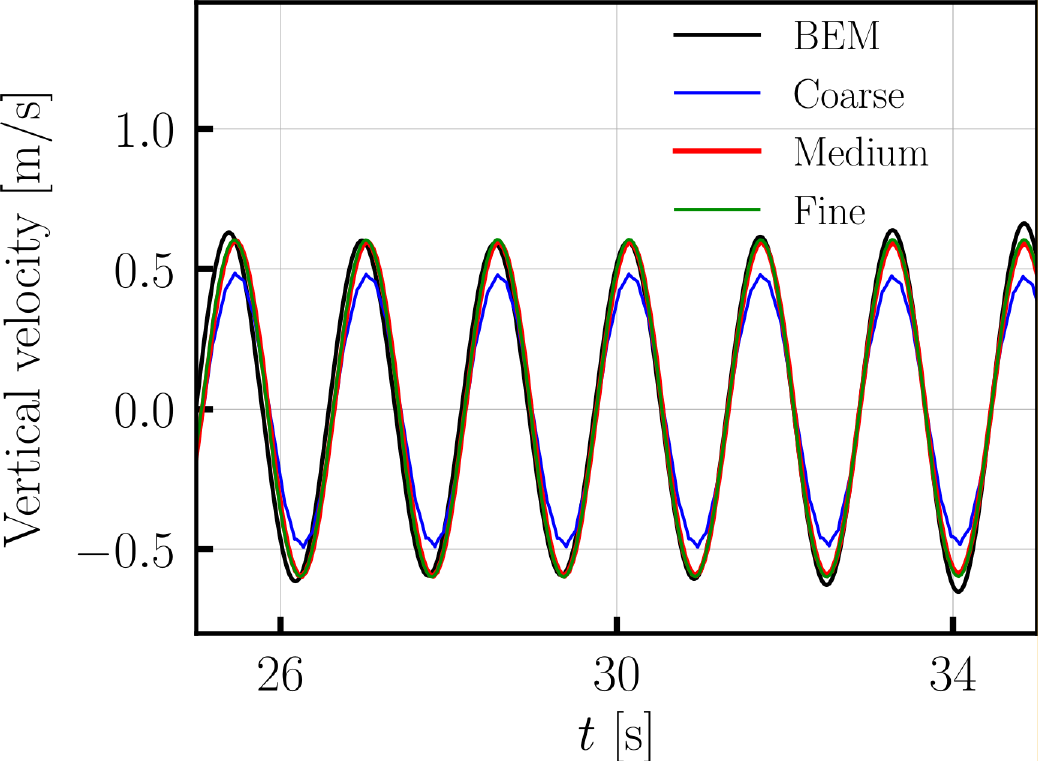}
	\label{fig_vcyl_velocity_grid_refinement}
   }  
    \caption{Temporal evolution of \subref{fig_vcyl_displacement_grid_refinement} vertical displacement and \subref{fig_vcyl_velocity_grid_refinement} vertical velocity of the cylinder for three grid sizes: Coarse, Medium, and Fine on first order water waves of $\mathcal{H} = 0.1$ m and $\mathcal{T}_\text{p} = 1.5652$ s compared with the BEM simulation results.}
   \label{fig_vcyl_refined_grid_reg_waves}
\end{figure}

Initially, the center of mass position of the vertical cylinder is at $(\lambda + 5R_\text{cyl}, d, 6R_\text{cyl})$. The density and viscosity of water are $\rho_\text{w} = 1025$ kg/$\text{m}^3$ and $\mu_\text{w} = 10^{-3}$ Pa$\cdot$s, and that of air are $\rho_\text{a} = 1.225$ kg/$\text{m}^3$ and $\mu_\text{a} = 1.8\times10^{-5}$ Pa$\cdot$s. To accurately resolve the device dynamics, a grid convergence study is conducted. Three grid sizes are considered: coarse, medium, and fine as listed in Table~\ref{tab_vcyl_grid_convergence}. The computational grid is made up of a heirarchy of $\ell$ grid levels. There are $N_x \times N_y  \times N_z$ cells in the entire computational domain $\Omega$ at the coarsest level. The coarsest grid level is then locally refined $(\ell - 1)$ times using an integer refinement ratio of $n_\text{ref}$. This refinement is done so that the air-water interface and WEC device are always on the finest grid level. The grid spacing on the finest level is calculated as $\Delta x = \Delta x_0 / n_\text{ref}^{\ell-1}$, $\Delta y = \Delta y_0 / n_\text{ref}^{\ell-1}$, and $\Delta z = \Delta z_0 / n_\text{ref}^{\ell-1}$, in which $\Delta x_0, \Delta y_0$, and $\Delta z_0$ are the grid spacings in the three directions on the coarsest level. Fig.~\ref{fig_vcyl_locally_refined_grid} shows the locally refined Cartesian grid with two grid levels, and Fig.~\ref{fig_vcyl_on_reg_waves} illustrates the wave structure interaction at a representative time instant $t =12.9$ s. The vertical displacement and velocity evolution of the cylinder for the three grids are compared in Figs.~\ref{fig_vcyl_displacement_grid_refinement} and \ref{fig_vcyl_velocity_grid_refinement}. We compare these solutions to those generated by an in-house boundary element method code. For more details on the BEM solver, see \cite{Khedkar2022}. As can be observed, a medium grid resolution is sufficient to resolve the WSI accurately. Results later in the section are based on medium grid resolution.

\begin{table}[]
 \centering
 \caption{Grid refinement parameters used for the WEC problem.}
 \begin{tabular}{|c|c|c|c|}
  \hline
 Parameters & Coarse & Medium & Fine \\ \hline
  $n_{\text{ref}}$ & 4   & 4   & 4  \\ \hline
  $\ell$         & 2   & 2   & 2         \\ \hline
 $N_{x}$       & 60 & 120 & 180   \\ \hline
 $N_{y}$       & 15 & 30 & 45   \\ \hline
 $N_{z}$       & 22  & 44  & 66   \\ \hline
 $\Delta x_0 = \Delta y_0 = \Delta z_0$ (m)  & 0.2 & 0.1  & 0.0667   \\ \hline
 $\Delta x = \Delta y = \Delta z$ (m)       & 0.05  & 0.025  & 0.0166   \\ \hline
 $\Delta t_\text{max}$ (s)   & $5\times 10^{-3}$ & $2.5\times 10^{-3}$ & $1.5\times 10^{-3}$ \\ \hline
 \end{tabular}
 \label{tab_vcyl_grid_convergence}
\end{table}

\begin{figure}[]
   \centering
   \subfigure[Vertical dispacement vs time]{
   	\includegraphics[scale= 0.38]{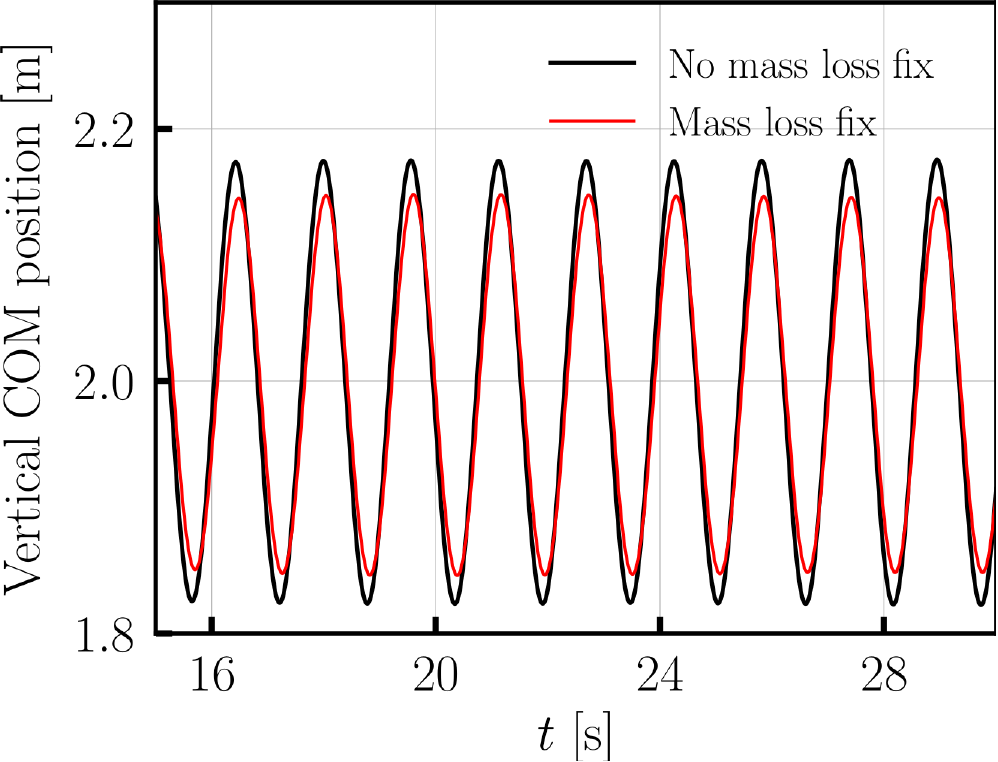}
	\label{fig_vcyl_displacement_mass_comparison}
   }
      \subfigure[Vertical velocity vs time]{
   	\includegraphics[scale= 0.38]{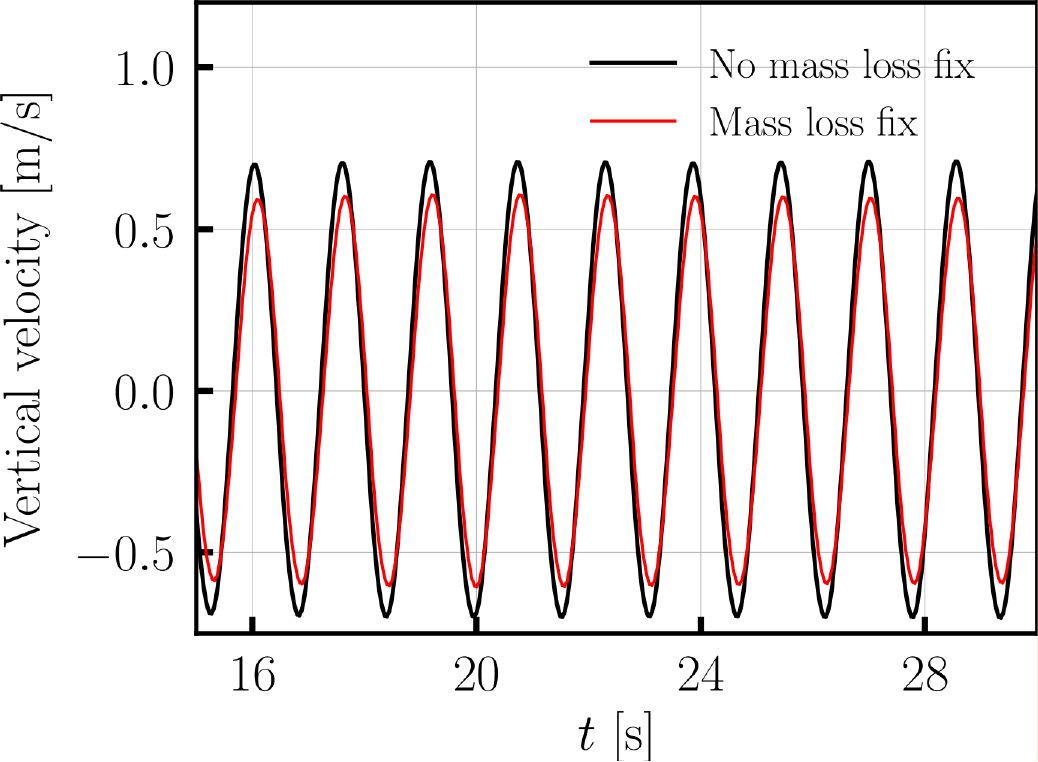}
	\label{fig_vcyl__velocity_mass_comparison}
   }  
    \caption{Temporal evolution of \subref{fig_vcyl_displacement_mass_comparison} vertical displacement and \subref{fig_vcyl__velocity_mass_comparison} vertical velocity of the cylinder with or without the mass loss fix.}
   \label{fig_vcyl_mass_fix_comparison}
\end{figure}

Because the domain is large, we anticipate that the mass loss fix method is not crucial to obtaining the correct WSI dynamics for the WEC device in this problem as well. Figs.~\ref{fig_vcyl_displacement_mass_comparison} and ~\ref{fig_vcyl__velocity_mass_comparison} compare the vertical center of mass position and velocity of the cylinder over time with and without the mass fix. The results confirm that WEC dynamics remain the same with or without the fix. 

\section{Conclusions}
\label{sec_conclusions}
In this work we critically analyzed the reasons for mass loss with the standard level set method. It is primarily due to the use of smoothed Heaviside and delta functions. However, level set reinitialization errors can still contribute to mass loss, and existing fixes can help address these (secondary) issues. To prevent mass loss, we proposed a novel variational approach that introduces a Lagrange multiplier within the standard level set method. This variational analysis was applied to both two-phase (two fluids) and three-phase (two fluids and one solid) flows. The exact Lagrange multiplier ensures mass conservation but disrupts the signed distance property of the level set function. To address this, we developed approximate Lagrange multipliers that achieve both properties. In the context of three-phase flows, we also presented an immersed formulation of the level set equation. This allows us to simulate fluid-structure interaction (FSI) problems using the fictitious domain Brinkman penalization method. It was demonstrated that the differential treatment of the Brinkman penalty leads to incorrect FSI dynamics, wherein the two-fluid interface is artificially repelled by the solid surface.

Mass/volume conservation constraints considered here are in integral form. Eq.~\eqref{eq_2phase_pw_nconsv_proof} suggests that pointwise mass conservation is not possible with the level set technique. This is because there will always be errors associated with $\bu \ne \u$. This means that a signed distance function based on level set reinitialization is fundamentally incompatible with pointwise mass conservation. However, for many practical applications, including those in this work, integral/average mass conservation is sufficient. For problems requiring local mass conservation, a different interface tracking method would be necessary.


\section*{Acknowledgements}

A.P.S.B. acknowledges support from NSF awards OAC 1931368 and CBET CAREER 2234387. N.A.P. acknowledges support from NSF grants OAC 1450374 and 1931372. Compute time on SDSU’s high performance computing cluster Fermi is greatly acknowledged


\appendix
\section{Analytical calculations for the floating rectangular block problem}
\label{sec_analytical_rectangle_calculations}

\begin{figure}[]
   \centering
   	\includegraphics[scale= 0.4]{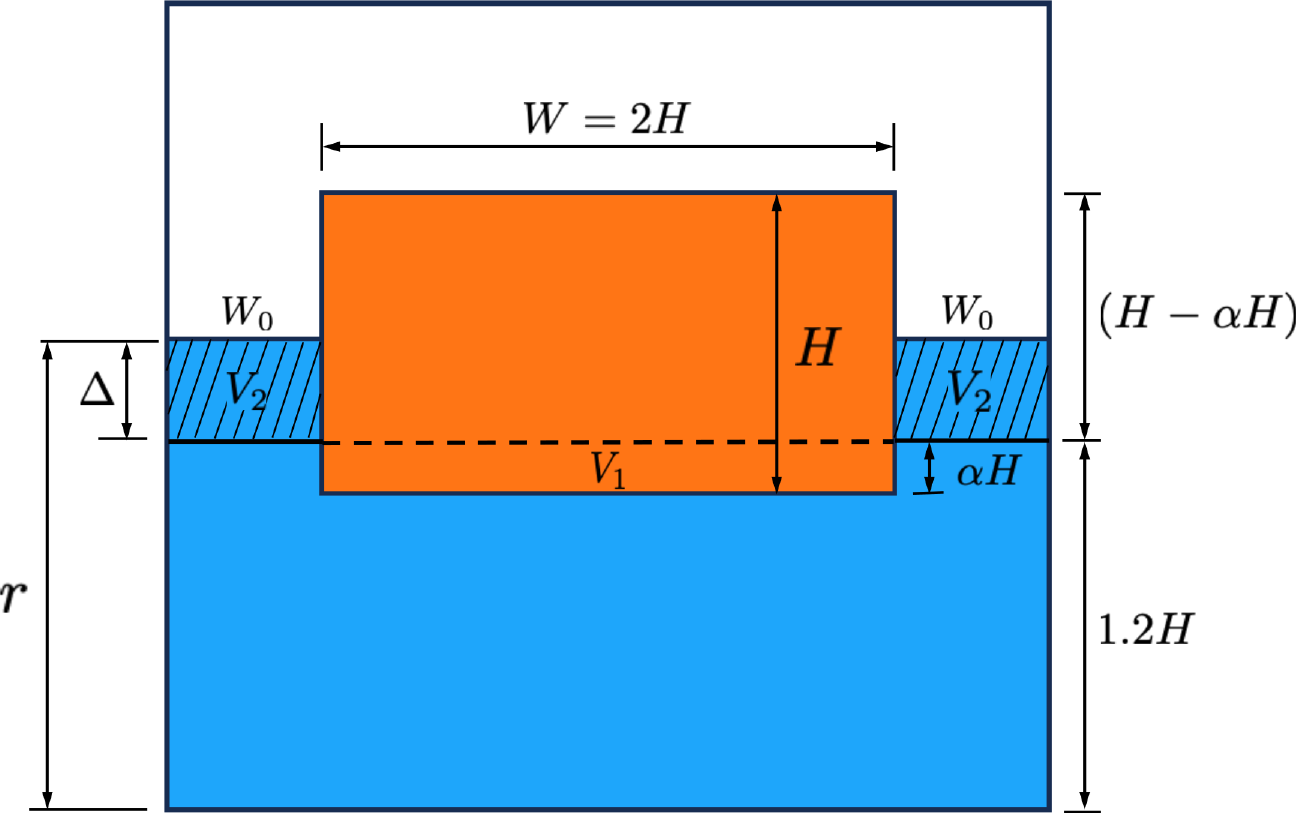}
    \caption{Equilibrium conditions for the floating rectangular block problem.}
   \label{fig_rect_analytical_schematic}
\end{figure}

Consider a rectangular block of dimensions $W \times H = 2H\times H$ released from a small height above the air-water interface as depicted in Fig.~\ref{fig_rect_schematic}. In equilibrium (hydrostatic conditions) the block settles as shown in Fig.~\ref{fig_rect_analytical_schematic}. The water level rises by an amount of $\Delta$  from its initial level. The amount of displaced water is shown by hashed lines in the figure. Its volume ($2V_2$) is equal to the volume of the submerged part of the block ($V_1$) below the initial water level, which is shown by the dashed line in Fig.~\ref{fig_rect_analytical_schematic}. Equating these volumes gives a relation
\begin{align}
  V_1 &= 2 \times V_2 \nonumber \\
	 W \times \alpha  H &= 2\times \Delta \times W_0, \nonumber  \\
	\hookrightarrow \Delta &= \frac{\alpha  H \times W}{2W_0}.
	\label{eq_rect_volume_displaced}
\end{align}
At equilibrium the weight of the rectangle is balanced by the buoyancy force, which yields the condition 
\begin{align}
	\rho_\text{s} \times W \times H &\times g = \rho_\text{w} \times (\Delta + \alpha H) \times W \times g, \nonumber \\
	\hookrightarrow \alpha &= \frac{\rho_\text{s} \times H - \rho_\text{w} \times \Delta}{\rho_\text{w} \times H}.
	\label{eq_rect_mass_displaced}
\end{align}
Substituting Eq.~\eqref{eq_rect_volume_displaced} into Eq.~\eqref{eq_rect_mass_displaced} and rearranging the terms, we get
\begin{align}
	\alpha = \frac{\rho_\text{s}}{\rho_\text{w}\left( 1 + \frac{W}{2W_0}\right)}.
	\label{eq_rect_alpha_value}
\end{align}
For a very wide domain, i.e., when $W_0 \rightarrow \infty$, we get the well-known result 
$ \alpha_\infty = \frac{\rho_\text{s}}{\rho_\text{w}} $ .

Substituting problem specific parameters of Sec.~\ref{subsubsec_floating_rect}, we get $\alpha = 0.1428$ and $\Delta = 0.02678$ m. The position of the new water level from the bottom of the container is $r = \Delta + 1.2H$ = 0.1167 m. 

\section{Analytical calculations for the floating cylinder problem}
\label{sec_analytical_cylinder_calculations}

\begin{figure}[]
   \centering
   	\includegraphics[scale= 0.45]{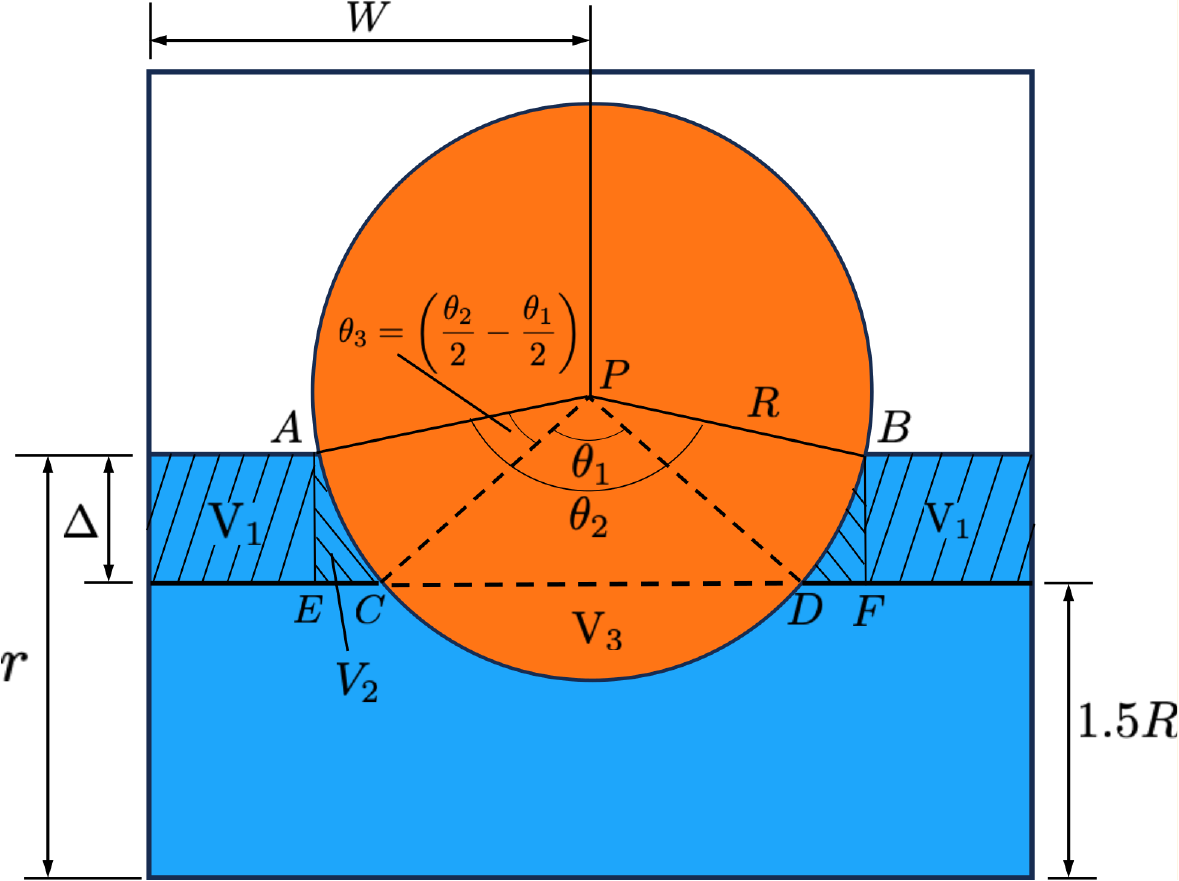}
    \caption{Equilibrium conditions for the floating cylinder problem.}
   \label{fig_cyl_analytical_schematic}
\end{figure}

\begin{figure}[]
   \centering
   	\includegraphics[scale= 0.5]{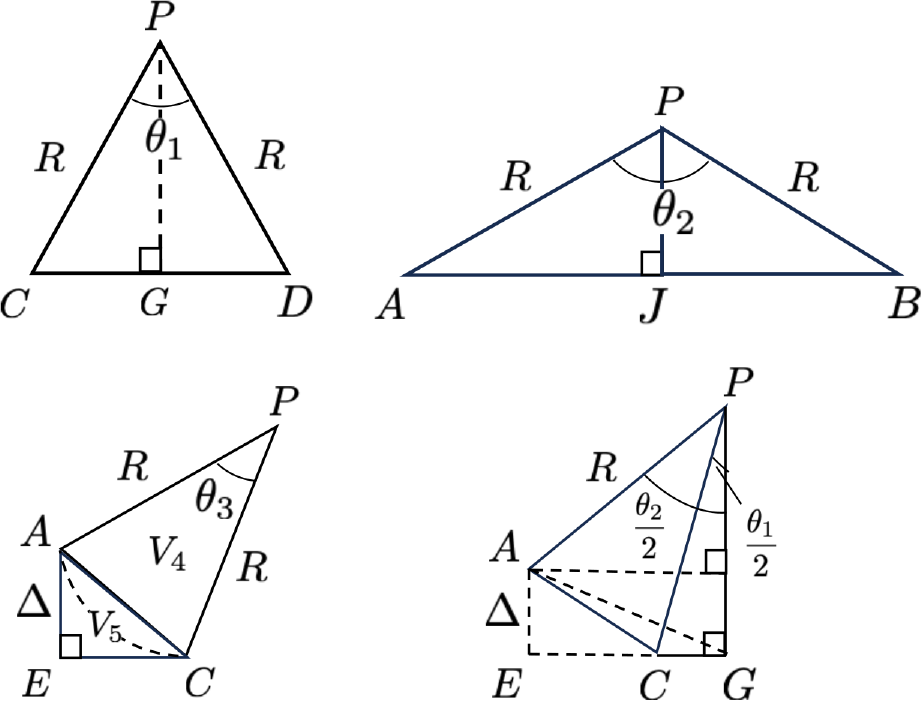}
    \caption{Sub-geometries derived from Fig.~\ref{fig_cyl_analytical_schematic} for the analytical calculation of rise in the water level for the floating cylinder problem.}
   \label{fig_cyl_triangles}
\end{figure}

Consider a cylinder of radius $R$ released from a small height above the air-water interface as depicted in Fig.~\ref{fig_cyl_schematic}. In equilibrium (hydrostatic conditions) the cylinder settles as shown in Fig.~\ref{fig_cyl_analytical_schematic}.  To compute the amount of water level rise $\delta$ we first use the volume conservation principle to equate
\begin{align}
V_3 = 2(V_1 + V_2),
\label{eq_cyl_vol_conserve}
\end{align}
in which, $V_3$ is the volume of the submerged part of the cylinder below the initial water level (horizontal dotted line) as illustrated in Fig.~\ref{fig_cyl_analytical_schematic}. 
At equilibrium, the weight of the cylinder is balanced by the buoyancy force, which gives the condition
\begin{align}
	\rho_\text{s} \times \pi R^2 \times g &= \left[ \left(\frac{\theta_2}{2\pi}\right) \times \pi R^2 - \frac{R^2}{2} \times \sin \theta_2 \right] \times \rho_\text{w} g \nonumber \\
	\hookrightarrow \theta_2 - \sin \theta_2 &= \frac{2\pi \rho_\text{s}}{\rho_\text{w}}.
	\label{eq_beta}
\end{align}

The following geometric relations are derived using Fig.~\ref{fig_cyl_triangles}, which are used to calculate volume $V_3$,  
\begin{align}
	CG &= R\times \sin\left( \frac{\theta_1}{2}\right), \nonumber \\
	PG &= R\times \cos\left( \frac{\theta_1}{2}\right), \nonumber \\
	\hookrightarrow V_3 &= \frac{\theta_1}{2\pi} \times \pi R^2 - \frac{1}{2} \times (2 \times CG) \times PG \nonumber \\
    &= \frac{R^2}{2} \left( \theta_1 - \sin \theta_1 \right), 
	\label{eq_V3}
\end{align}
volume $V_1$
\begin{align}
	AJ &= R\sin\left( \frac{\theta_2}{2}\right), \nonumber \\
\hookrightarrow	V_1 &= \left(W - R\sin\left( \frac{\theta_2}{2}\right) \right)\delta,
	\label{eq_V1}
\end{align}
and volume $V_2$
\begin{align}
	V_4 &= \frac{R^2}{2} \times \sin\theta_3, \\
	V_5 &= \frac{1}{2}\times \Delta \times EC, \\
	EC &= \left(W - R\sin \left(\frac{\theta_1}{2}\right)\right) - \left(W - R\sin \left(\frac{\theta_2}{2}\right)\right) \nonumber \\ 
       &= R\left( \sin\left( \frac{\theta_2}{2}\right) - \sin \left( \frac{\theta_1}{2}\right)\right), \label{eq_EC} \\
   \theta_3 &= \frac{1}{2} (\theta_2 - \theta_1), \label{eq_theta3} \\
	\hookrightarrow V_2 &= V_4 + V_5 - \left( \frac{\theta_3}{2\pi}\right)\times \pi R^2 \nonumber \\
	 &= \frac{R^2}{2} \times \sin \theta_3 + \frac{1}{2} \times \Delta \times EC - \frac{\theta_3 R^2}{2}.
	\label{eq_V2}
\end{align}
The rise in the water level can be expressed in terms of the angles $\theta_1$ and $\theta_2$ as
\begin{align}
	\Delta = R\cos \left( \frac{\theta_1}{2}\right) - R \cos \left( \frac{\theta_2}{2}\right).
	\label{eq_delta}
\end{align}
Finally, Eq.~\eqref{eq_cyl_vol_conserve} simplifies to yield
\begin{align}
	V_3 &= 2(\text{V}_1 + \text{V}_2) \nonumber \\
	\hookrightarrow \frac{R^2}{2}(\theta_1 - \sin \theta_1) &= 2 \left[\left(W - R\sin\left( \frac{\theta_2}{2}\right) \right)\Delta + \frac{R^2}{2} \sin \theta_3 + \frac{\Delta \times EC}{2} - \frac{\theta_3 R^2}{2} \right].
	\label{eq_theta1}
\end{align}

Substituting \eqref{eq_beta}, \eqref{eq_EC}, \eqref{eq_theta3} and \eqref{eq_delta} into Eq.~\eqref{eq_theta1} along with the problem specific parameters of Sec.~\ref{subsec_floating_cylinder}, we get $\theta_1 = 126.66^\circ$, $\Delta$ = 0.02692 m and $r = 0.1169$ m.

\section{Bibliography}
\begin{flushleft}
 \bibliography{LS_mass_fix}
\end{flushleft}

\end{document}